\documentclass[prc,aps,showpacs,floatfix]{revtex4}
\usepackage{graphicx}
\begin{document}

\title{Extended Partial-Wave Analysis of $\pi N$ 
       Scattering Data}

\author{R.~A.~Arndt}
\author{W.~J.~Briscoe}
\author{I.~I.~Strakovsky}
\author{R.~L.~Workman}
\affiliation{Center for Nuclear Studies, Department of
        Physics, \\
        The George Washington University, Washington,
        D.C. 20052, U.S.A.}

\begin{abstract}
We present results from a comprehensive partial-wave 
analysis of $\pi^\pm p$ elastic scattering and 
charge-exchange data, covering the region from threshold 
to 2.6~GeV in the lab pion kinetic energy, employing a 
coupled-channel formalism to simultaneously fit 
$\pi^-p\to\eta n$ data to 0.8~GeV.  Our main result, 
solution SP06, utilizes a complete set of forward and 
fixed-t dispersion relation constraints applied 
to the $\pi N$ elastic amplitude.  The results of these 
analyses are compared with previous solutions in terms 
of their resonance spectra and preferred values for 
couplings and low-energy parameters.
\end{abstract}

\maketitle

\section{Introduction}
\label{sec:intro}

Most $N$ and $\Delta$ resonances, listed as 3- and 4-star
states in the Review of Particle Properties (RPP)
~\cite{rpp}, have had their existence, masses, and widths 
determined through single-channel fits to scattering data, 
with $\pi N$ elastic scattering being the predominant 
source. The most comprehensive $\pi N$ analyses have been
performed by the Karlsruhe-Helsinki (KH)~\cite{kh80}, 
Carnegie-Mellon$-$Berkeley (CMB)~\cite{cmb}, and George 
Washington (GW)~\cite{fa02} groups. 

All of these studies essentially agree on the existence 
and (most) properties of the 4-star states. For the 3-star 
and lower states, however, even a statement of existence 
is problematic. Many states claimed in the KH and CMB 
fits have not been found in recent GW analyses.  This 
discrepancy clearly impacts the ``missing resonance" 
problem, which has more quark model states predicted 
than observed. If many 3-star and lower-rated states are 
not observed in $\pi N$ scattering data (where they were 
first identified) then many more states are either 
``missing" or weakly coupled to the $\pi N$ channel. 

These problems have motivated a reexamination of the KH 
analysis~\cite{KHnew} and further improvements to the 
ongoing GW studies.  We have recently added data from the 
reaction $\pi N\to\eta N$, in order to better describe 
the $\pi N$ S-wave and the $N(1535)S_{11}$ resonance
which have a significant coupling to the $\eta N$ final 
state~\cite{km_paper}. In the present study, we have 
extended the energy upper limit from 2.1~GeV to 2.6~GeV, 
in the lab pion kinetic energy, in order to cover the 
resonance region more completely. This extended energy 
range will be carried over to our fits of pion 
photoproduction and electroproduction data, which are 
parameterized in terms of the $\pi N$ scattering 
amplitudes. The extended energy range for photoproduction 
should allow us to fit all single-pion photoproduction 
data expected from the present generation of Jefferson 
Lab experiments. 

A description of the coupled-channel analysis of $\pi N$ 
elastic and $\eta N$ production data, constrained by 
dispersion relations, is given in Ref.~\cite{fa02} and 
will not be repeated here.  In this report, we will 
concentrate on new features seen over our extended energy 
region and changes to the fit below 2.1~GeV. These are
discussed in Section~\ref{sec:output}. Changes to 
the database are described in Section~\ref{sec:expt}. We 
have mainly added data from 2.1 to 2.6~GeV, but have 
also included new measurements at very low energies. 
Some of these require special attention. Finally, in 
Section~\ref{sec:conc}, we summarize our results.

\section{Database}
\label{sec:expt}

The first two decades (1957 through 1979) of experiments
focused on the $\pi N$ system and non-strange baryon 
resonances produced a large amount of data below 2.6~GeV 
(9932 $\pi^+p$, 9637 $\pi^-p$, and 1569 charge-exchange 
data).  These data were used in the canonical KH~\cite{kh80} 
and CMB~\cite{cmb} analyses.  In the present study, we 
have fitted 13344 $\pi^+p$, 11967 $\pi^-p$, 2933 
charge-exchange, and 257 $\eta$-production data.  This 
increase is primarily due to a second generation of $\pi N$ 
measurements (both unpolarized and polarized) carried out
at high-intensity facilities such as LAMPF, TRIUMF, and PSI 
(former SIN).  These more recent measurements generally 
have small statistical and systematic uncertainties 
and, therefore, have a significant influence on fits to the 
full database.

The evolution of our database is summarized in 
Table~\ref{tbl1}.  Over the course of five previous 
pion-nucleon analyses~\cite{fa84,sm90,fa93,sm95,fa02}, our 
energy range was extended from 1.1 to 2.1~GeV in the lab 
pion kinetic energy.  Here we have incorporated missed 
measurements below 2.1~GeV and the existing database to
2.6~GeV (the $\eta$ production database was not extended) 
using the Durham RAL Database~\cite{hep}. 

Below, we list recent (post-2003) additions below 2~GeV 
for elastic scattering, charge-exchange scattering, and 
$\eta$-production.  As in previous fits, not all of the
available data have been used.  Some data with very large 
$\chi^2$ contributions have been excluded from our fits. 
Redundant data are also excluded. These include total 
elastic cross sections based on differential cross 
sections already contained in the database.  
Measurements of $P$ with uncertainties more than 
0.2 are not included as they have little influence in our fits.  
However, 
all available data have been retained in the database  
(the excluded data labeled as ``flagged"~\cite{flag}) so 
that comparisons can be made through our on-line 
facility~\cite{said}.  Some of the data, listed as new, 
were available in unpublished form at the time of our 
previous analysis~\cite{fa02}.  A complete description of 
the database and those data not included in our fits is 
available from the authors~\cite{said}.  

Most recent $\pi^\pm$p measurements have been performed 
at low energies, TRIUMF being the main source.  From 
this laboratory, we have added 274 $\pi^+p$ and 271 
$\pi^-p$ differential cross sections from 20 to 40~MeV.  
These data cover a broad angular range from 10$^\circ$ 
(including the Coulomb-nuclear interference region) to 
170$^\circ$~\cite{de06} and have allowed  us to extend 
our single-energy fits to very low energies (20~MeV) for 
the first time (see Table~\ref{tbl2}).  

\begin{figure}[th]
\centering{
\includegraphics[height=0.4\textwidth, angle=90]{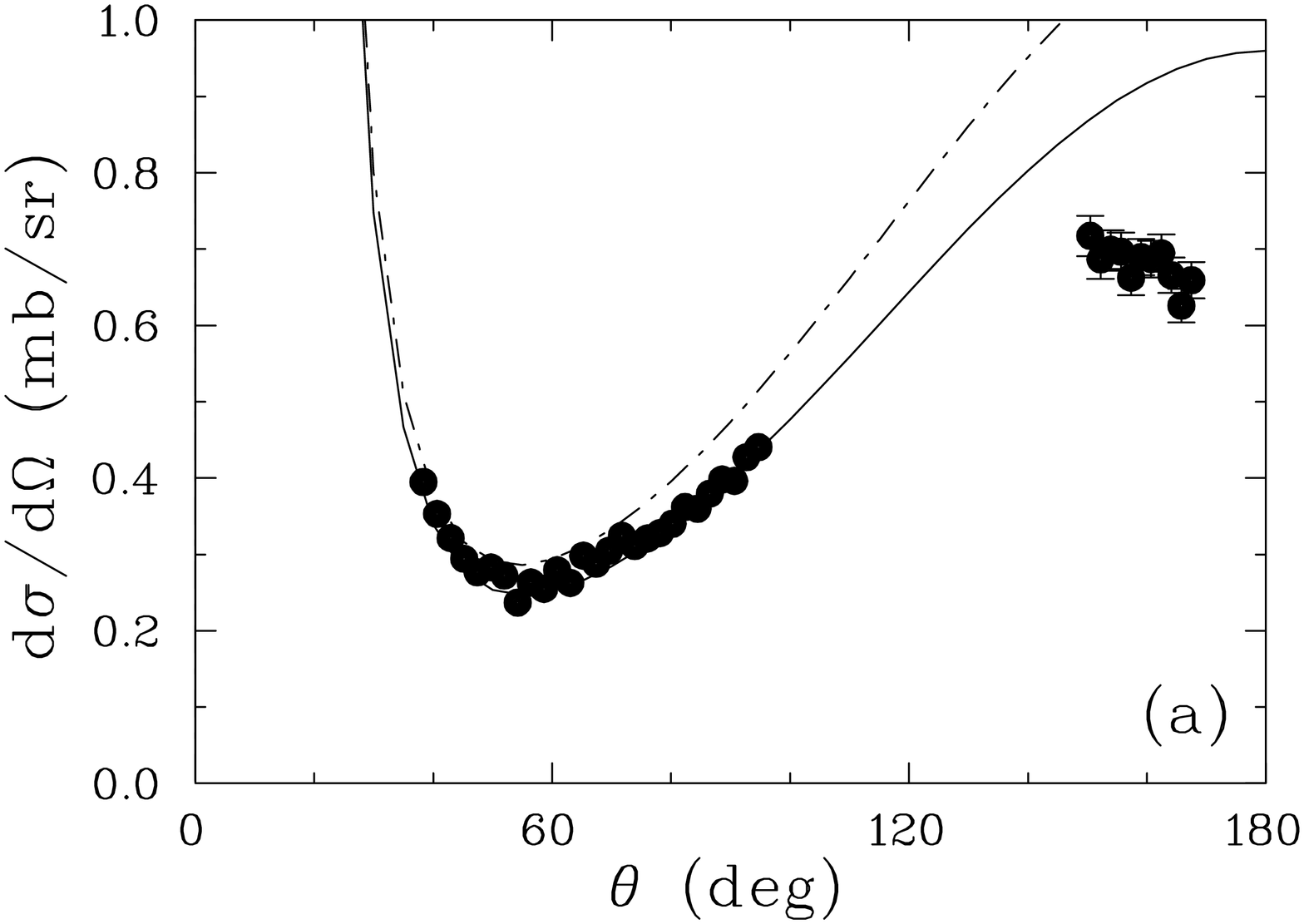}\hfill
\includegraphics[height=0.4\textwidth, angle=90]{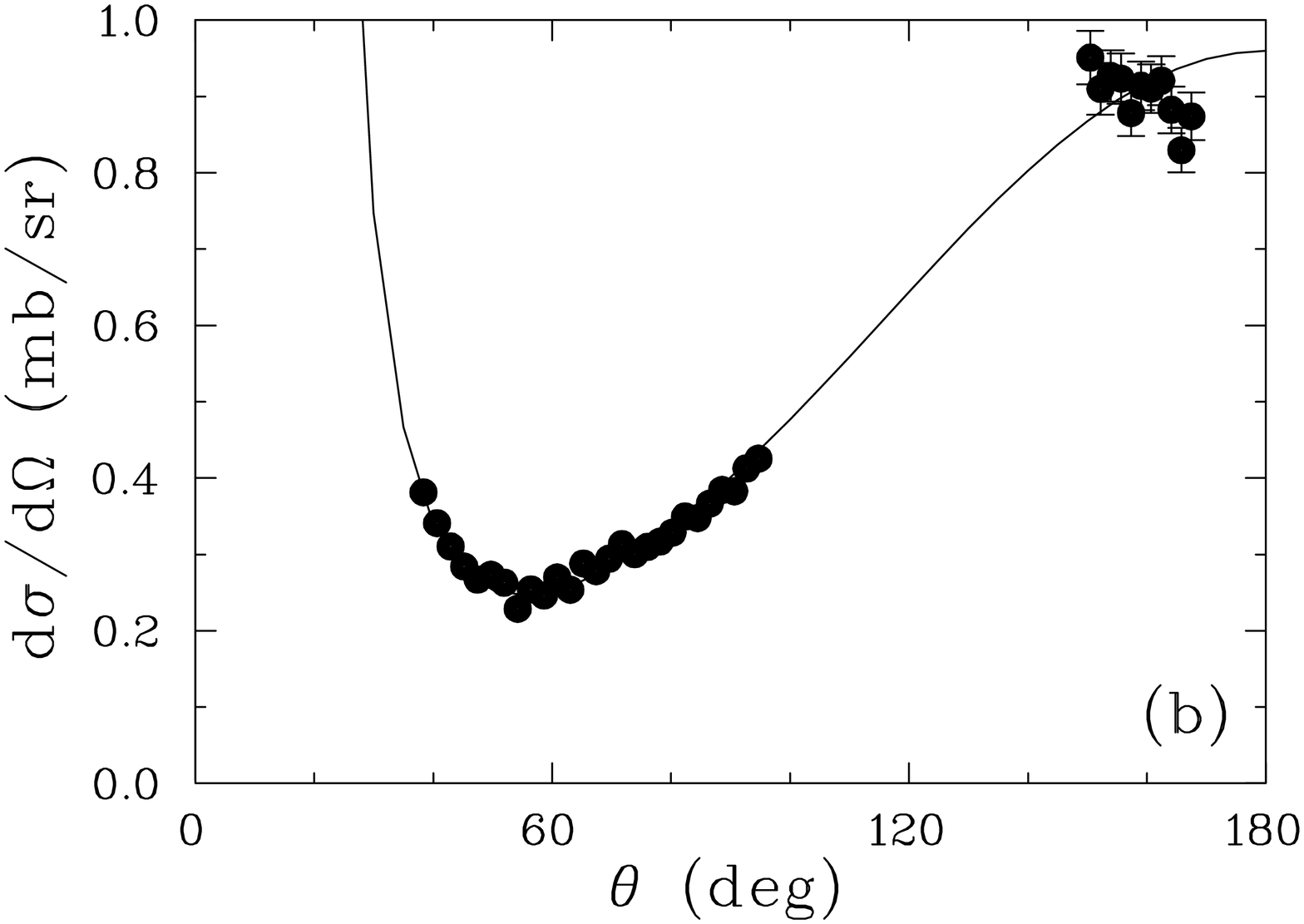}
}\caption{Differential cross sections for $\pi^+p$
          elastic scattering at 26~MeV:  (a)
          unnormalized and (b) normalized data.  
          The Karlsruhe KA84 prediction~\protect\cite{kh80} 
          is plotted as a dot-dashed line.  Data 
          are taken from Ref.~\protect\cite{de06}. 
          \label{fig:g1}}
\end{figure}

The TRIUMF cross sections for $\pi^\pm$ elastic scattering 
were measured simultaneously over the full angular range 
using the CHAOS facility. At low energies, however, the 
foward (backward) cross sections are determined from 
measurements of the charged pion (proton). We mention this 
because the full angular range is difficult to fit with a 
single systematic uncertainty.  The backward angle data 
disagree with both the KH predictions and predictions 
based on our FA02 solution. Including these data in our 
fit did not solve the problem, as can be seen in 
Fig.~\ref{fig:g1}(a).  To resolve the conflict between 
forward, medium and backward scattering measurements, we 
divided the data into two or three pieces and treated 
them independently [Fig.~\ref{fig:g1}(b)].  Clearly, the 
angular dependence at backward angles is not reproduced 
by SP06, nor was it reproduced by our single-energy fit. 
The reason for this conflict is unclear.

Further low-energy additions include 25 $\pi^+p$ and 3 
$\pi^-p$ $A_y$ data between 50 and 130~MeV, at medium 
scattering angles, measured at PSI~\cite{me04}.  New 
total cross sections for charge-exchange measurements 
between 40 and 250~MeV came from PSI recently~\cite{psi}.  
They have very little effect and seem quite well fitted 
by SP06 without any adjustment.  Two BNL--AGS experiments 
from the Crystal Ball Collaboration have also been 
analyzed and added to our database.  These include 648 
charge-exchange data between 520 and 620~MeV~\cite{st05} 
and 84 $\eta$-production data between 560 and 
620~MeV~\cite{pr05}.  The angular coverage was 30 to 
160$^\circ$ in both cases.  Results based on the 
inclusion of these $\eta$-production data are given in 
Ref.~\cite{km_paper}.

\begin{figure}[th]
\centering{
\includegraphics[height=0.5\textwidth, angle=90]{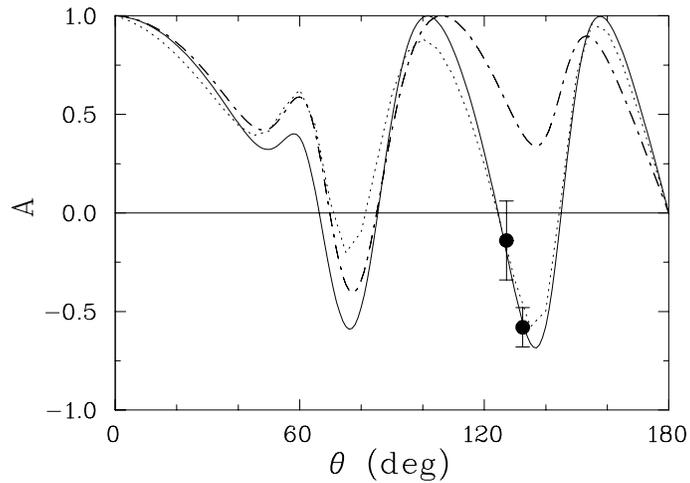}
}\caption{Spin-rotation parameter $A$ for 
          $\pi^+p$.  The original KA84 
          solution~\protect\cite{kh80}
          (dot-dashed line) compared to a
          Barrelet-transformed solution
          ~\protect\cite{alekseev} (dotted line)
          and our SP06 solution (solid line).
          Data are taken from Ref.
          ~\protect\cite{al95}. \label{fig:g2}}
\end{figure}

Finally, ITEP--PNPI $\pi^-p$ experiments have provided 3 
$P$ and 3 $A$ measurements at 1300~MeV in the backward 
direction~\cite{al05}.  Previous measurements of the 
$\pi^+p$ spin-rotation parameter $A$, by the same 
collaboration~\cite{al95}, allowed us to resolve a 
discrepancy between the GW and the CMB/KH predictions 
(Fig.~\ref{fig:g2}), using the method of Barrelet. These 
new measurements agree with predictions from our older 
FA02 and SM95 solutions.

\section{Results and Discussion}
\label{sec:output}

\subsection{SP06 versus the FA02 and KH fits}
\label{sec:sp06}

The main result of this work is an energy-dependent 
solution (SP06), fitting data from threshold to 2.6~GeV, 
and a set of single-energy solutions (SES) ranging from 
20~MeV to 2.575~GeV. Our present and previous 
energy-dependent solutions are compared in 
Table~\ref{tbl1}.  Results from the KH solutions are 
listed here as well. A comparison of SP06 and our 
previous solution FA02, up to the energy limit of FA02, 
shows that a fit to higher energies is possible 
without degrading the description of data below 2.1~GeV. 

As in previous analyses, we have used the systematic 
uncertainty as an overall normalization factor for 
angular distributions.  With each angular distribution, 
we associate the pair $(X,\epsilon_X)$: a normalization 
constant $(X)$ and its uncertainty $(\epsilon_X)$.  The 
quantity $\epsilon_X$ is generally associated with the 
systematic uncertainty (if known.)  The modified 
$\chi^2$ function, to be minimized, is then given by
\begin{eqnarray}
\chi^2 = \sum_i\left(
{{X\theta_i - \theta^{\rm exp}_i}\over{\epsilon_i}}
\right)^2 + \left( {{X-1}\over{\epsilon_X}}
\right)^2,
\label{3}\end{eqnarray}
where the subscript $i$ labels data points within the
distribution, $\theta^{\rm exp}_i$ is an individual
measurement, $\theta_i$ is the calculated value, and
$\epsilon_i$ is the statistical uncertainty.  For total 
cross sections and excitation data, we have combined 
statistical and systematic uncertainties in quadrature.  

Renormalization freedom significantly improves
our best-fit results, as shown in Table~\ref{tbl3}. 
This renormalization procedure was also applied
to the KH solutions. Here, however, only the normalization
constants were searched to minimize $\chi^2$ (no adjustment
of the partial waves was possible).  
In cases where the systematic uncertainty varies with 
angle, this procedure may be considered a first 
approximation.  Clearly, this procedure can significantly 
improve the overall $\chi^2$ attributed to a fit, and has 
been applied in calculating the $\chi^2$ values of 
Table~\ref{tbl1}. 

\begin{figure}[th]
\centering{         
\includegraphics[height=0.4\textwidth, angle=90]{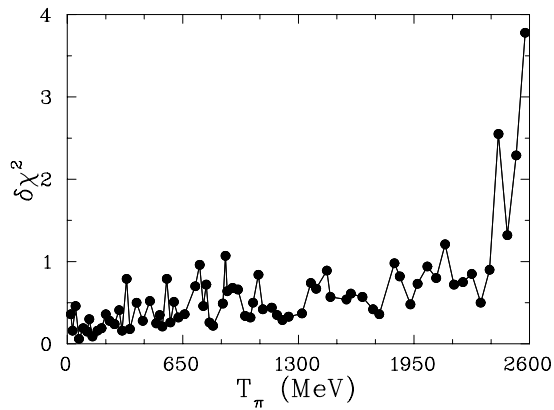}
}\caption{Comparison of the SES and 
          global SP06 fits via $\delta\chi^2 
          = {[}\chi^2(SP06)-\chi^2(SES){]}$/data 
          presented in Table~\protect\ref{tbl2}. 
          \label{fig:g3}} 
\end{figure}

In Table~\ref{tbl2}, we compare the energy-dependent and 
SES results over the energy bins used in each 
single-energy analysis. The quantity $\delta\chi^2$ 
computes ${[}\chi^2(SP06)-\chi^2(SES){]}$ divided by the
number of data in each single-energy bin, providing a
measure of the agreement between an individual SES and 
the global SP06 results (see Fig.~\ref{fig:g3}).  Also 
listed is the number of parameters varied in each SES. 
As was emphasized in Ref.~\cite{fa02}, the SES are 
generated mainly to search for missing structures in 
the global fit. 

\begin{figure}[th]
\centering{
\includegraphics[height=0.33\textwidth, angle=90]{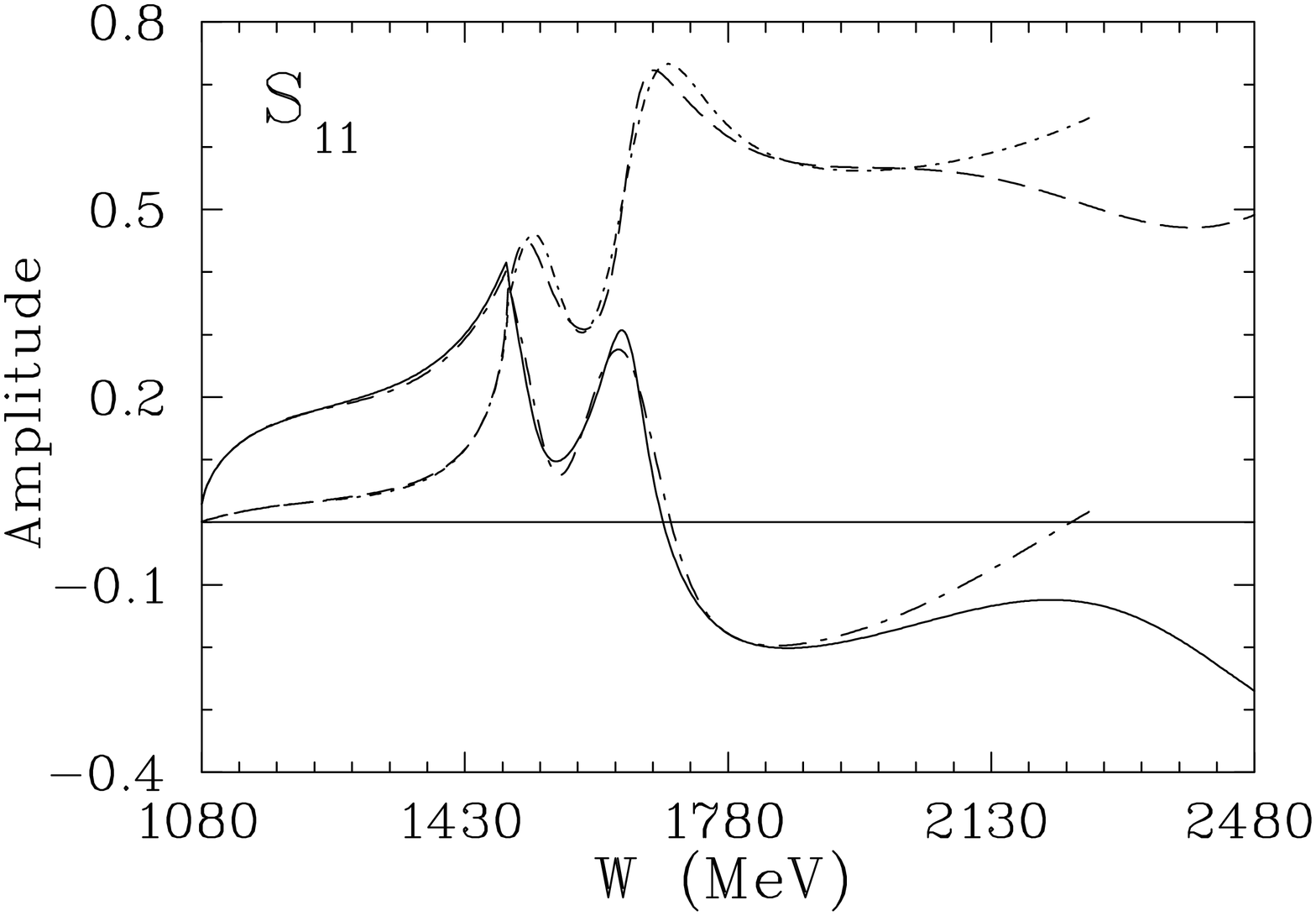}\hfill
\includegraphics[height=0.33\textwidth, angle=90]{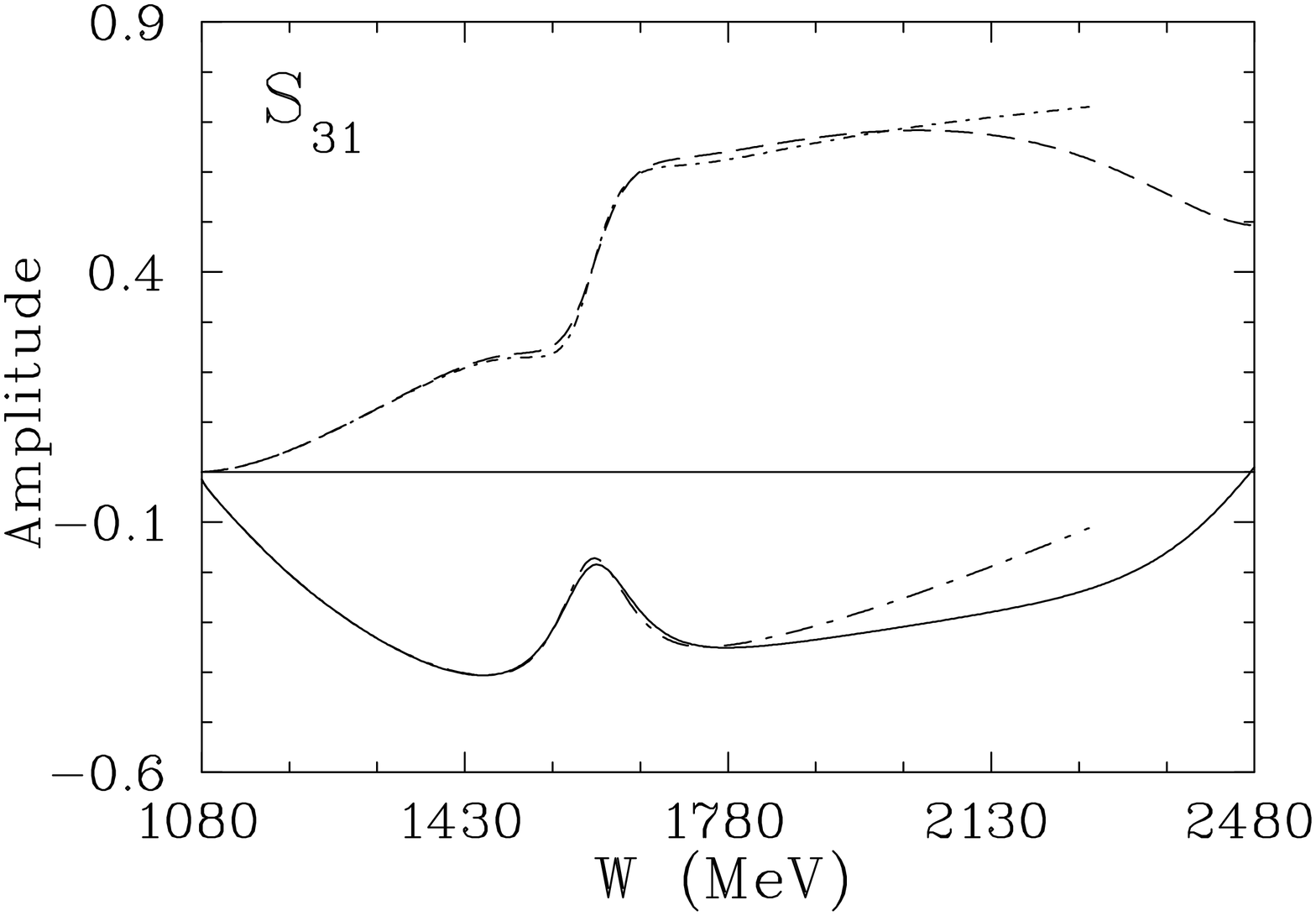}
\includegraphics[height=0.33\textwidth, angle=90]{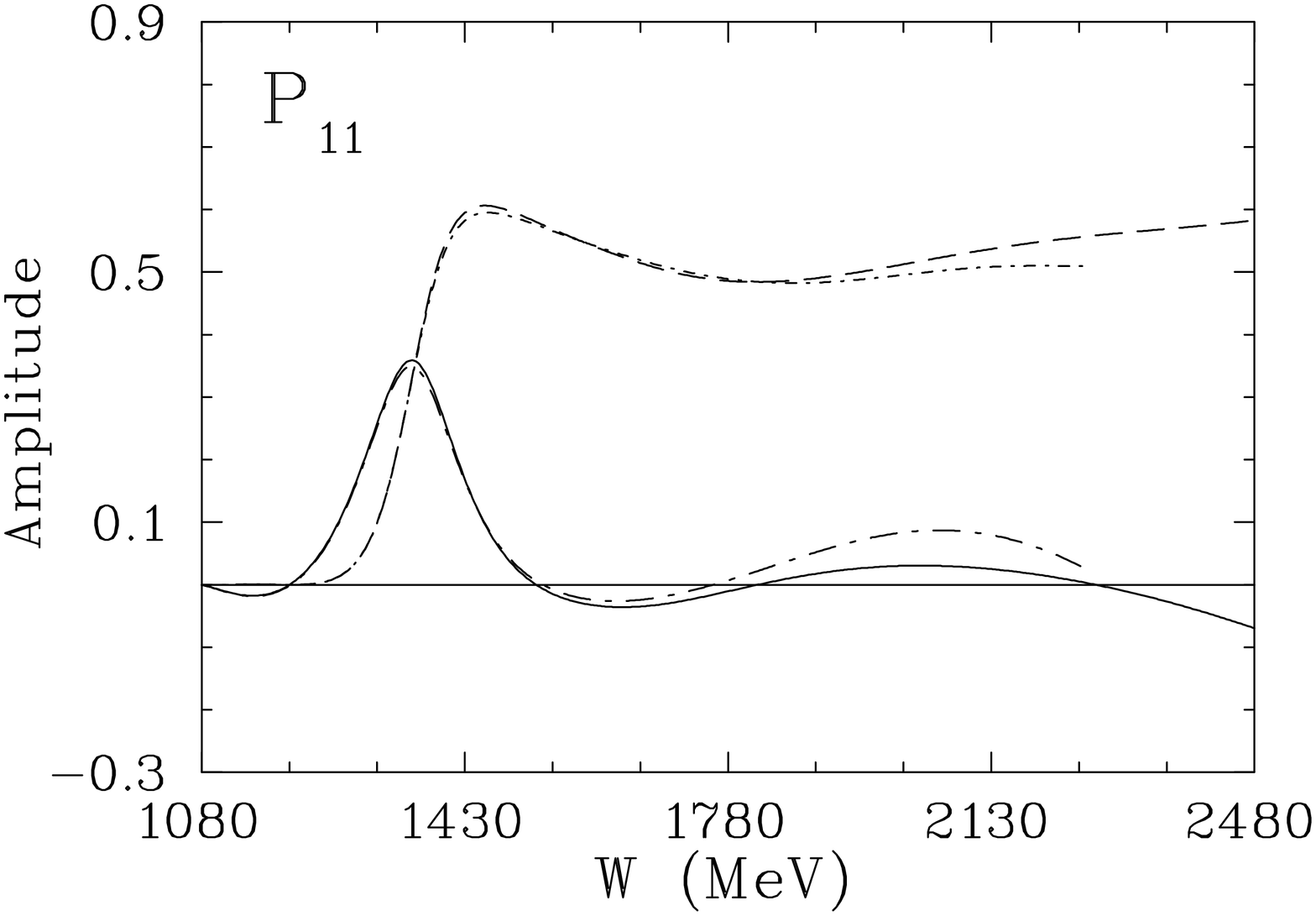}\hfill
\includegraphics[height=0.33\textwidth, angle=90]{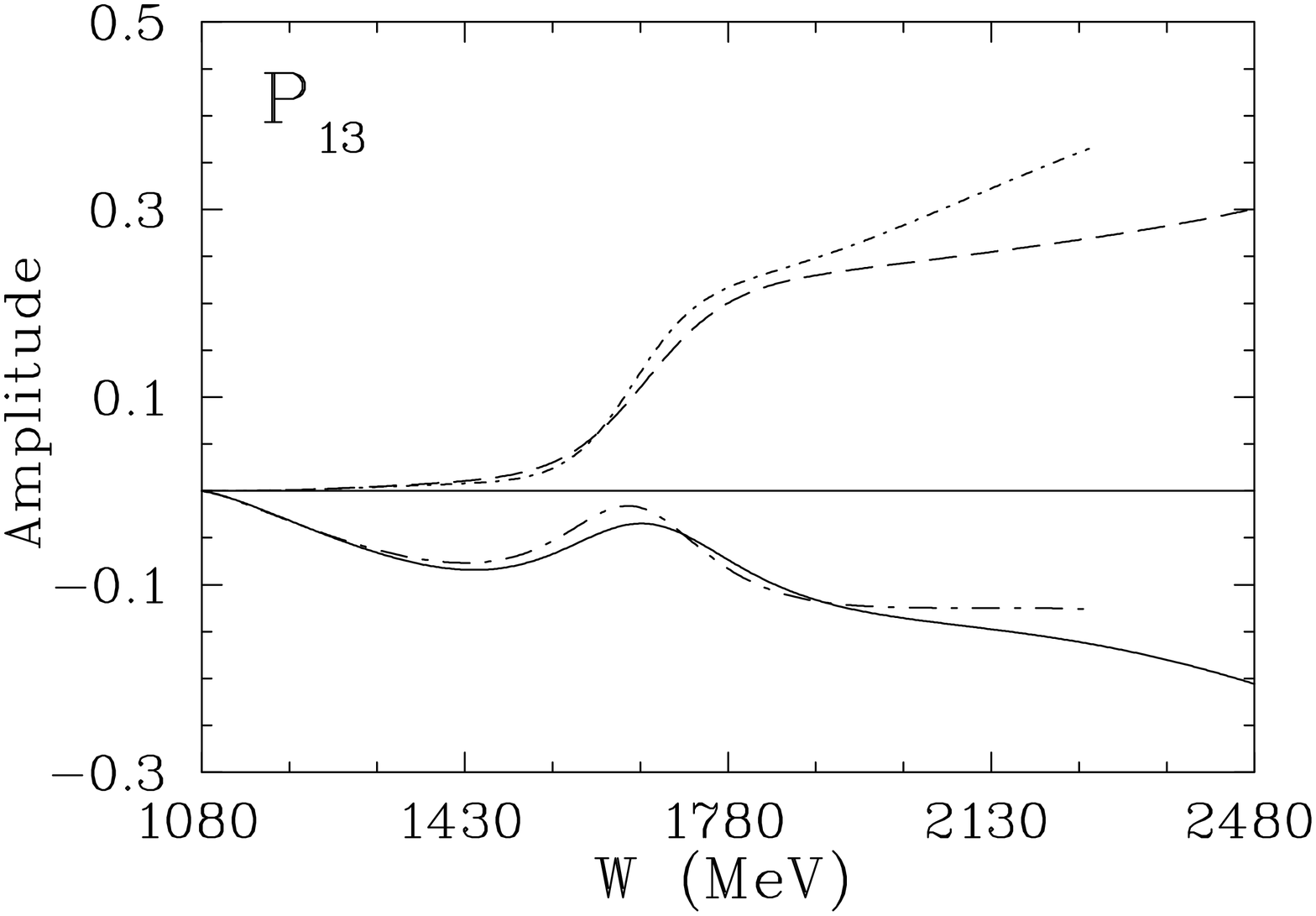}
\includegraphics[height=0.33\textwidth, angle=90]{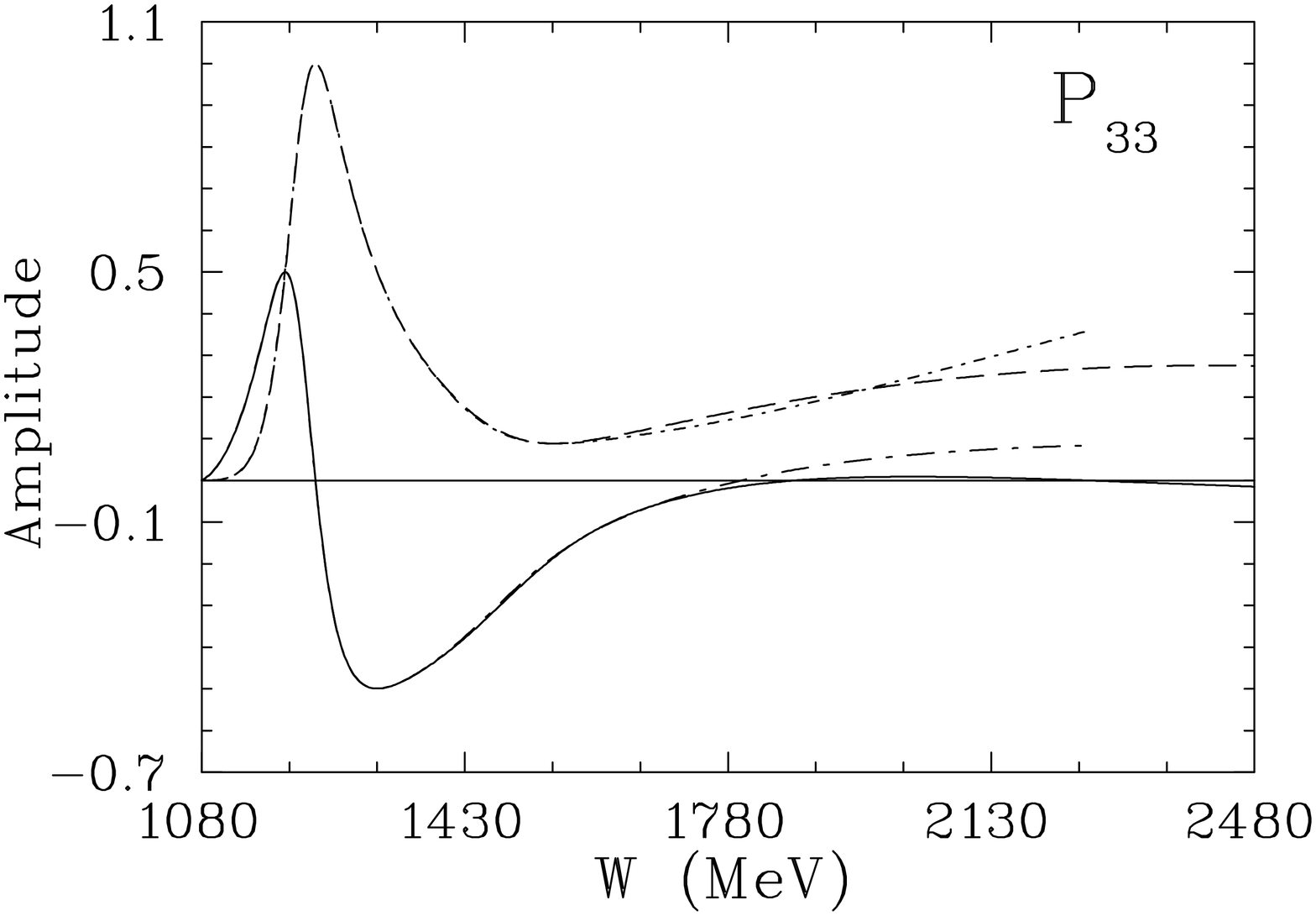}\hfill
\includegraphics[height=0.33\textwidth, angle=90]{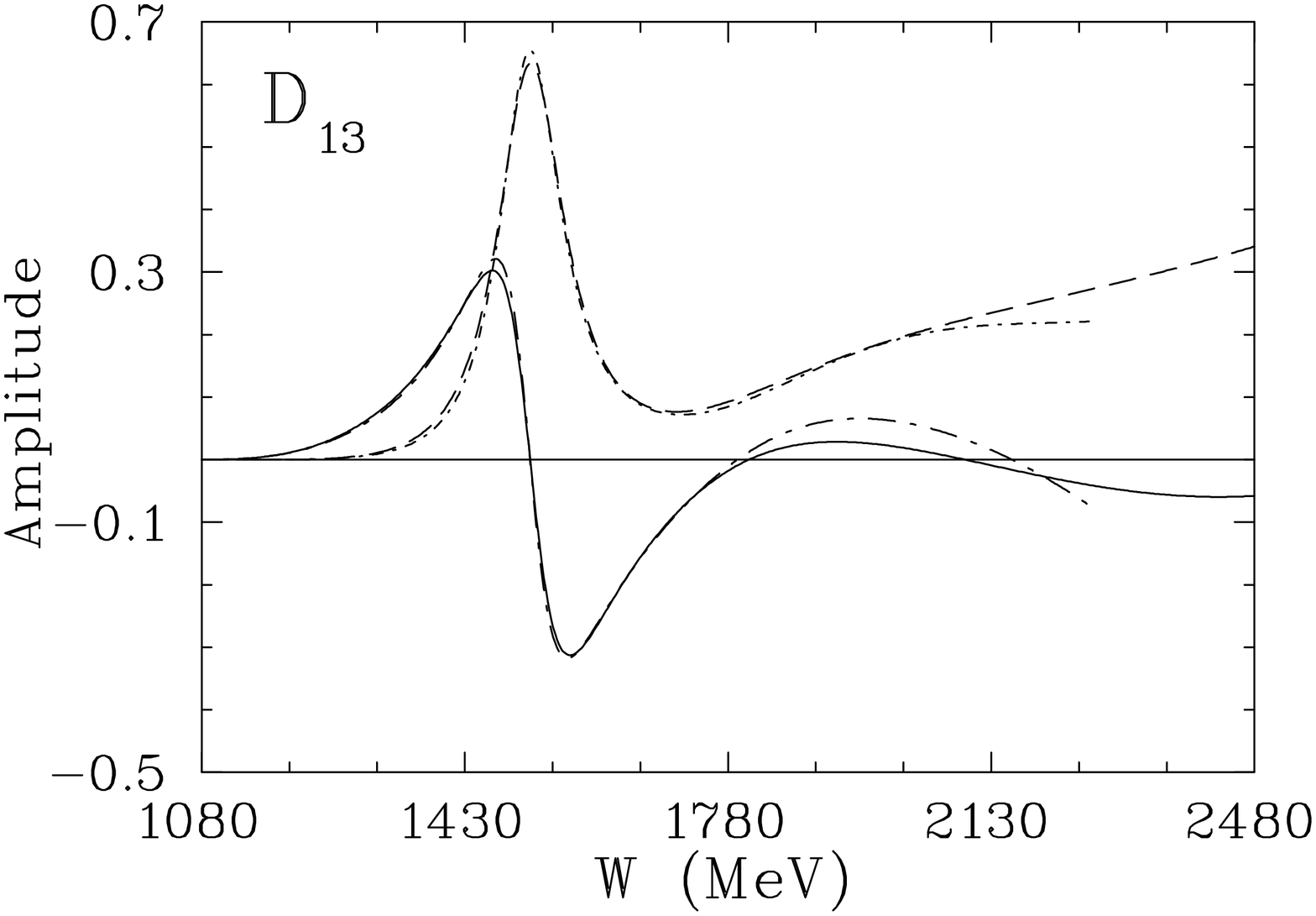}
\includegraphics[height=0.33\textwidth, angle=90]{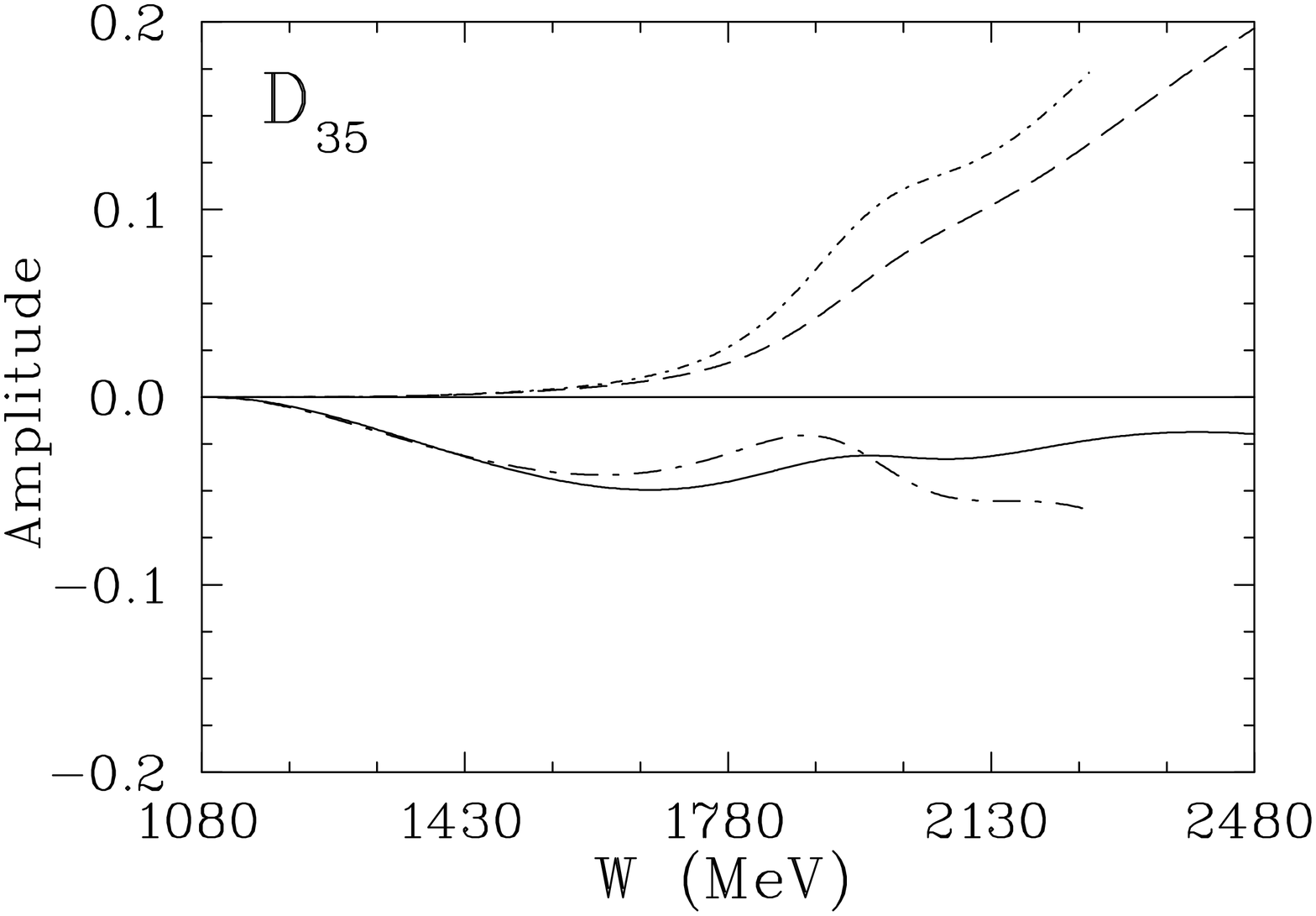}\hfill
\includegraphics[height=0.33\textwidth, angle=90]{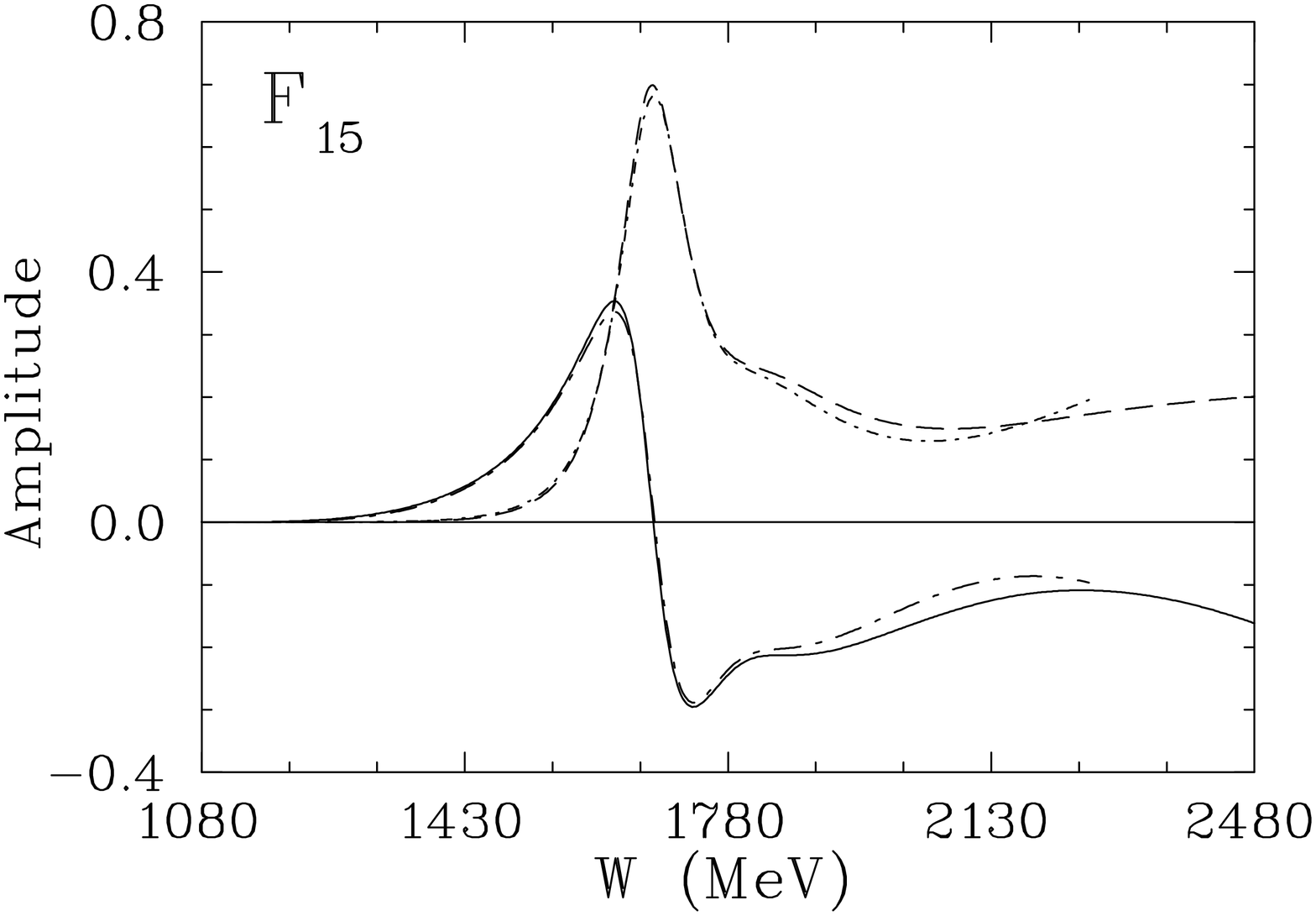}
\includegraphics[height=0.33\textwidth, angle=90]{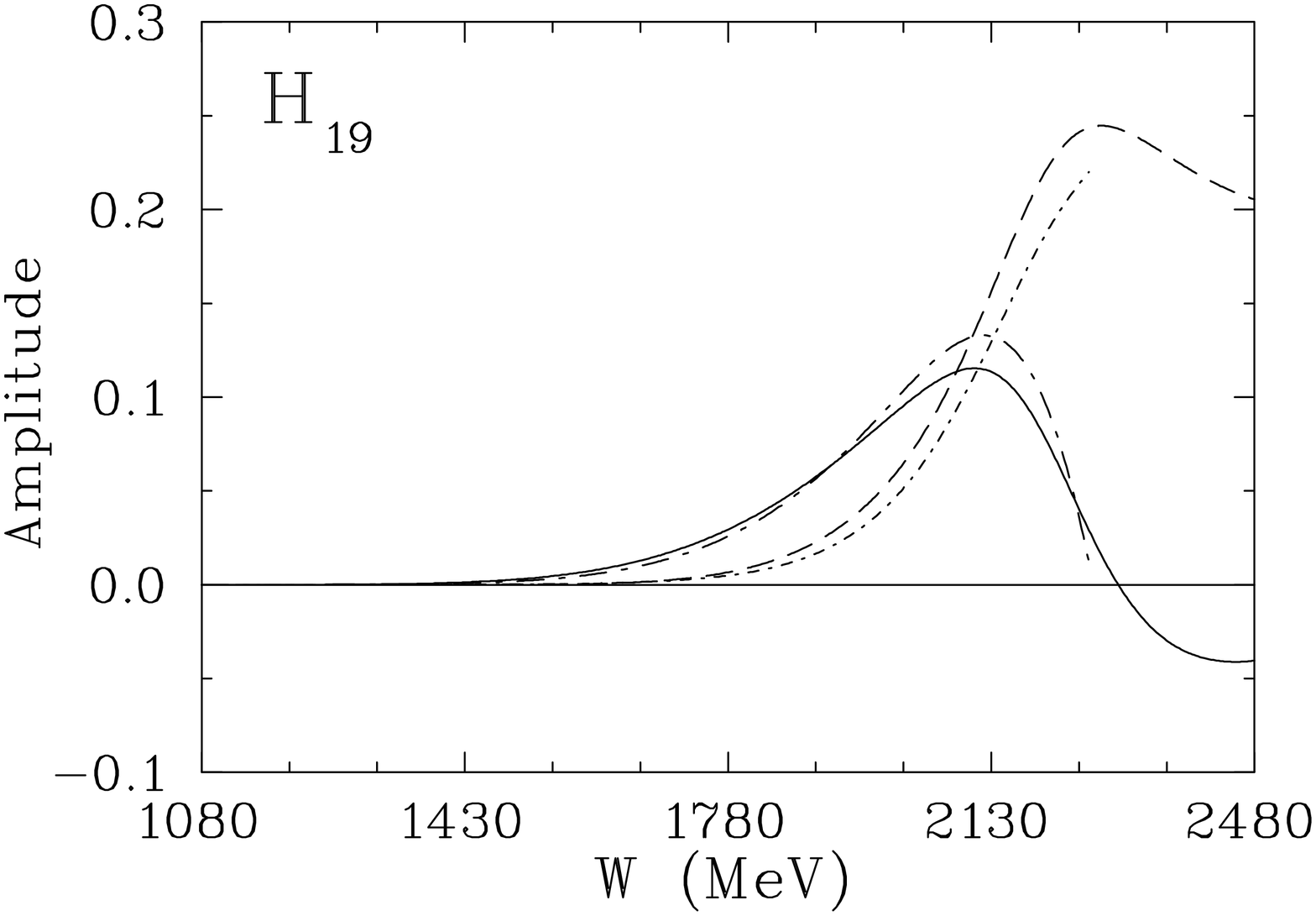}\hfill
}\caption{Comparison of isospin $1/2$ and $3/2$ 
          partial-wave amplitudes (L$_{2I, 2J}$) 
          from $T_{\pi}$ = 0 to 2.6~GeV.  Solid 
          (dashed) curves give the real (imaginary) 
          parts of the SP06 amplitudes.  The FA02 
          solution (valid to 
          2.1~GeV)~\protect\cite{fa02} is plotted 
          with long dash-dotted (real part) and 
          short dash-dotted (imaginary part) lines.  
          All amplitudes are dimensionless.  
          \label{fig:g8}}
\end{figure}

Figs.~\ref{fig:g4} through~\ref{fig:g7} compare the 
energy-dependent fits SP06 and KA84~\cite{kh80} over 
the SP06 energy range (KA84 is valid to 10~GeV/c).  The 
SP06 analysis has fitted waves up to $l=8$, compared to 
$l=7$ for FA02.  Deviations from the KA84 results are 
largest in the isospin $3/2$ amplitudes. One possible
explanation is illustrated in Fig.~\ref{fig:g2}, which 
compares the KA84 solution to a Barrelet-transformed 
version versus the double-polarization quantity $A$
for $\pi^+ p$ (I = 3/2) scattering. 
This exercise and resulting changes to the KA84 isospin 
$3/2$ amplitudes were discussed in Ref.~\cite{alekseev} 
(see also the comments in Ref.~\cite{pdgXX}).  The 
agreement between SP06 and 
KA84 for $\pi^-p$ $A$ data~\cite{al05} is much closer, 
suggesting the absence of a Barrelet ambiguity in the 
isospin $1/2$ amplitudes. Deviations from FA02 are 
visible mainly near the end point of the FA02 analysis 
(some examples are given in Fig.~\ref{fig:g8}).

\subsection{Resonance Parameter Extraction}
\label{sec:res}

The resonance spectrum of our fit has been extracted  
in terms of poles and residues found by continuing into 
the complex energy plane.  These are compiled in 
Tables~\ref{tbl4} and~\ref{tbl5}.  Zeros can be found 
in a similar manner and have been listed in a previous 
paper~\cite{fa02}. The location of a zero is not 
directly related to resonance properties, but
the close proximity of zeros and poles may 
indicate cases where a simple Breit-Wigner parameterization 
is questionable. 

\begin{figure}[th]
\centering{
\includegraphics[height=0.5\textwidth, angle=90]{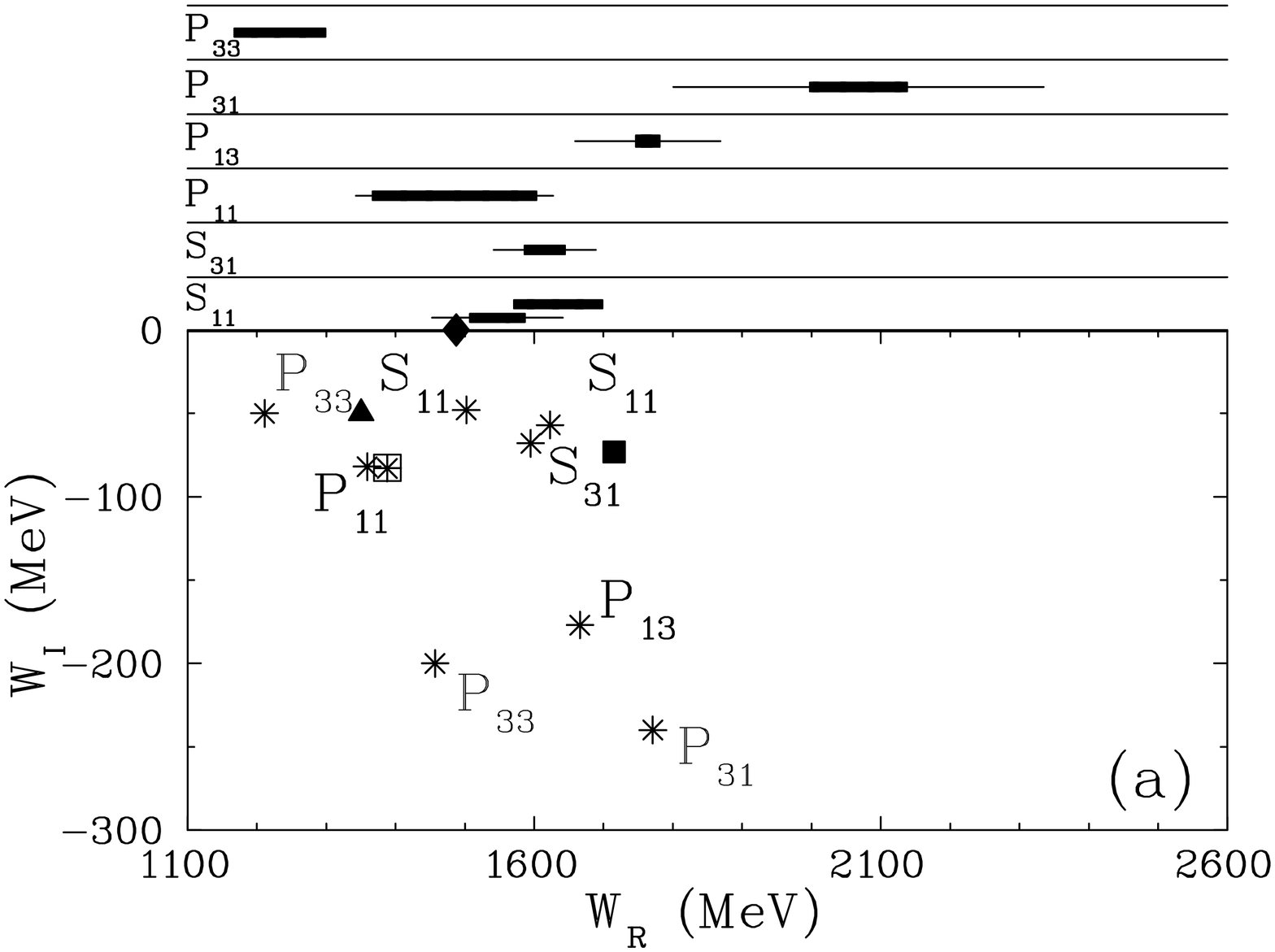}\hfill
\includegraphics[height=0.5\textwidth, angle=90]{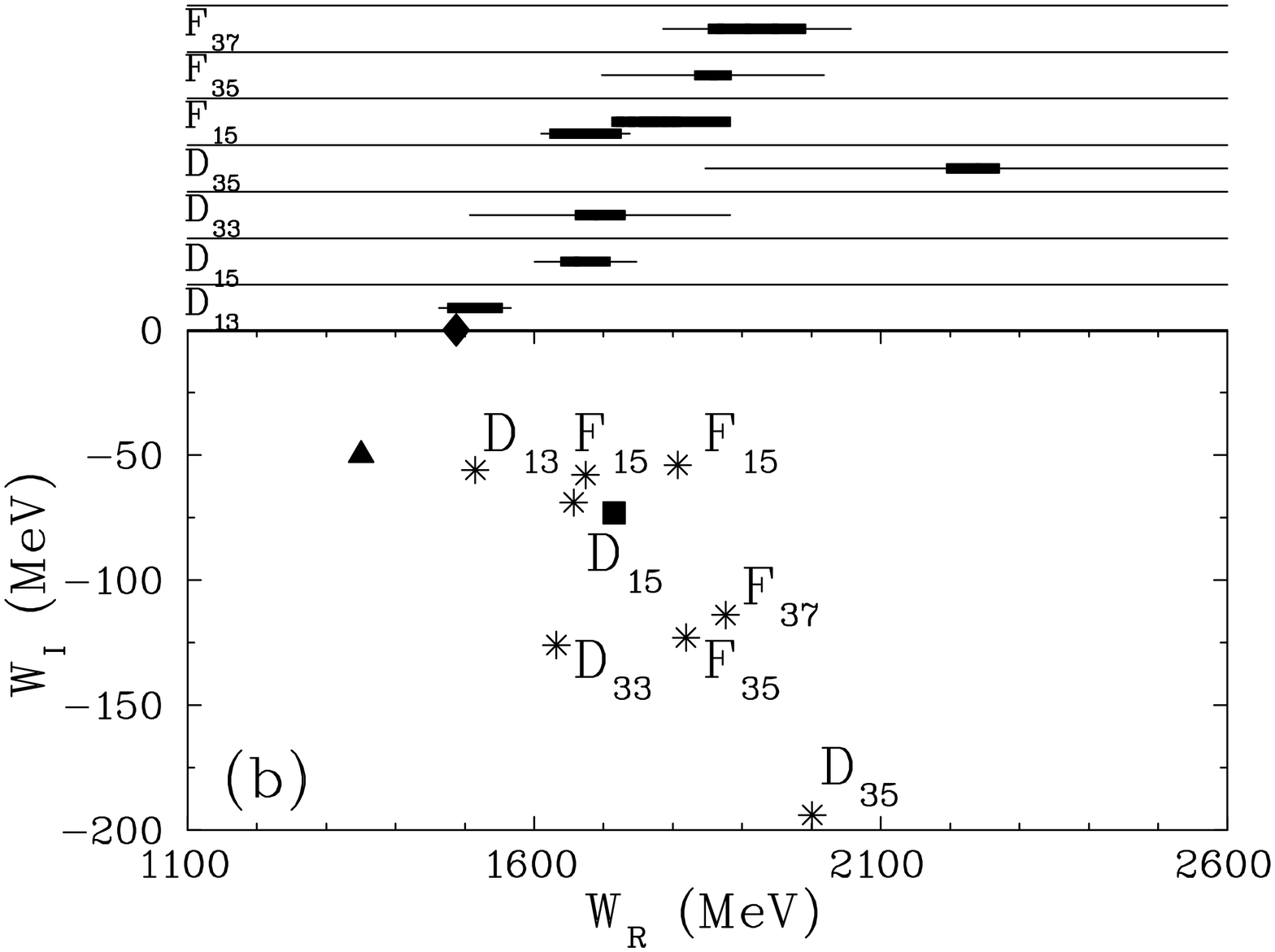}
\includegraphics[height=0.5\textwidth, angle=90]{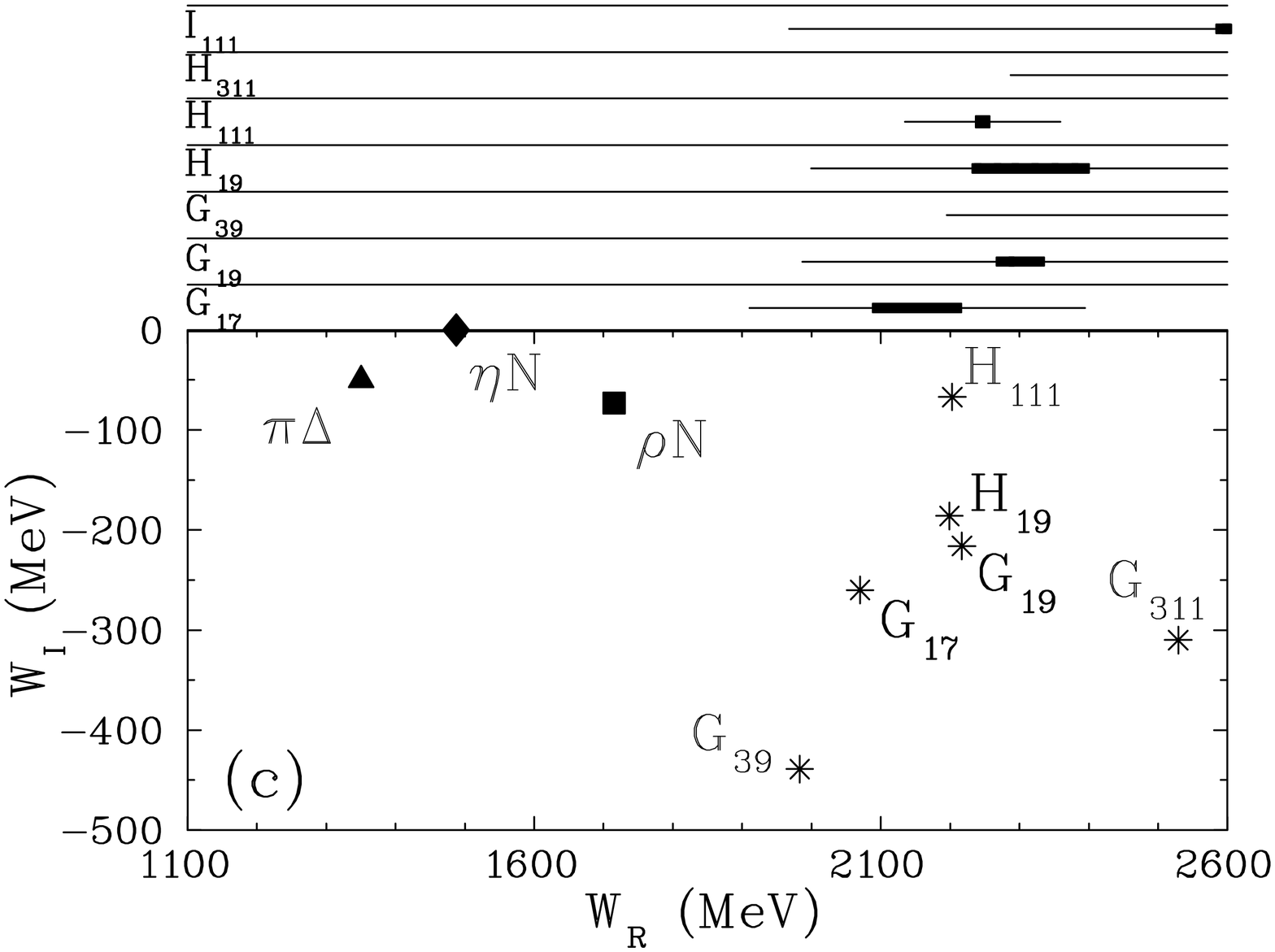}\hfill
}\caption{Comparison of complex plane (bottom 
          panel) and Breit-Wigner (top panel) 
          parameters for resonances found in the 
          SP06 solution.  Plotted are the result 
          for (a) S- and P-wave resonances, (b) 
          D- and F-wave resonances, and (c) G-, H, 
          and I-wave resonances.  Complex plane 
          poles are shown as stars (the boxed 
          star denotes a second-sheet pole).  
          $W_R$ and $W_I$ give real and imaginary 
          parts of the center-of-mass energy.  
          The full ($\pi N$ partial) widths are 
          denoted by thin (thick) bars for each 
          resonance.  The branch point for 
          $\pi \Delta (1232)$, 1350 - i50~MeV, 
          is represented as a solid triangle.  
          The branch points for $\eta N$, 1487 
          -i0~MeV, and  
          $\rho N$, 1715 - i73~MeV, thresholds 
          are shown as a solid diamond and solid 
          square, respectively.  \label{fig:g9}}
\end{figure}

The more commonly used, and more model-dependent, 
Breit-Wigner parameters for resonances are listed in 
Tables~\ref{tbl6} and~\ref{tbl7}. Here, in the FA02 
and SM95 fits, a unitary Breit-Wigner plus background 
form was assumed for the resonant partial wave.  Data 
within an energy bin were then fitted using this 
representation.  The remaining waves were fixed to 
values found in the full global analysis.  Energy 
ranges over which fits were performed, and $\chi^2$ 
comparisons are given in Tables~\ref{tbl8} 
and~\ref{tbl9}.  This method is more directly linked 
to data than a fit to the SES.  However, the resulting 
parameter uncertainties tend to be small, reflecting 
the statistical error but not the (possibly large) 
systematic error associated with a separation of 
resonance and background contributions. The pole and 
Breit-Wigner representations are compared in 
Fig.~\ref{fig:g9}.

The onset of resonant behavior, seen in the FA02 
$G_{17}$, $G_{19}$, and $H_{19}$ partial waves, is 
fully developed in SP06, the extension by 500~MeV in 
$T_{\pi}$ corresponding to a 200~MeV increase in center 
of mass energy. We can now also see resonant behavior 
in the $G_{39}$, $H_{3,11}$, and $I_{1,11}$ waves. A 
possible resonance is seen in $H_{1,11}$, though the 
SES scatter is large and the amplitude is small in 
magnitude. 

We have tried to associate each state with its 
corresponding PDG designation. In some cases, this 
resulted in a resonance mass far from that of a ``named" 
resonance.  One such case is the $N(2000)F_{15}$.  We 
find evidence for a second $F_{15}$ state closer to 
1800~MeV. The KH analysis also finds a mass near 
1880~MeV, which suggests a name change for this 2-star 
state may be in order. In SP06, this second $F_{15}$
resonance was found by scanning each partial wave for 
small structures. Its resonance parameters have been
determined through a fit to the full database and 
are quoted without errors.
          
The $\Delta(1910)P_{31}$ is 
also problematic. We find only a single $P_{31}$ 
state, with a pole position more in line with the 
(1-star) $\Delta(1750)P_{31}$ than the (4-star) 
$\Delta(1910)P_{31}$.  As can be seen in 
Fig.~\ref{fig:g6}, the $P_{31}$ resonance signature 
is particularly subtle for a 4-star state. Small 
changes in the KH and CMB amplitudes, due to the 
Barrelet ambiguity, could explain this mass shift.  
Our Breit-Wigner fits to this structure yeilded 
spurious results, with a mass several hundred
MeV above the pole position and a width exceeding 1~GeV, 
if data were fitted around the assumed 4-star state mass. 
More reasonable values were obtained when the fitted 
energy range was expanded. This fit to a Breit-Wigner 
form is questionable for states with poles so far from 
the physical axis (see Fig.~\ref{fig:g9}).

\begin{figure}[th]
\centering{
\includegraphics[height=0.47\textwidth, angle=90]{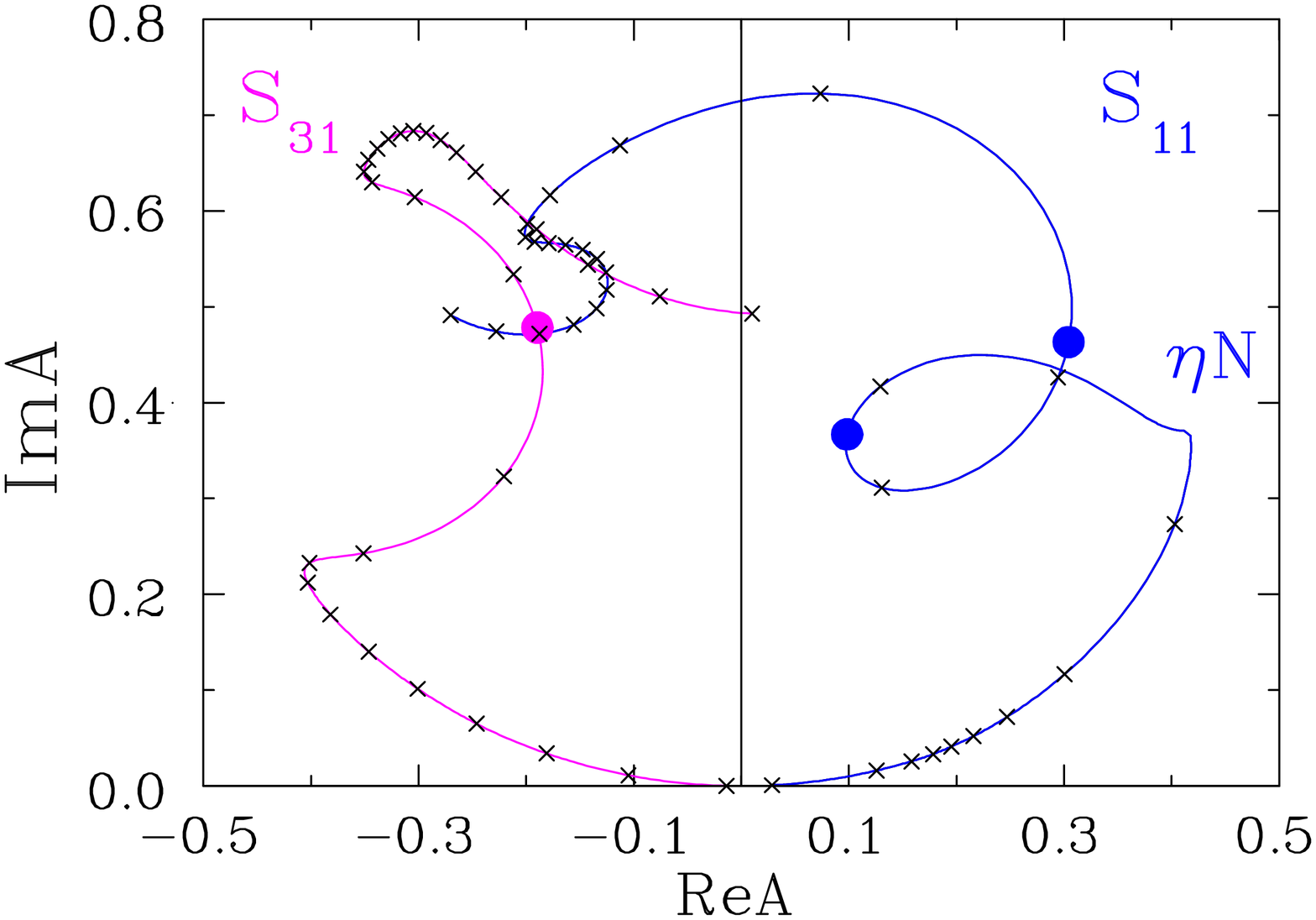}\hfill
\includegraphics[height=0.47\textwidth, angle=90]{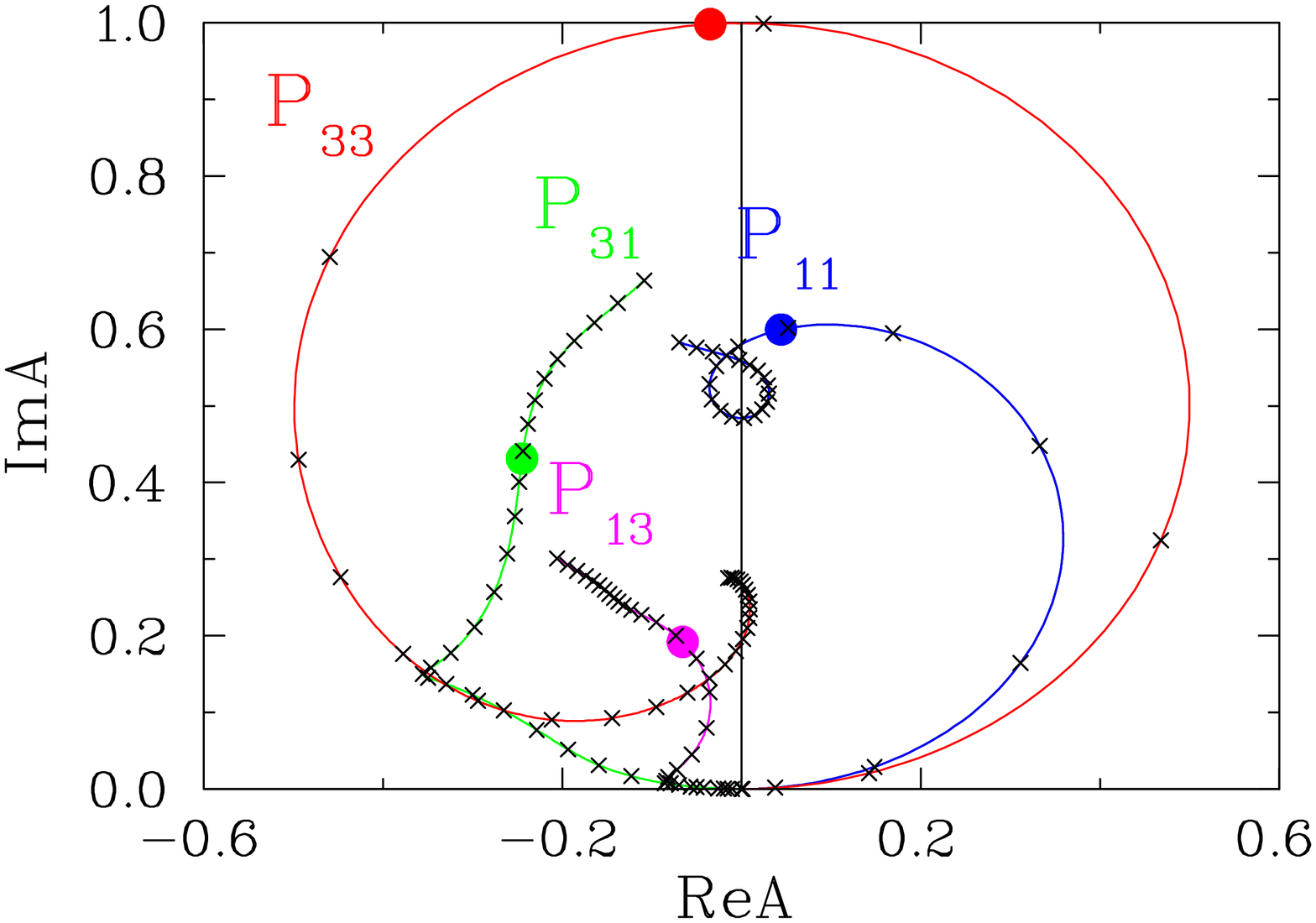}
\includegraphics[height=0.47\textwidth, angle=90]{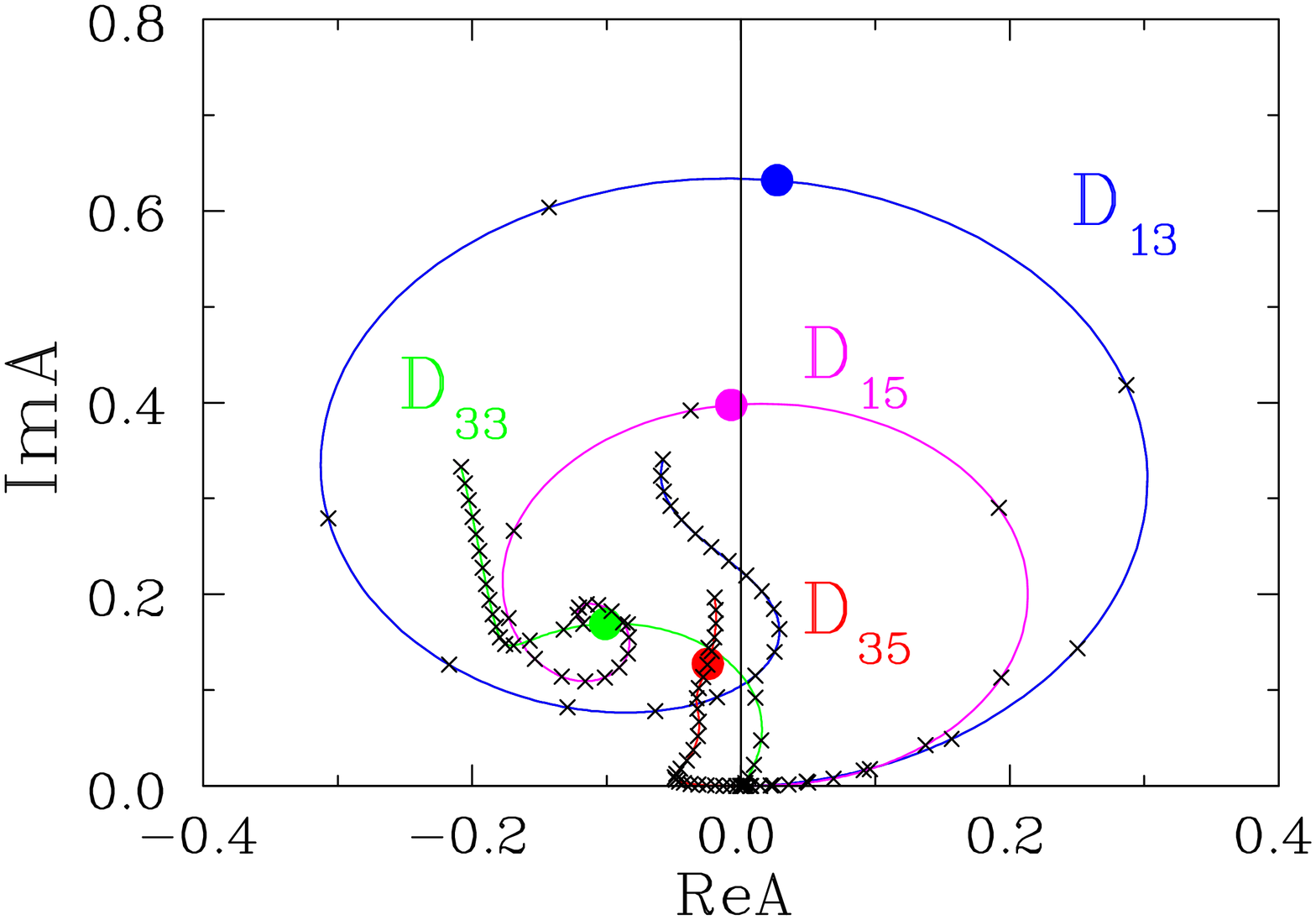}\hfill
\includegraphics[height=0.47\textwidth, angle=90]{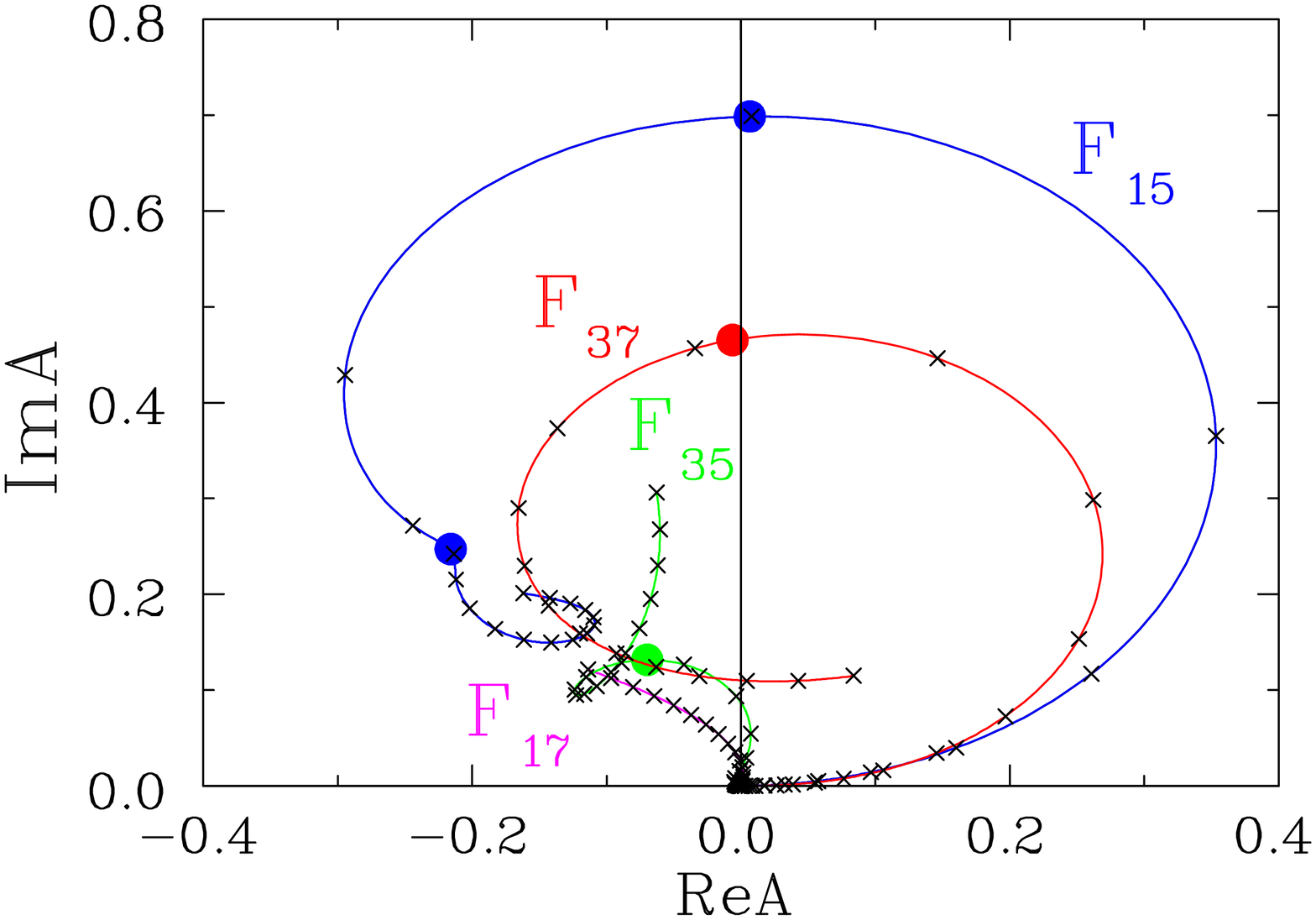}
\includegraphics[height=0.47\textwidth, angle=90]{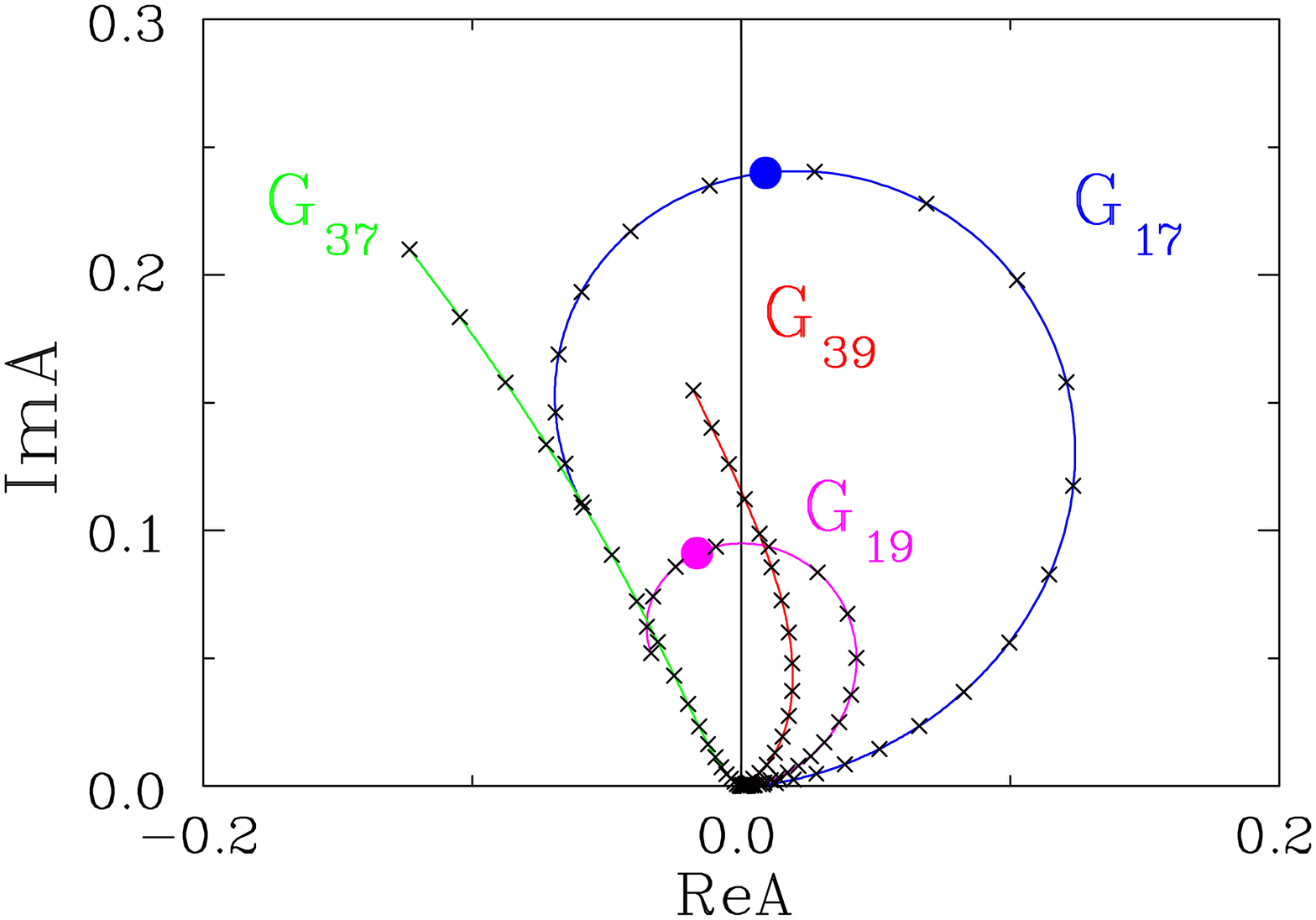}\hfill
\includegraphics[height=0.47\textwidth, angle=90]{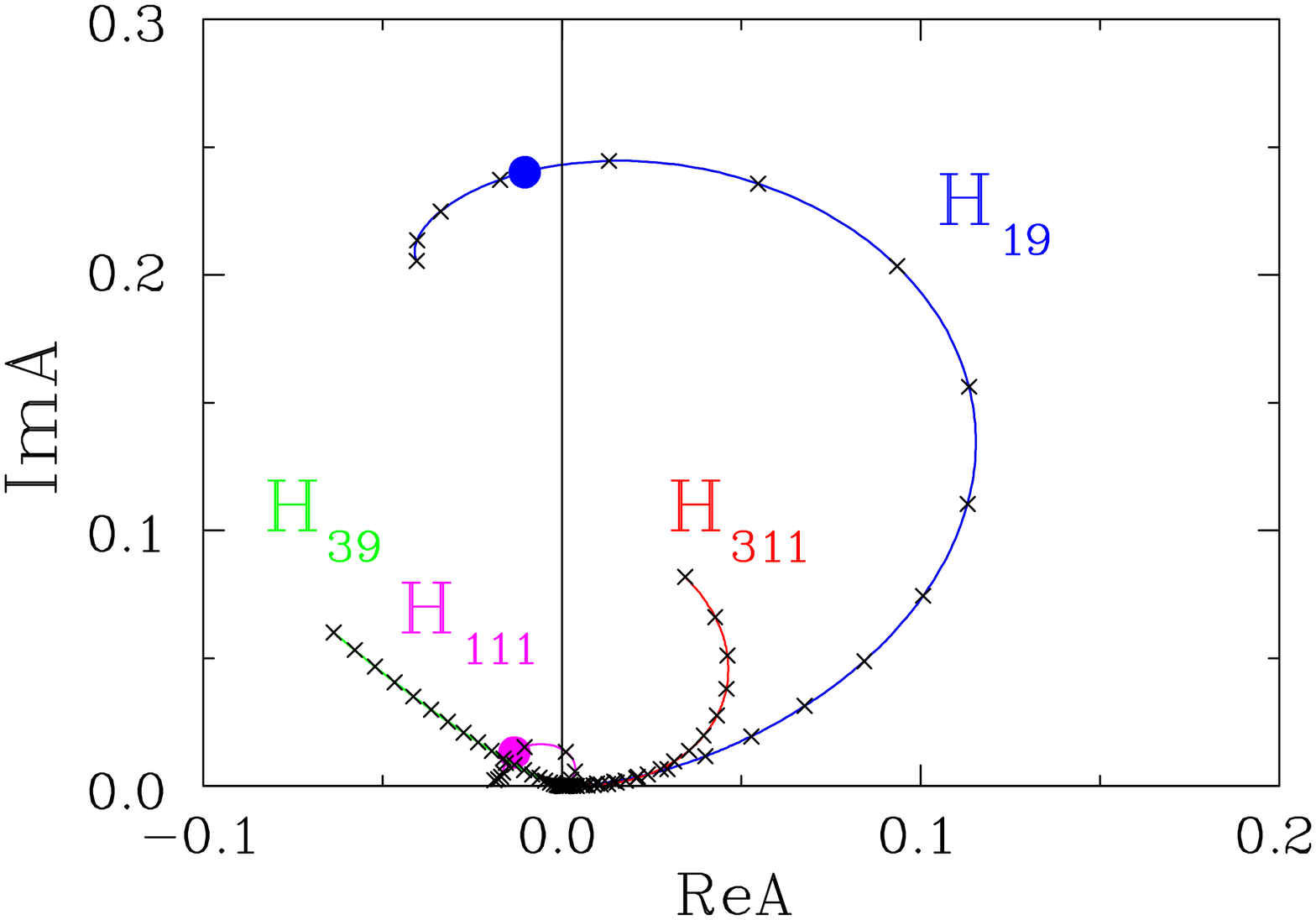}
}\caption{Argand plots for partial-wave amplitudes
          from threshold (1080~MeV) to W = 2.5~GeV. 
          Crosses indicate 50~MeV steps in W.  
          Solid circles correspond to BW $W_R$ 
          determination presented in 
          Tables~\protect\ref{tbl6} 
          and~\protect\ref{tbl7}. \label{fig:g10}}
\end{figure}

The $P_{11}$ partial wave of KA84 and the SES associated 
with SP06 agree reasonably well over the full range of
SP06. However, this does not lead to agreement on the
resonance content. The prominent $N(1440)P_{11}$ resonance 
is clearly evident in both analyses, but occurs very near
the $\pi \Delta$ threshold, making a Breit-Wigner fit
questionable~\cite{pdgXX}. Above this energy, the 
$P_{11}$ partial wave wraps around the center of the
Argand diagram (Fig.~\ref{fig:g10}).  As a result, small 
changes in the amplitude can produce large changes in the 
phase, though these changes have little influence on the 
fit to data.  States above the $N(1440)P_{11}$ should be 
established in reactions where they are more clearly 
required. 

\section{Summary and Conclusions}
\label{sec:conc}

We have fitted the existing $\pi N$ elastic scattering
and charge-exchange database 
to 2.6~GeV ($\eta N$ data included to 800~MeV), 
employing a complete set of dispersion relation 
constraints, up to T$_{\pi}$ = 1~GeV and $t = 
-0.4~(GeV/c)^2$.  This extension in $T_{\pi}$ has 
allowed us to search an addition 200~MeV  
of the resonance region (in center of mass energy).

\begin{figure}[th] 
\centering{ 
\includegraphics[height=0.48\textwidth, angle=90]{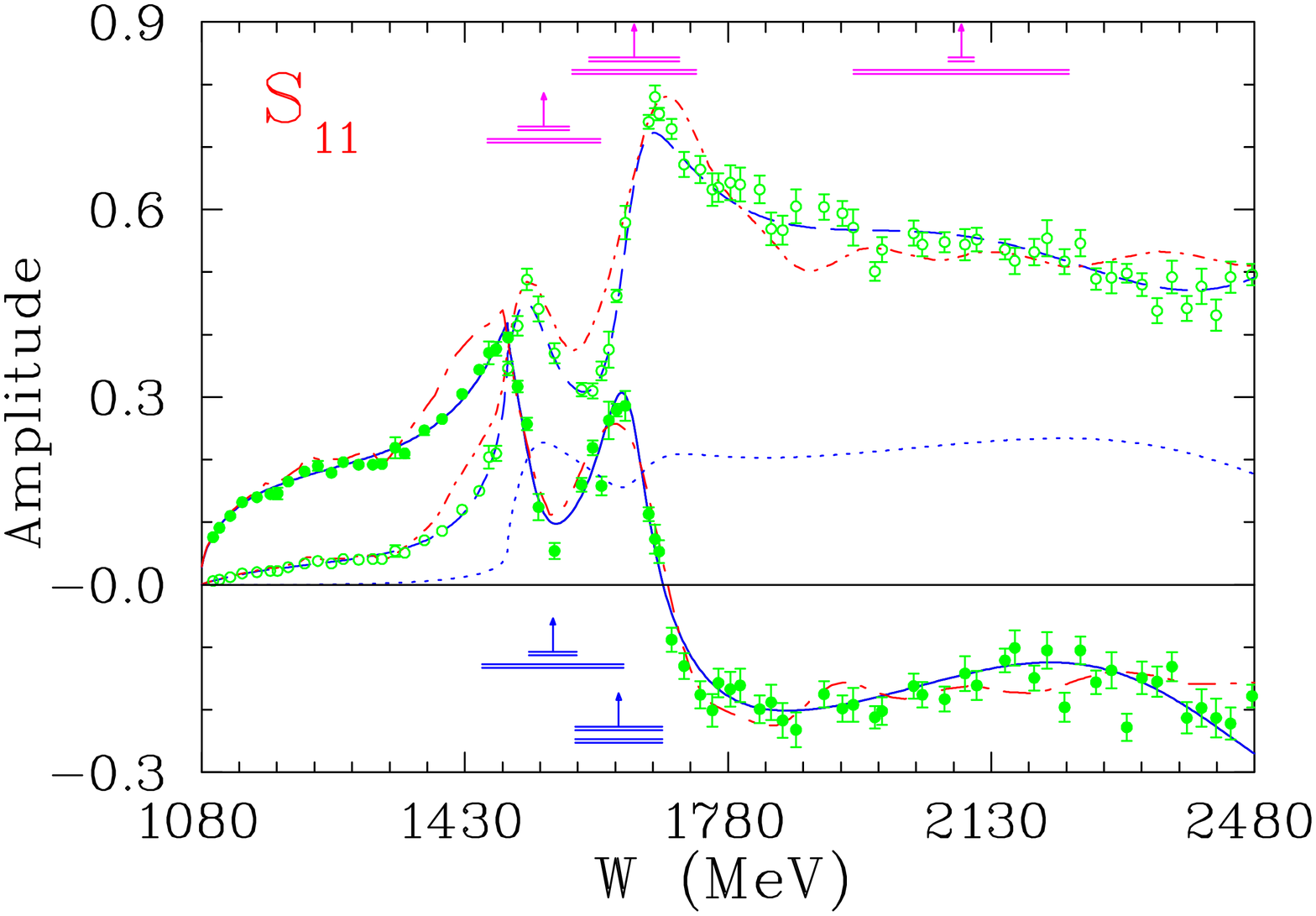}\hfill 
\includegraphics[height=0.48\textwidth, angle=90]{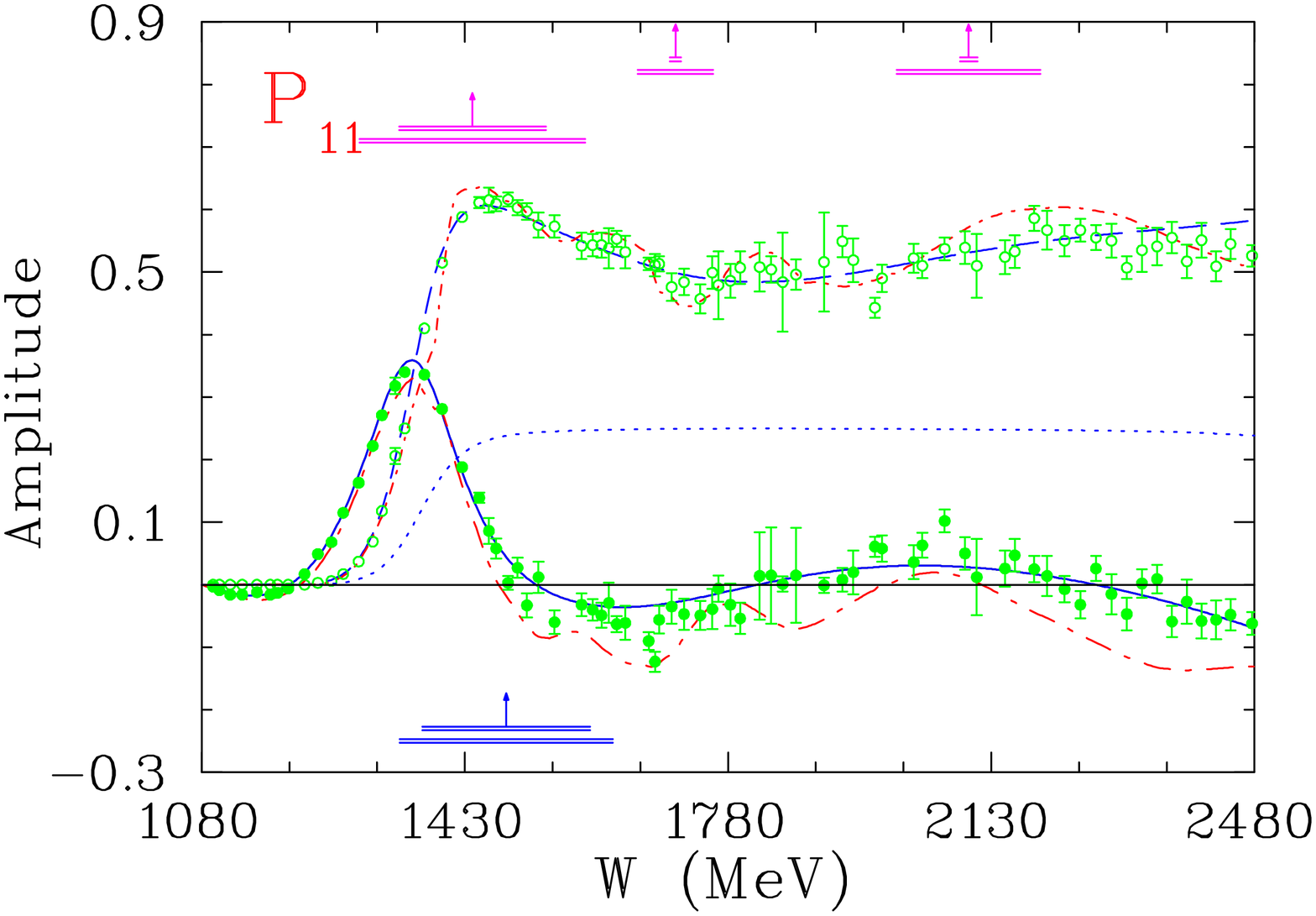} 
\includegraphics[height=0.48\textwidth, angle=90]{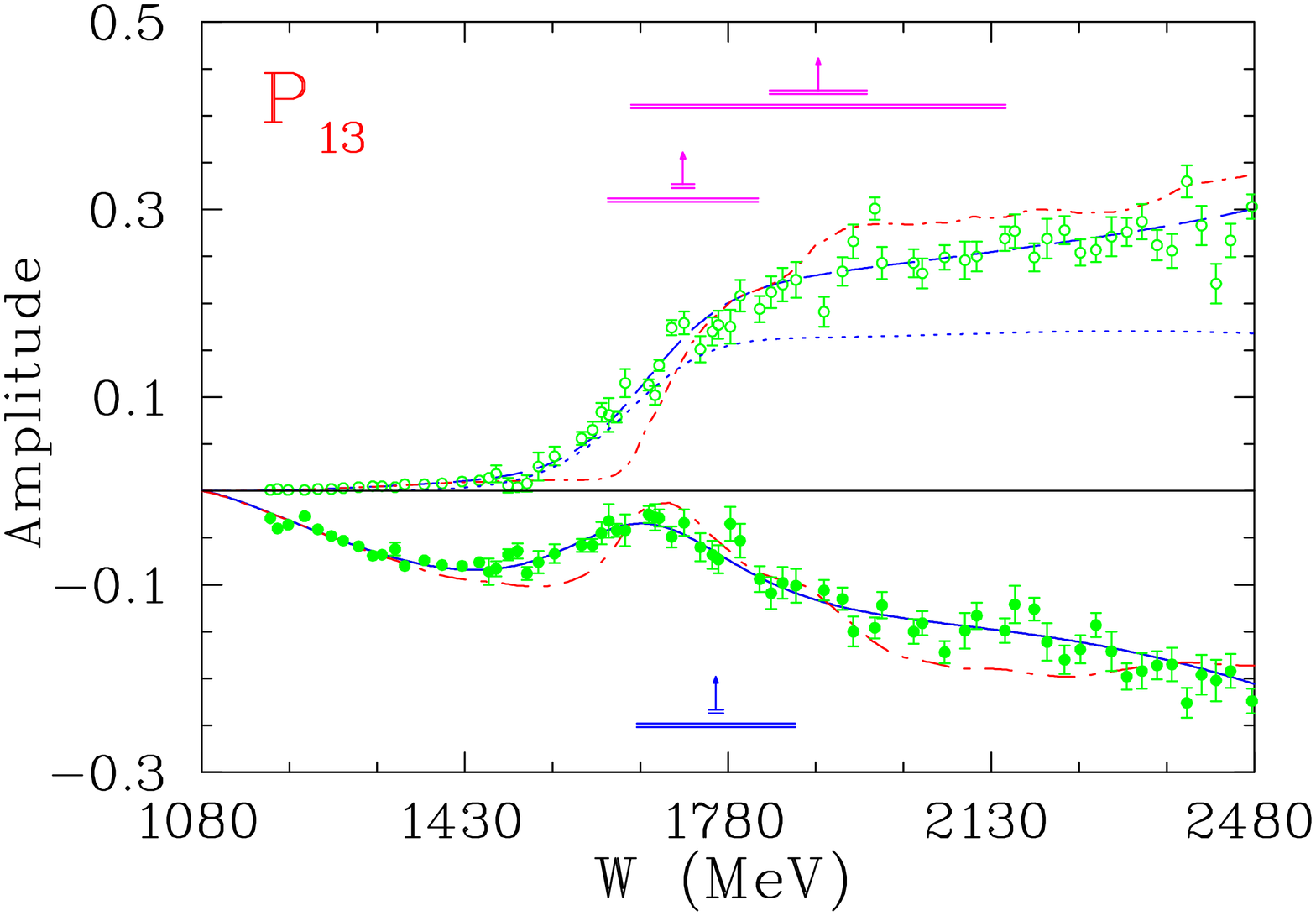}\hfill 
\includegraphics[height=0.48\textwidth, angle=90]{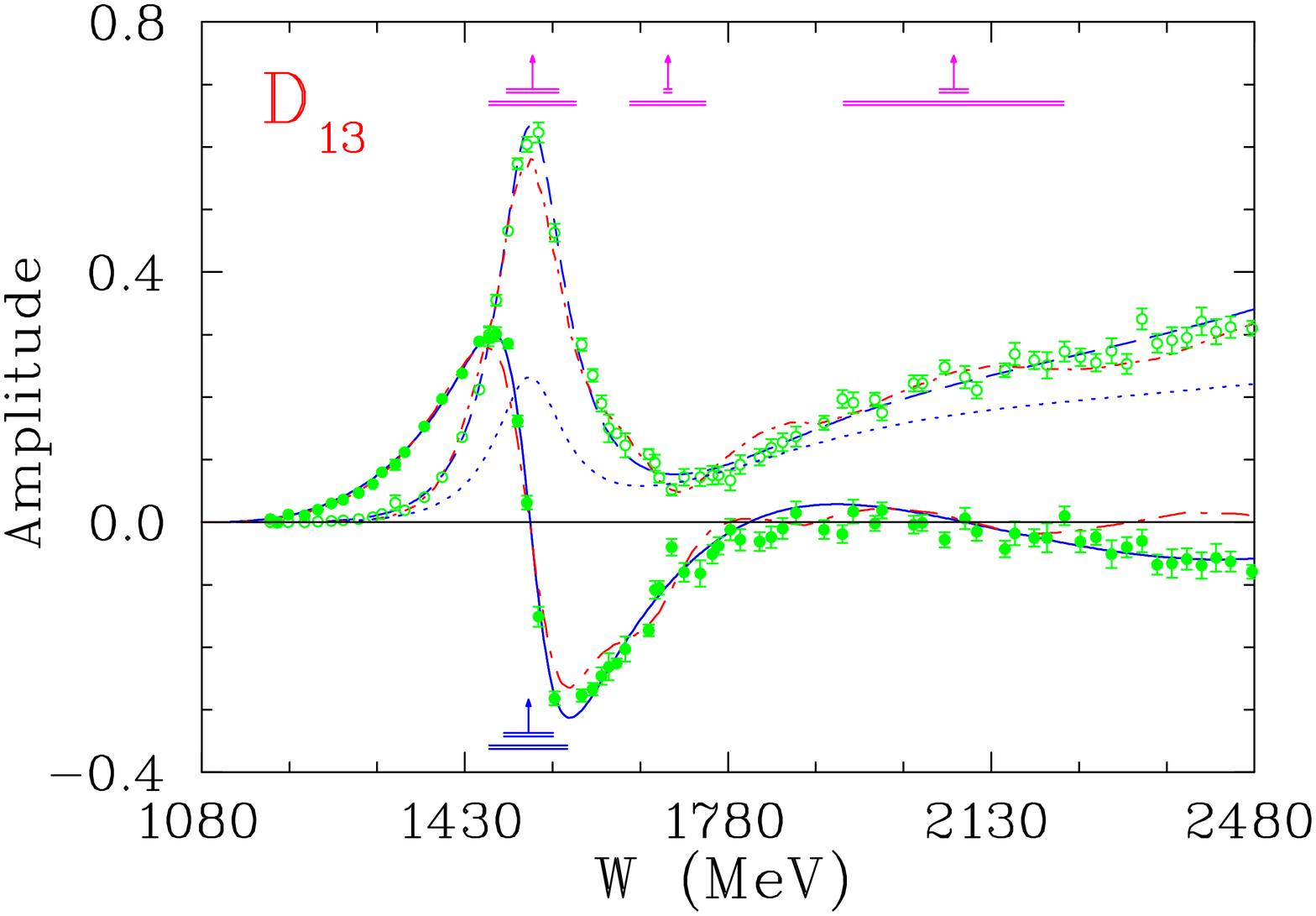} 
\includegraphics[height=0.48\textwidth, angle=90]{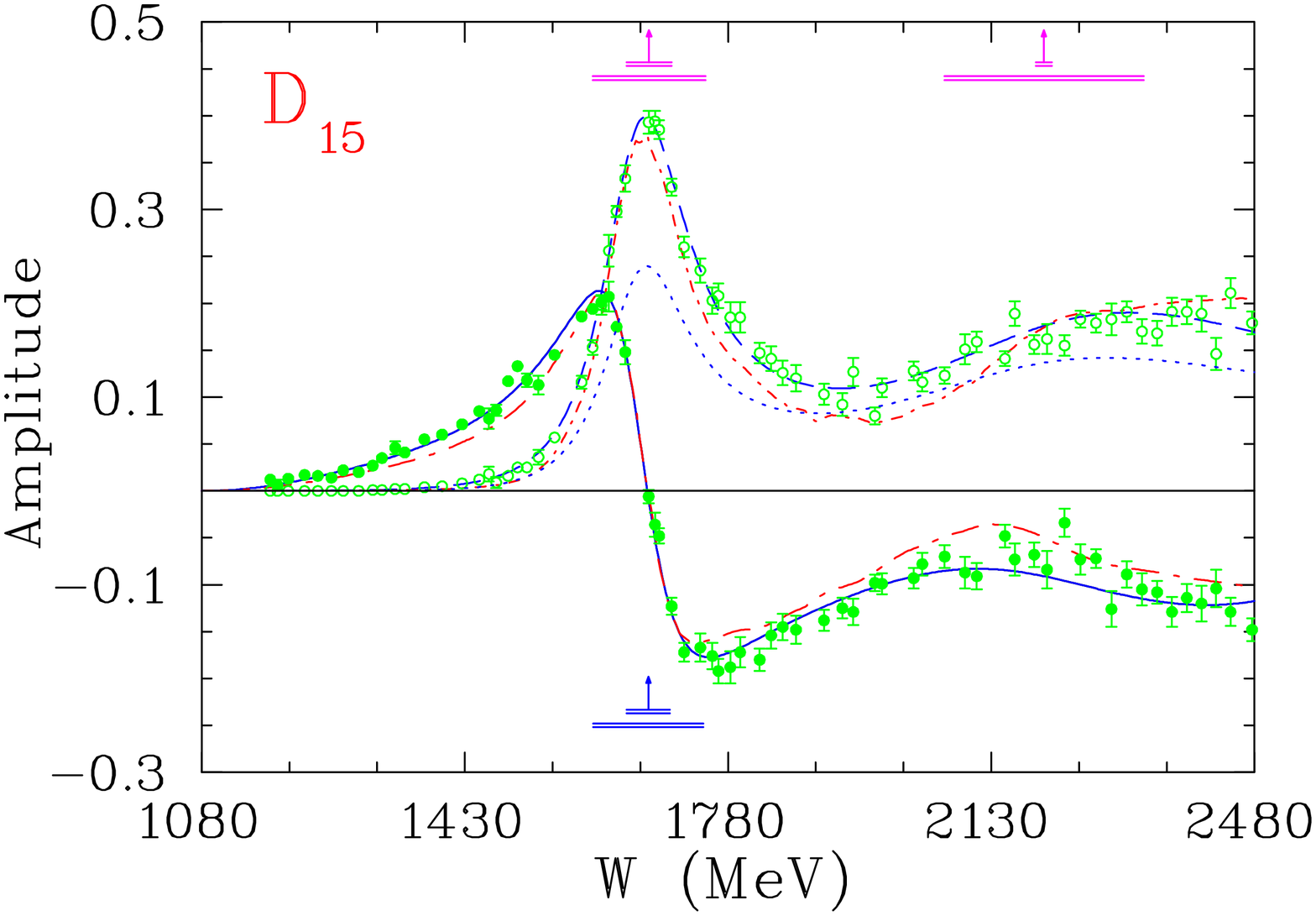}\hfill 
\includegraphics[height=0.48\textwidth, angle=90]{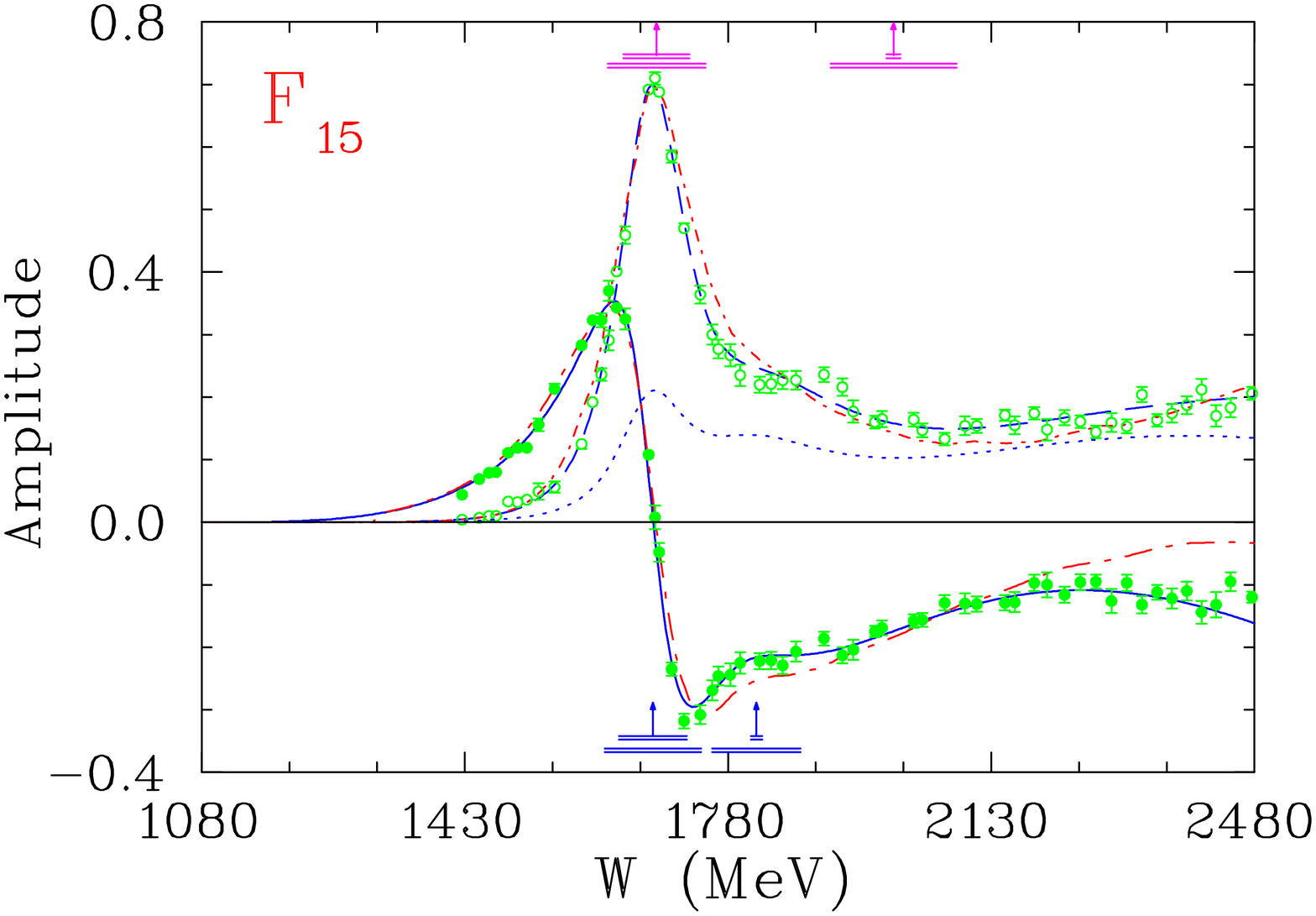} 
}\caption{Isospin $1/2$ partial-wave amplitudes $J < 3$
          (L$_{2I, 2J}$) from $T_{\pi}$ = 0 to 2.6~GeV.  
          Solid (dashed) curves give the real (imaginary) 
          parts of amplitudes corresponding to the 
          SP06 solution.  The real (imaginary) parts 
          of single-energy solutions are plotted as 
          filled (open) circles.  The dotted curve 
          gives the unitarity limit ($ImT - 
          T^{\ast}T$) from SP06.  The Karlsruhe KA84 
          solution~\protect\cite{kh80} is plotted 
          with long dash-dotted (real part) and 
          short dash-dotted (imaginary part) lines.  
          All amplitudes are dimensionless.  Vertical 
          arrows indicate resonance $W_R$ values 
          and horizontal bars show full $\Gamma$ and 
          partial widths for $\Gamma_{\pi N}$.  The 
          lower BW resonance symbols are associated 
          with the SP06 values of Table
          ~\protect\ref{tbl6}; upper symbols give 
          RPP~\protect\cite{rpp} values. 
          \label{fig:g4}}
\end{figure}

\begin{figure}[th]
\centering{
\includegraphics[height=0.48\textwidth, angle=90]{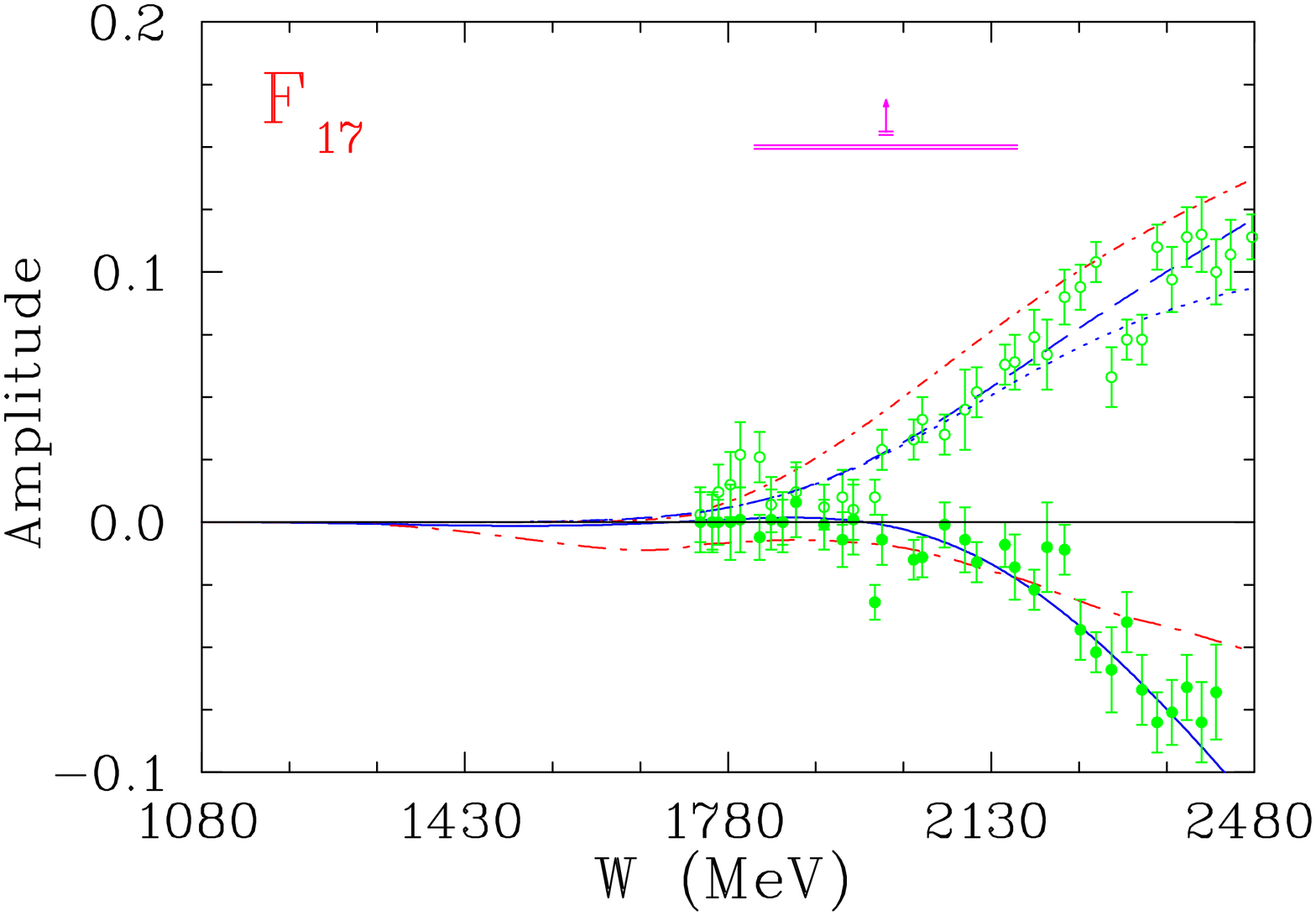}\hfill
\includegraphics[height=0.48\textwidth, angle=90]{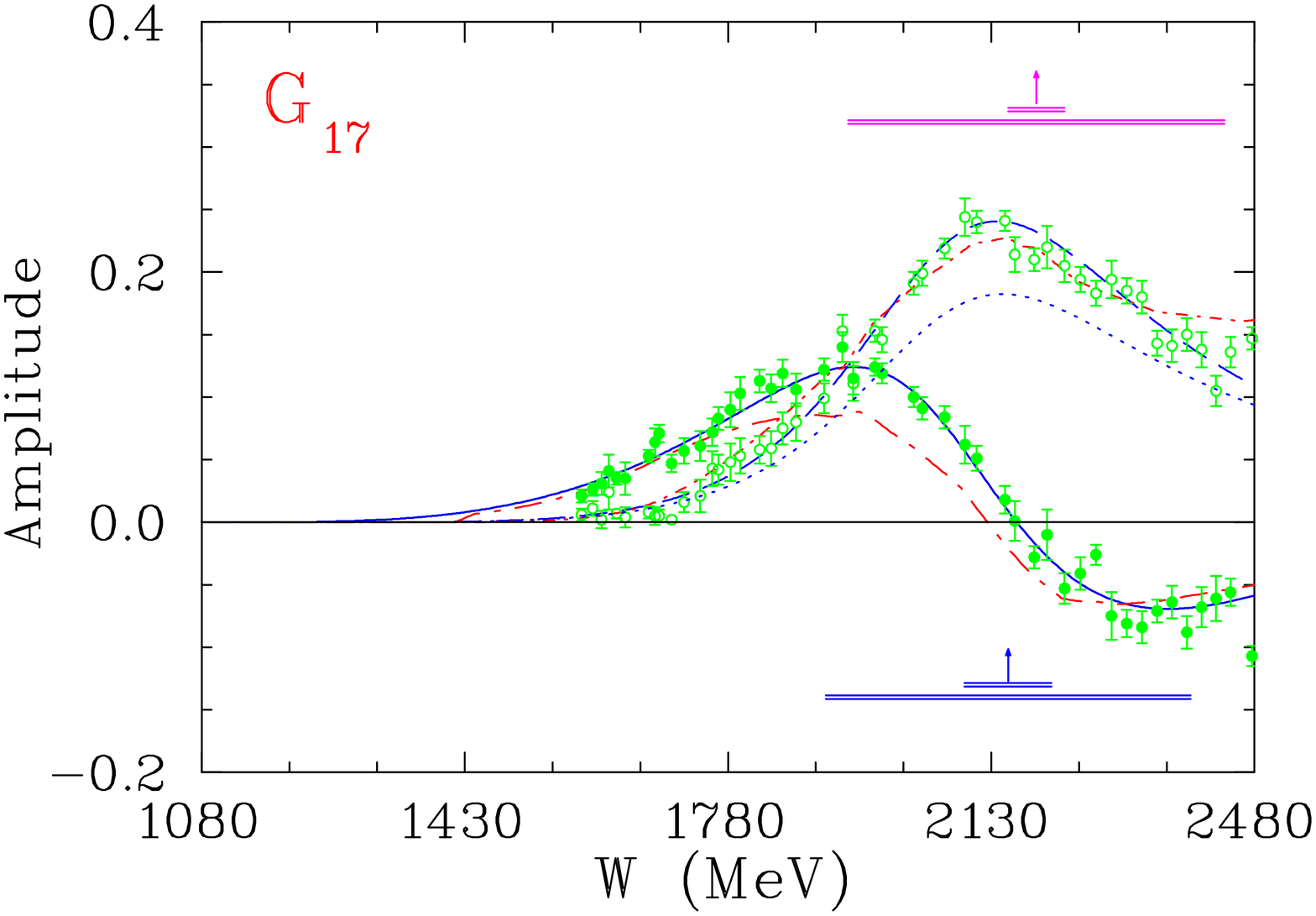}
\includegraphics[height=0.48\textwidth, angle=90]{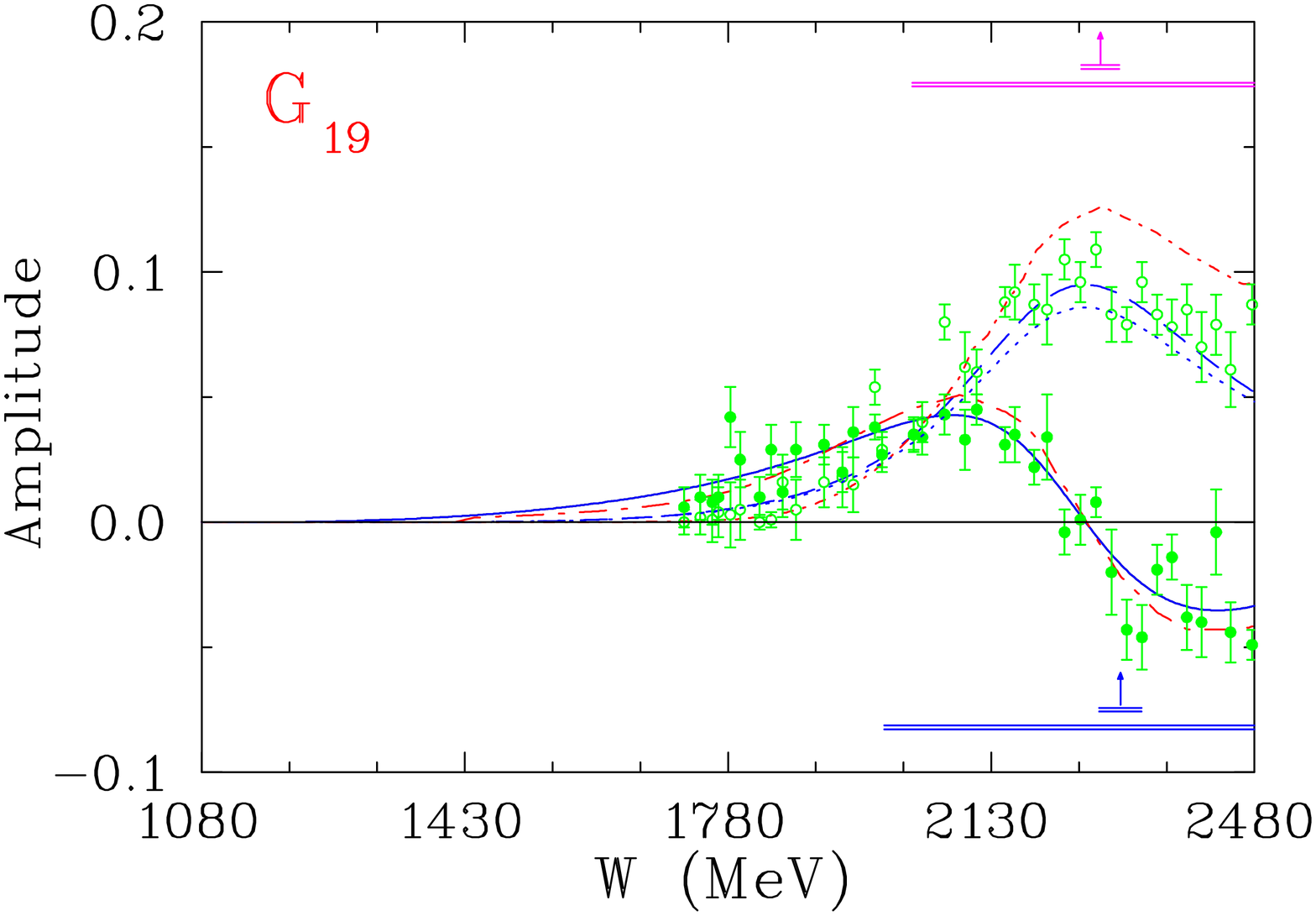}\hfill
\includegraphics[height=0.48\textwidth, angle=90]{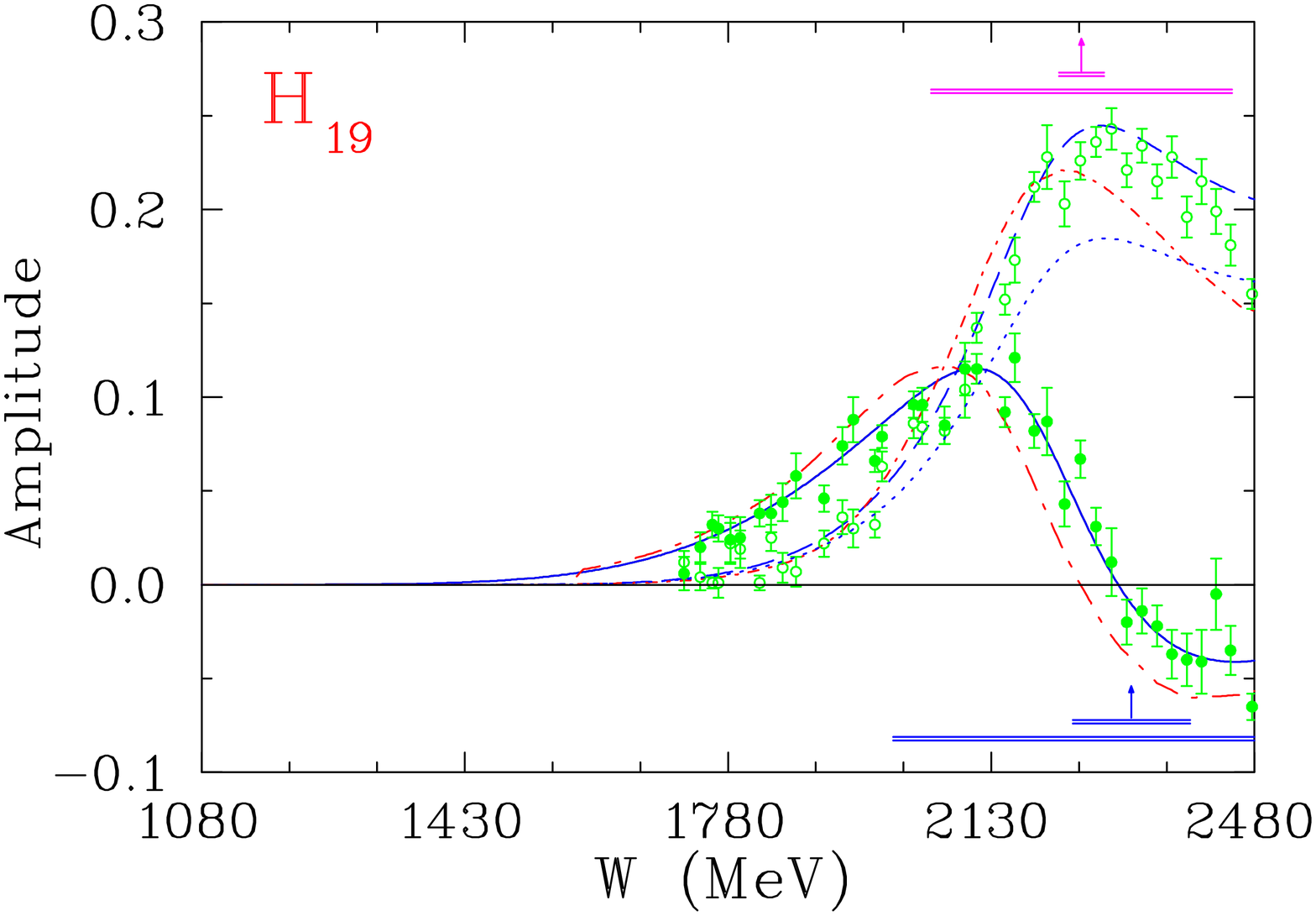}
\includegraphics[height=0.48\textwidth, angle=90]{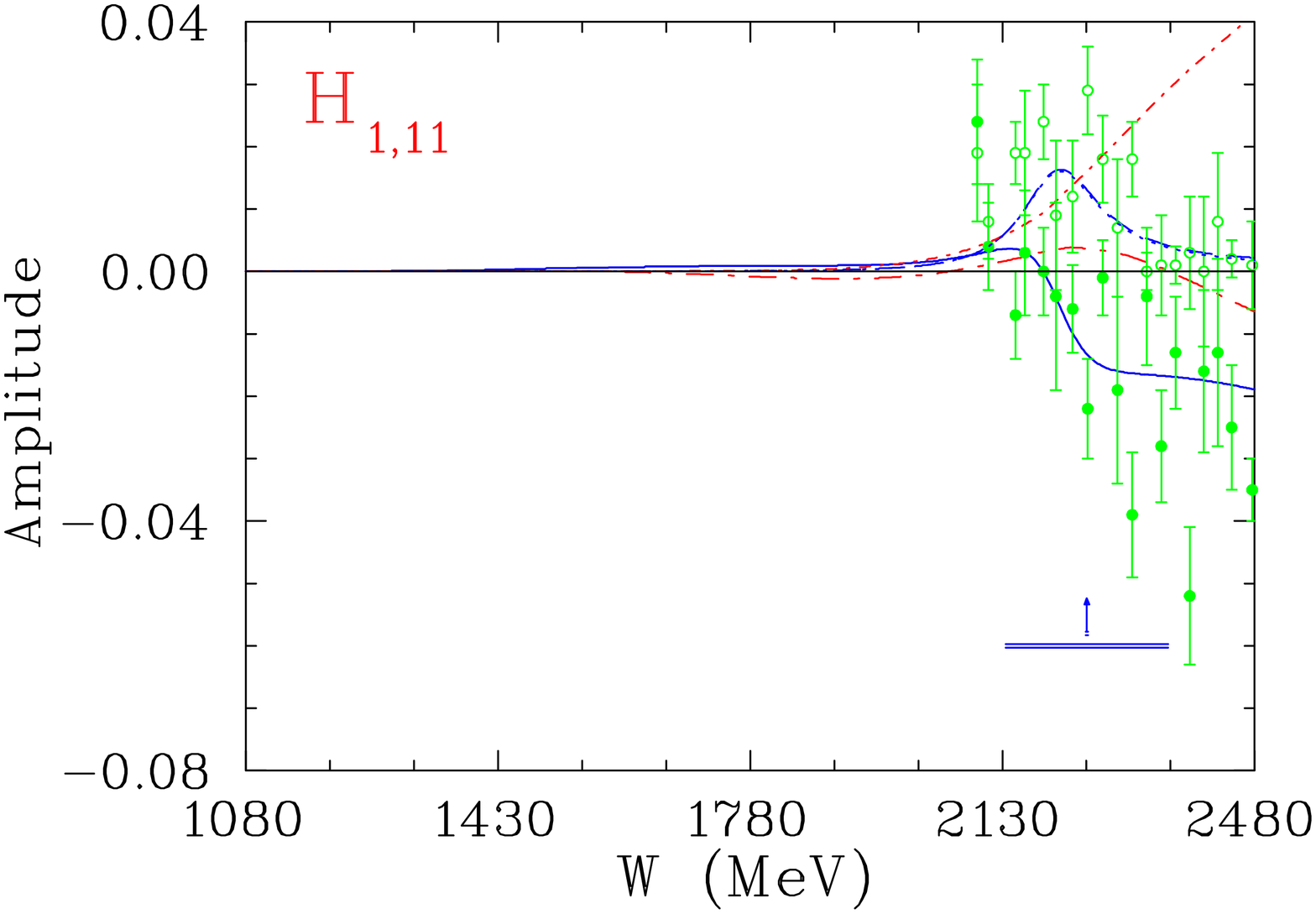}\hfill
\includegraphics[height=0.48\textwidth, angle=90]{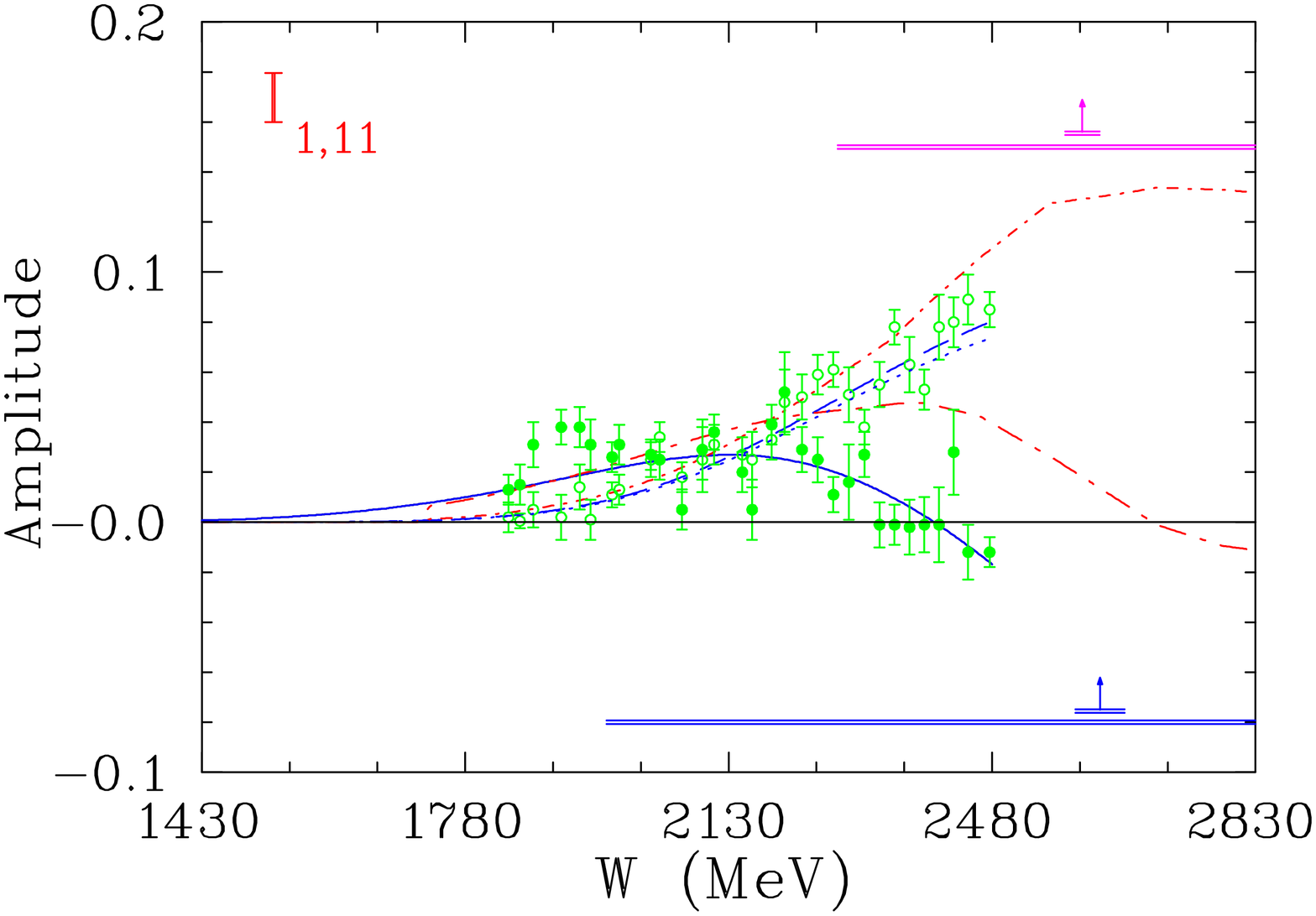}
}\caption{Isospin $1/2$ partial-wave amplitudes $J > 3$
          (L$_{2I, 2J}$) from $T_{\pi}$ = 0 to 2.6~GeV.
          Notation as in Fig.~\protect\ref{fig:g4}.
          \label{fig:g5}}
\end{figure}

\begin{figure}[th]
\centering{
\includegraphics[height=0.48\textwidth, angle=90]{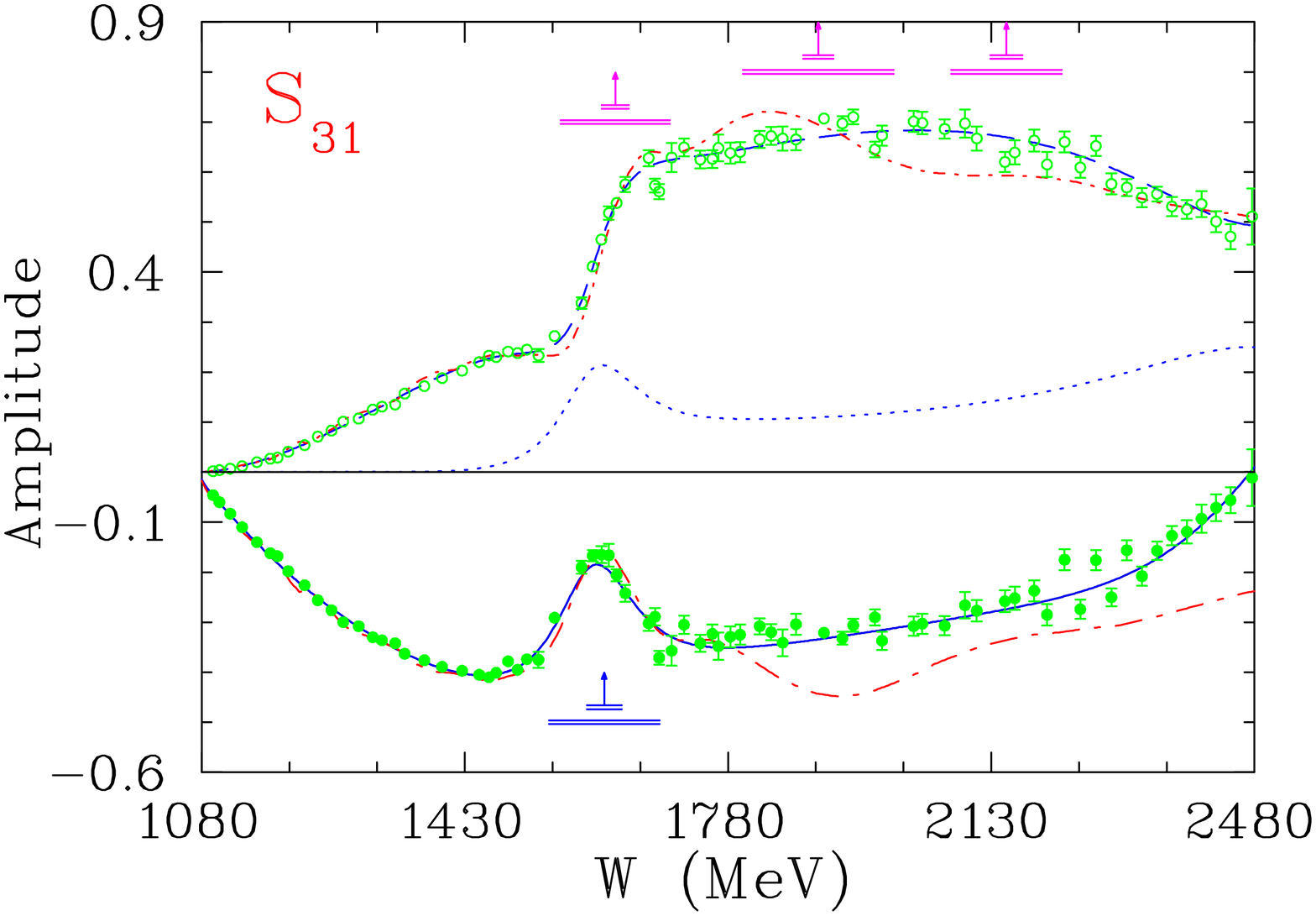}\hfill
\includegraphics[height=0.48\textwidth, angle=90]{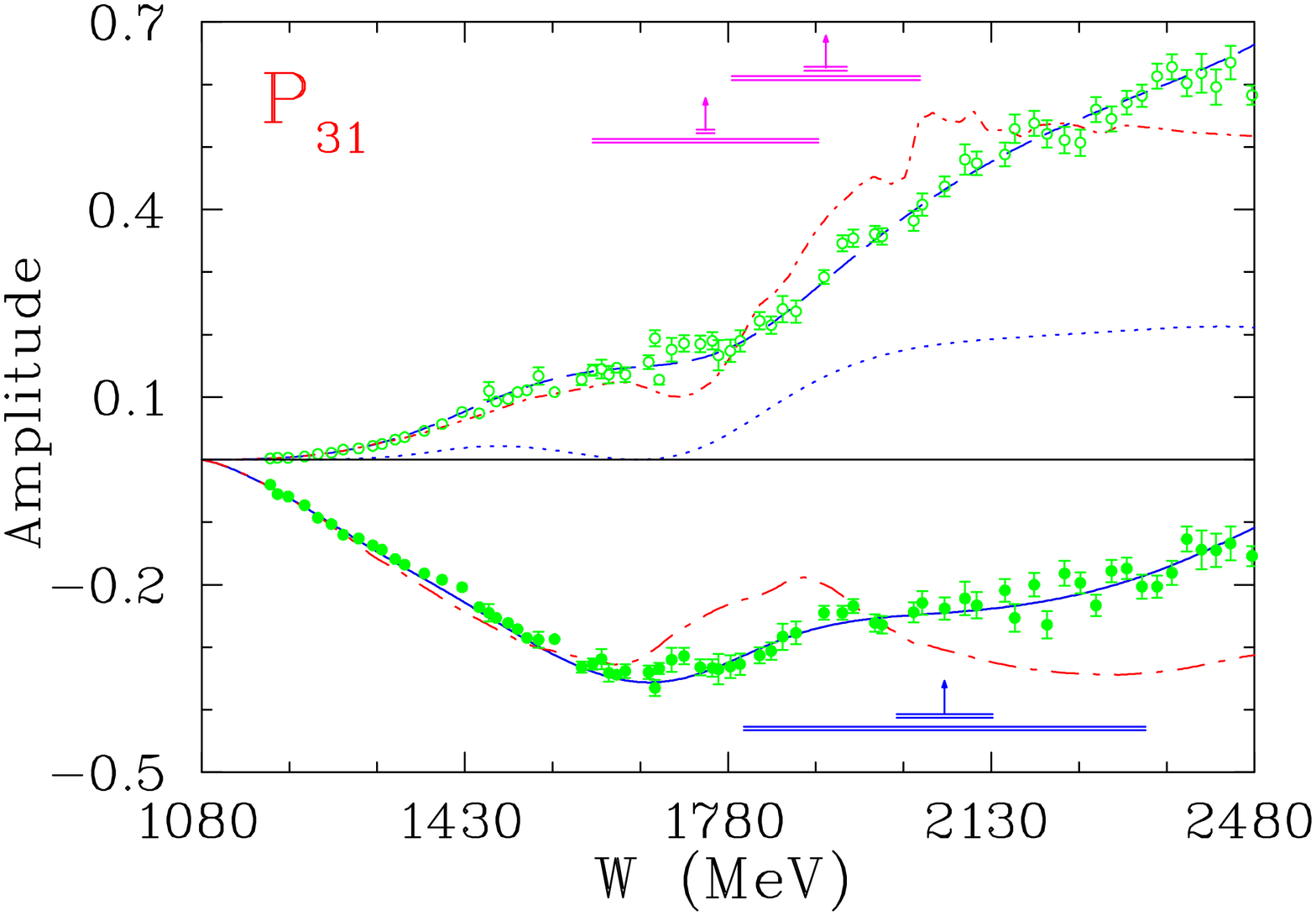}
\includegraphics[height=0.48\textwidth, angle=90]{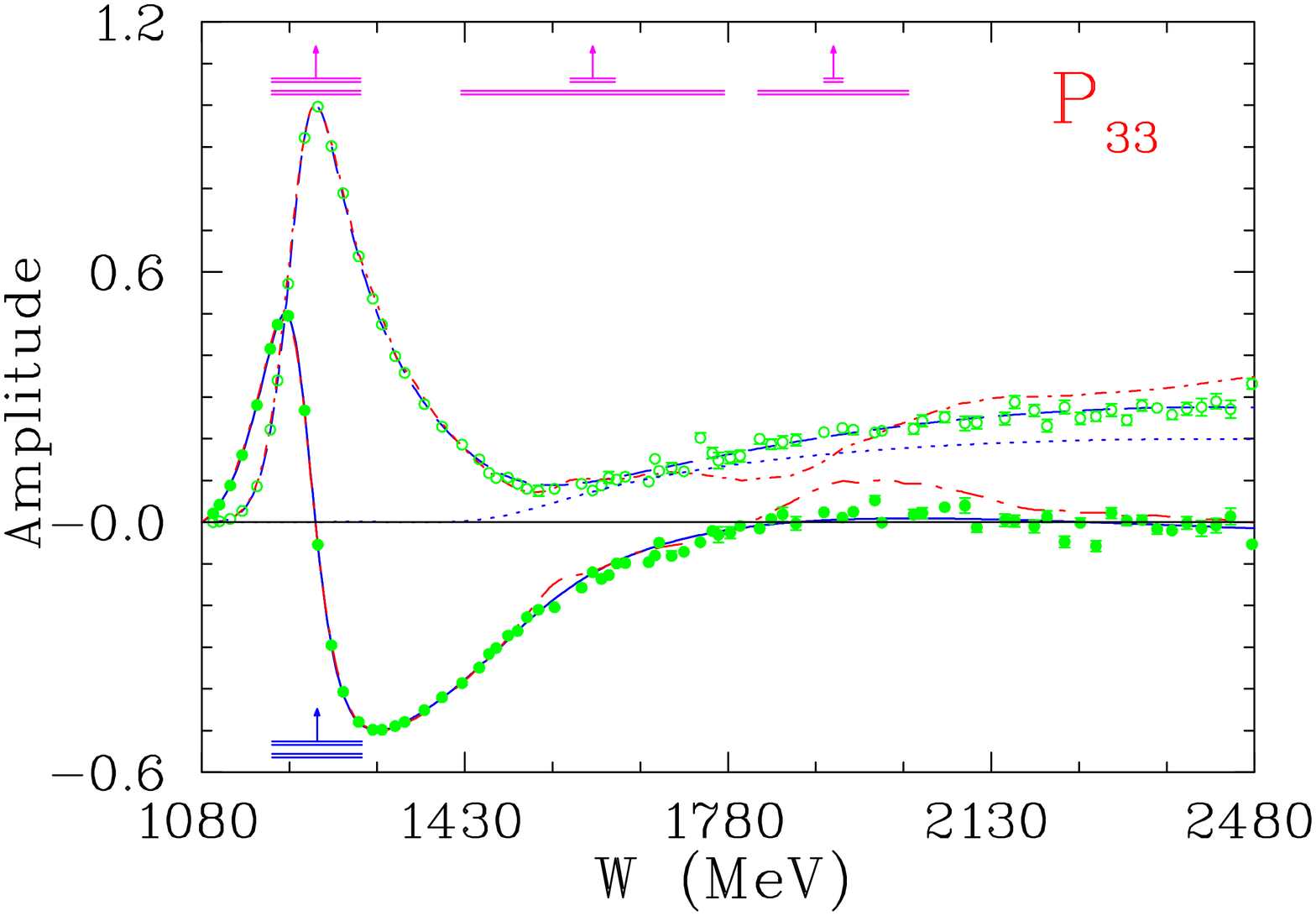}\hfill
\includegraphics[height=0.48\textwidth, angle=90]{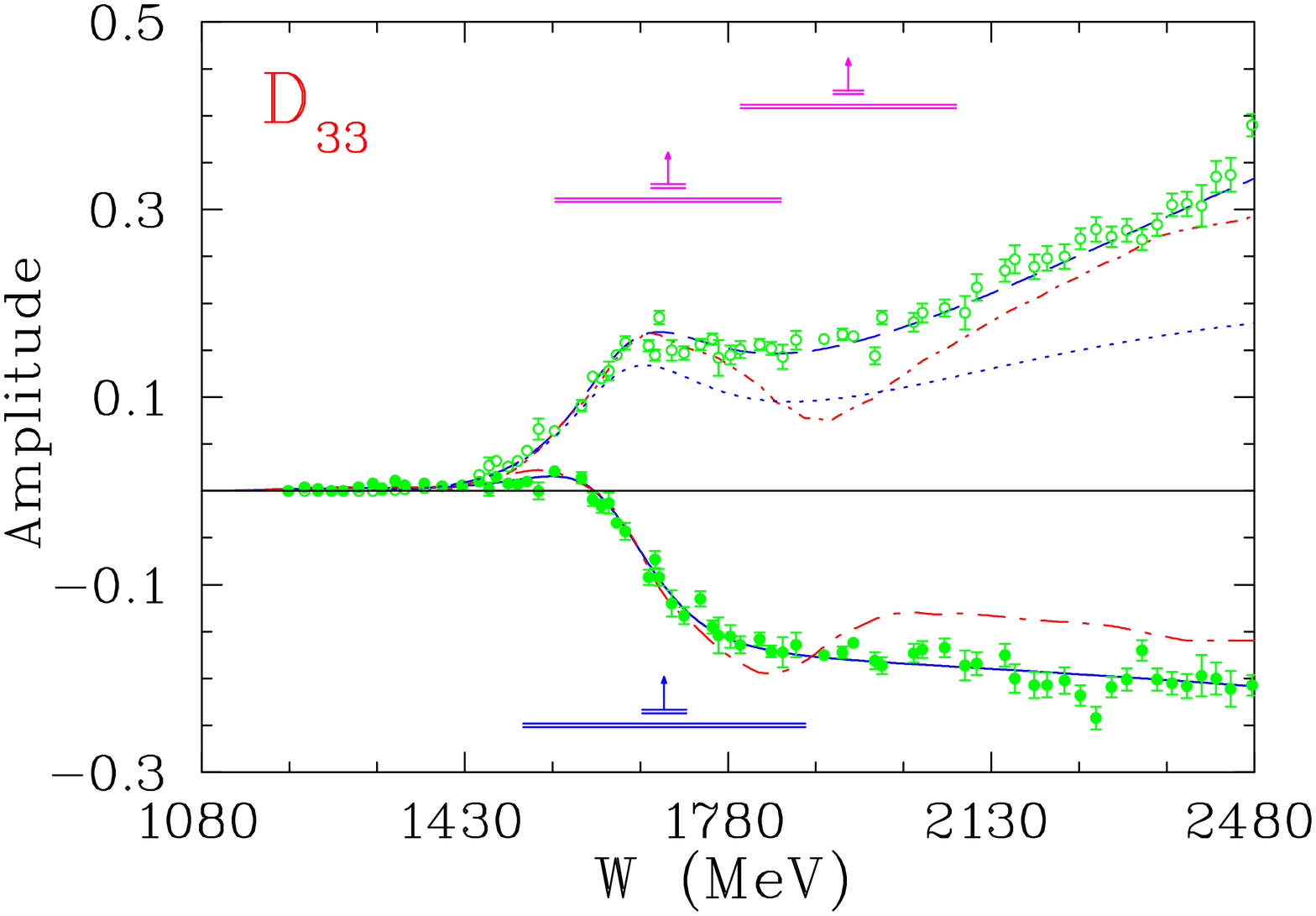}
\includegraphics[height=0.48\textwidth, angle=90]{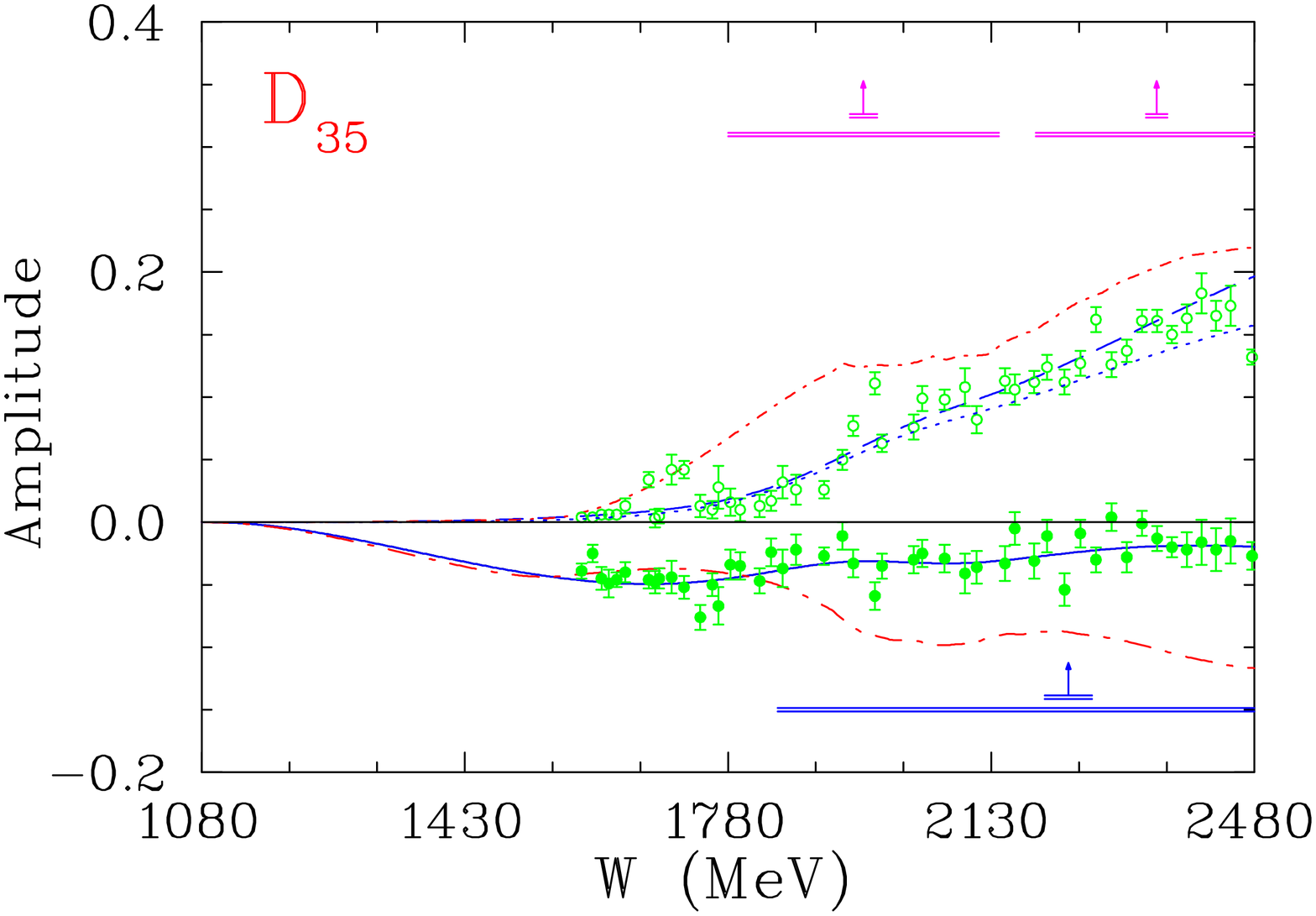}\hfill
\includegraphics[height=0.48\textwidth, angle=90]{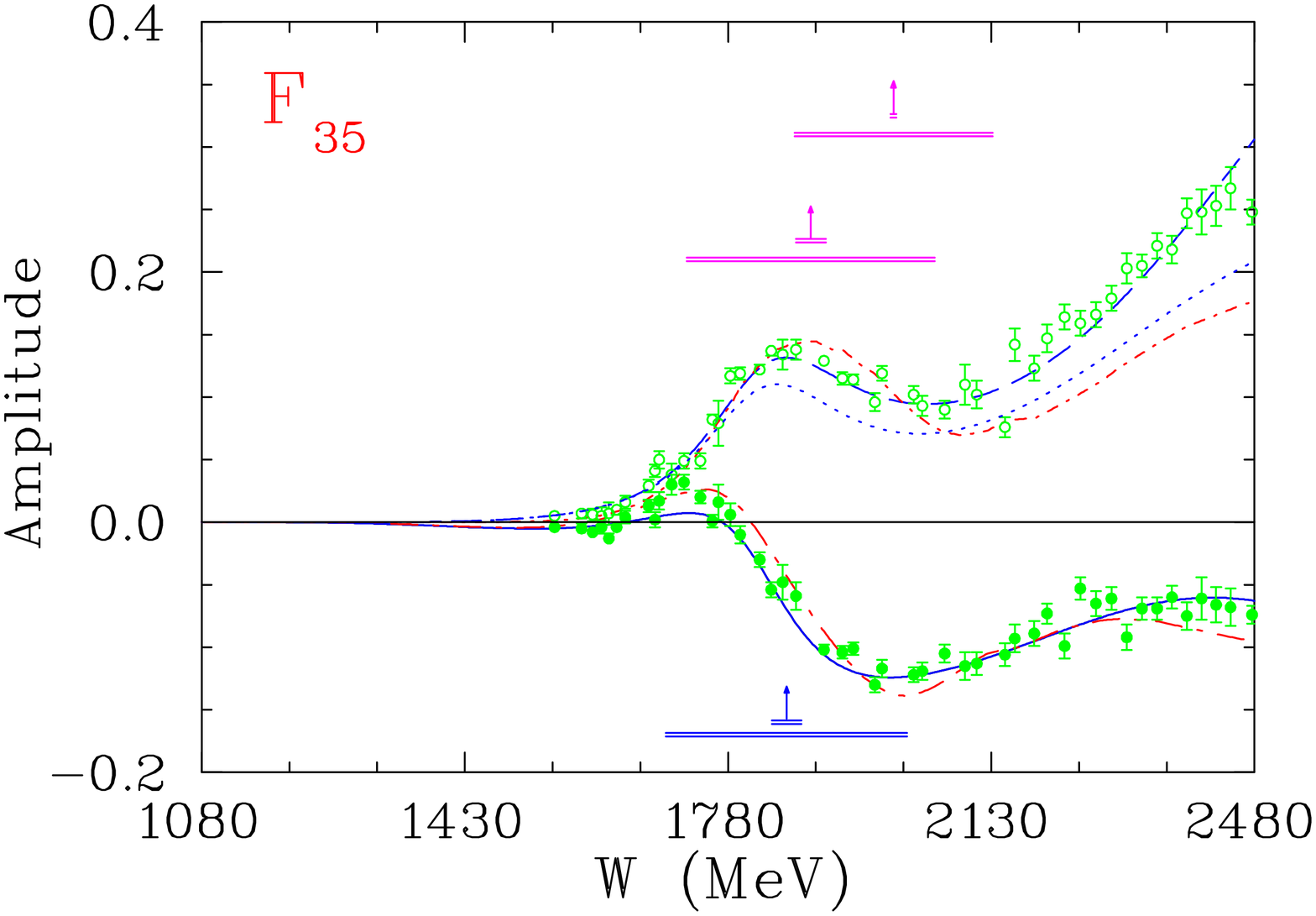}
}\caption{Isospin $3/2$ partial-wave amplitudes 
          $J < 3$ ($L_{2I, 2J}$) from $T_{\pi}$ = 0 
          to 2.6~GeV.  The lower BW resonances are
          associated with the SP06 values of 
          Table~\protect\ref{tbl6}; upper symbols 
          give RPP~\protect\cite{rpp} values. 
          Notation as in Fig.~\protect\ref{fig:g4}.   
          \label{fig:g6}}
\end{figure}

\begin{figure}[th]
\centering{
\includegraphics[height=0.42\textwidth, angle=90]{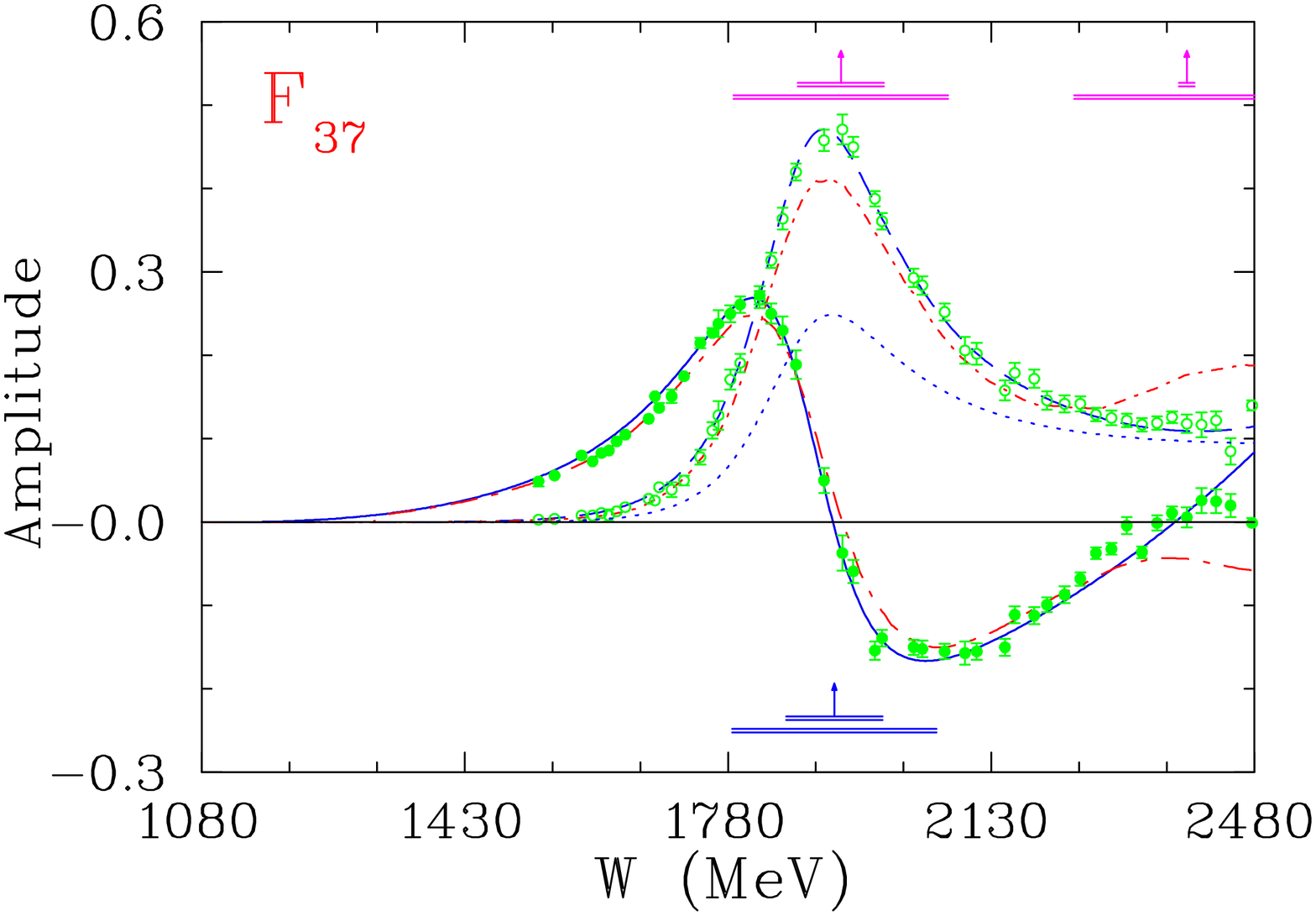}\hfill
\includegraphics[height=0.42\textwidth, angle=90]{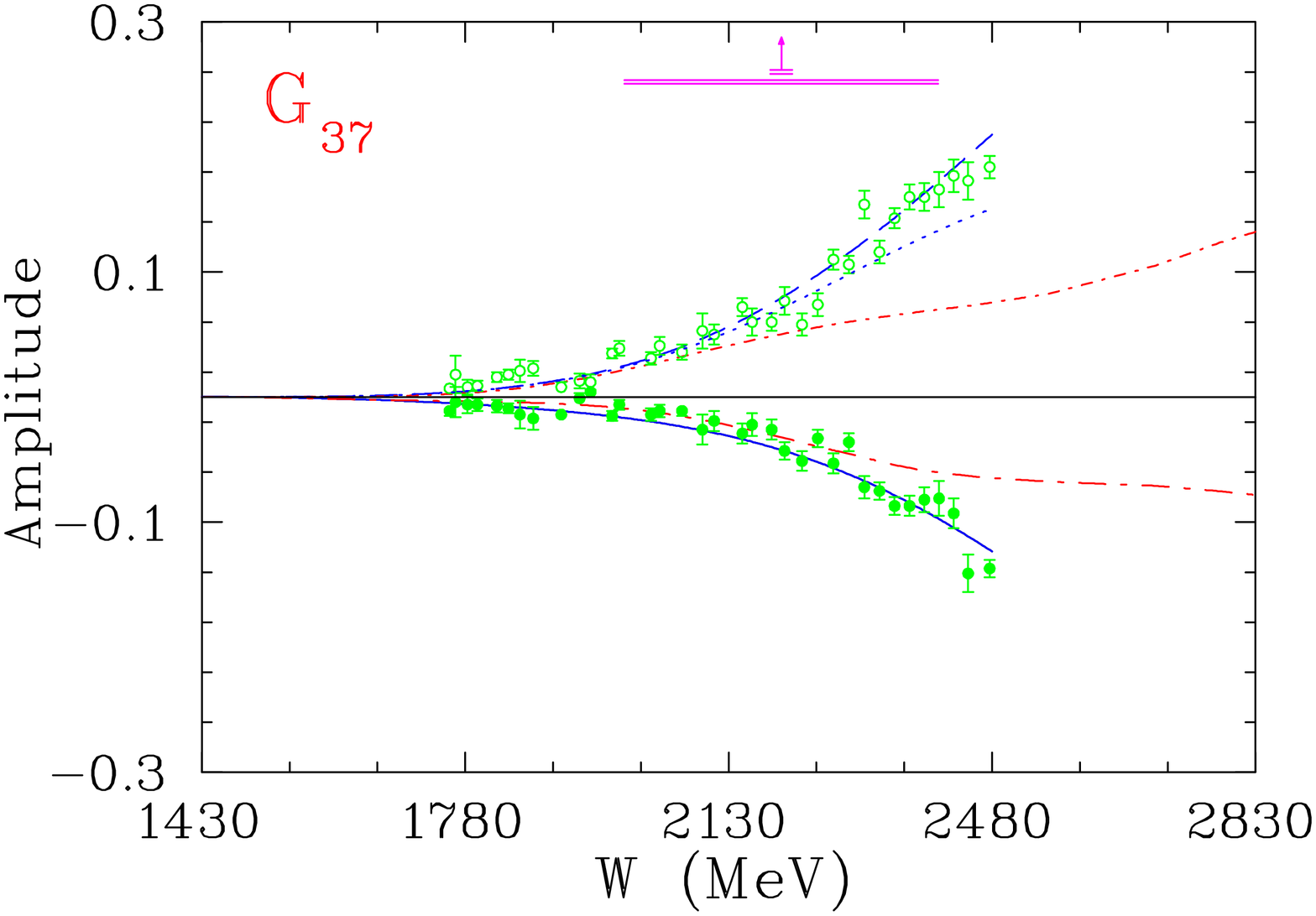}
\includegraphics[height=0.42\textwidth, angle=90]{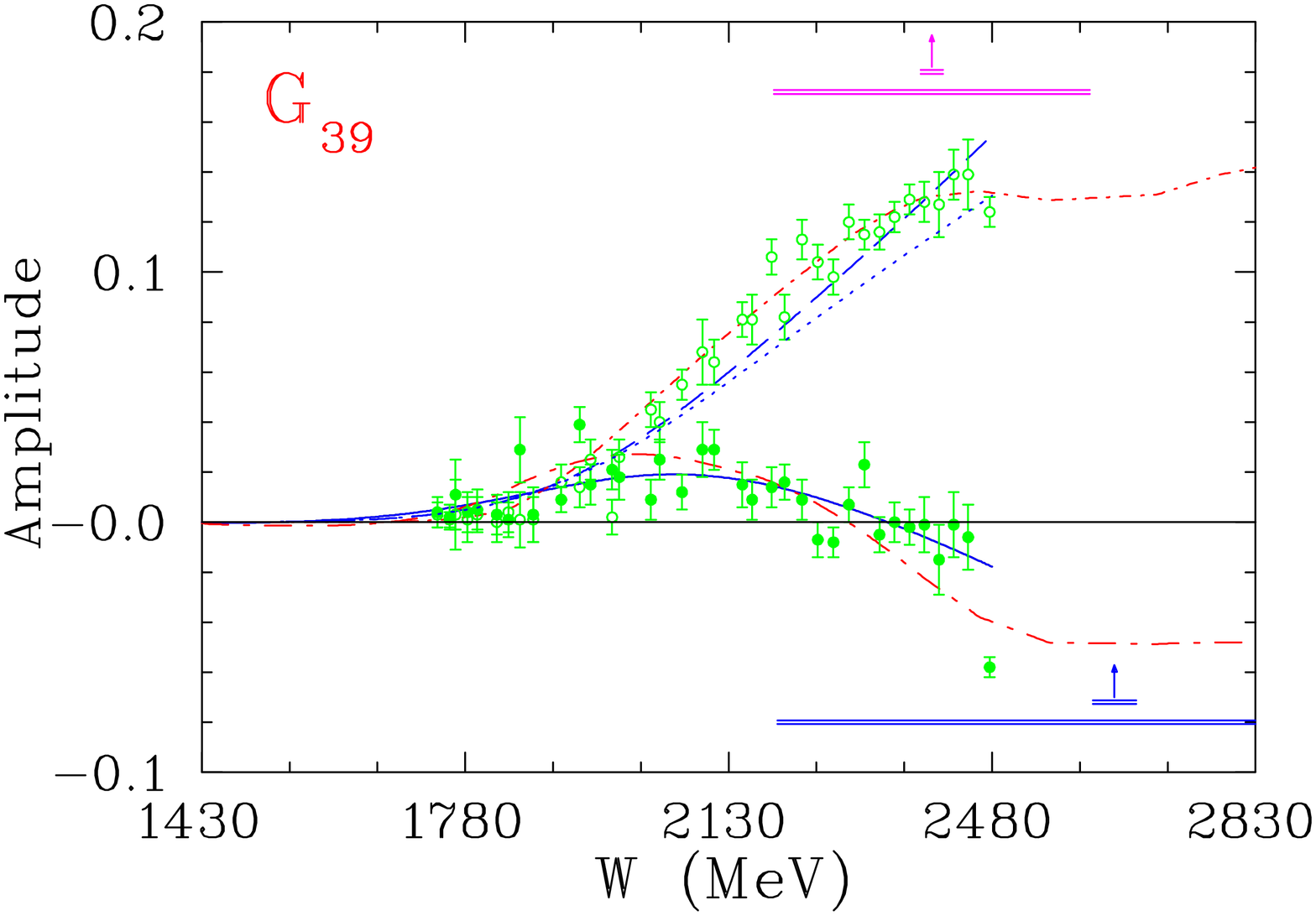}\hfill
\includegraphics[height=0.42\textwidth, angle=90]{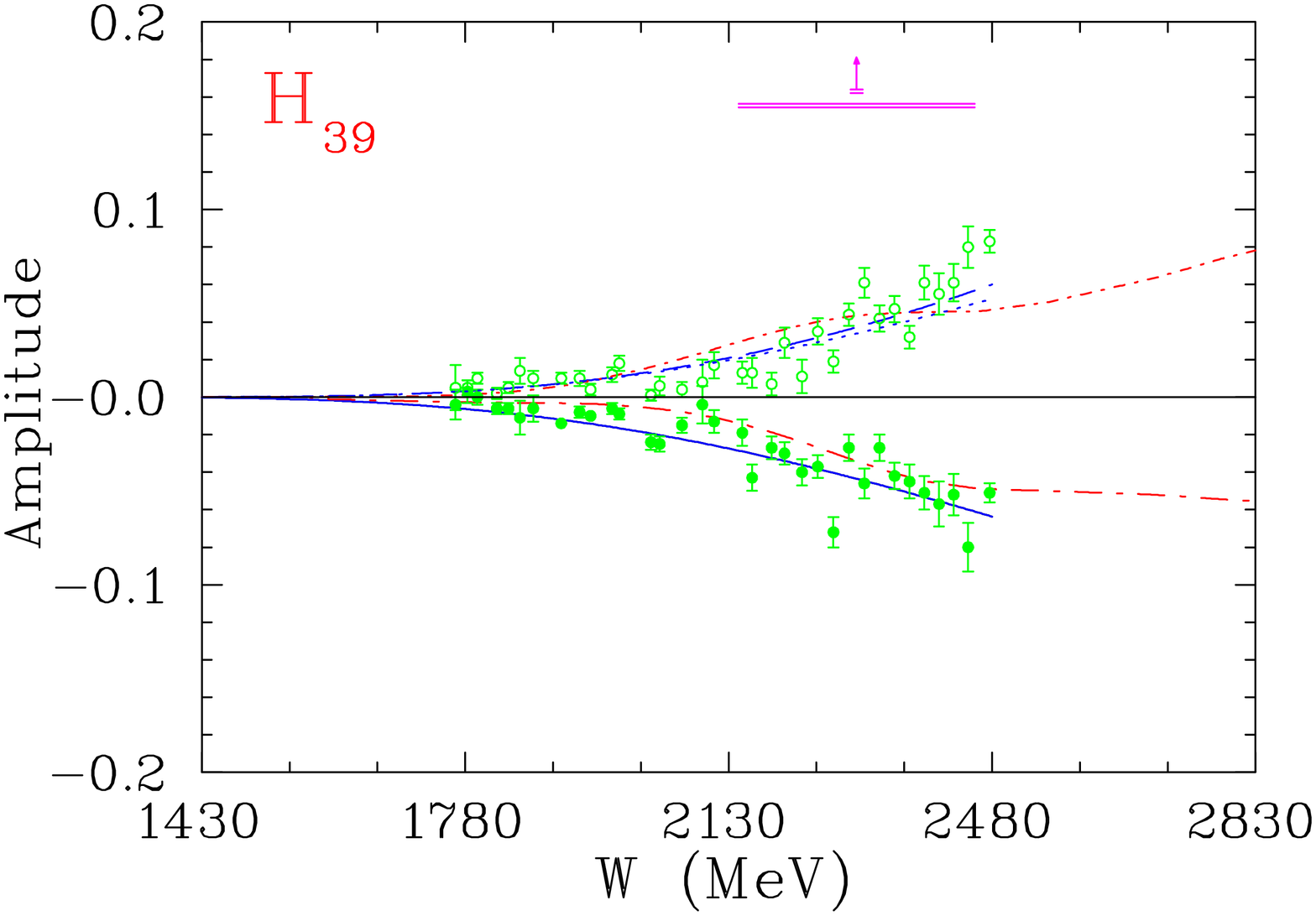}
\includegraphics[height=0.42\textwidth, angle=90]{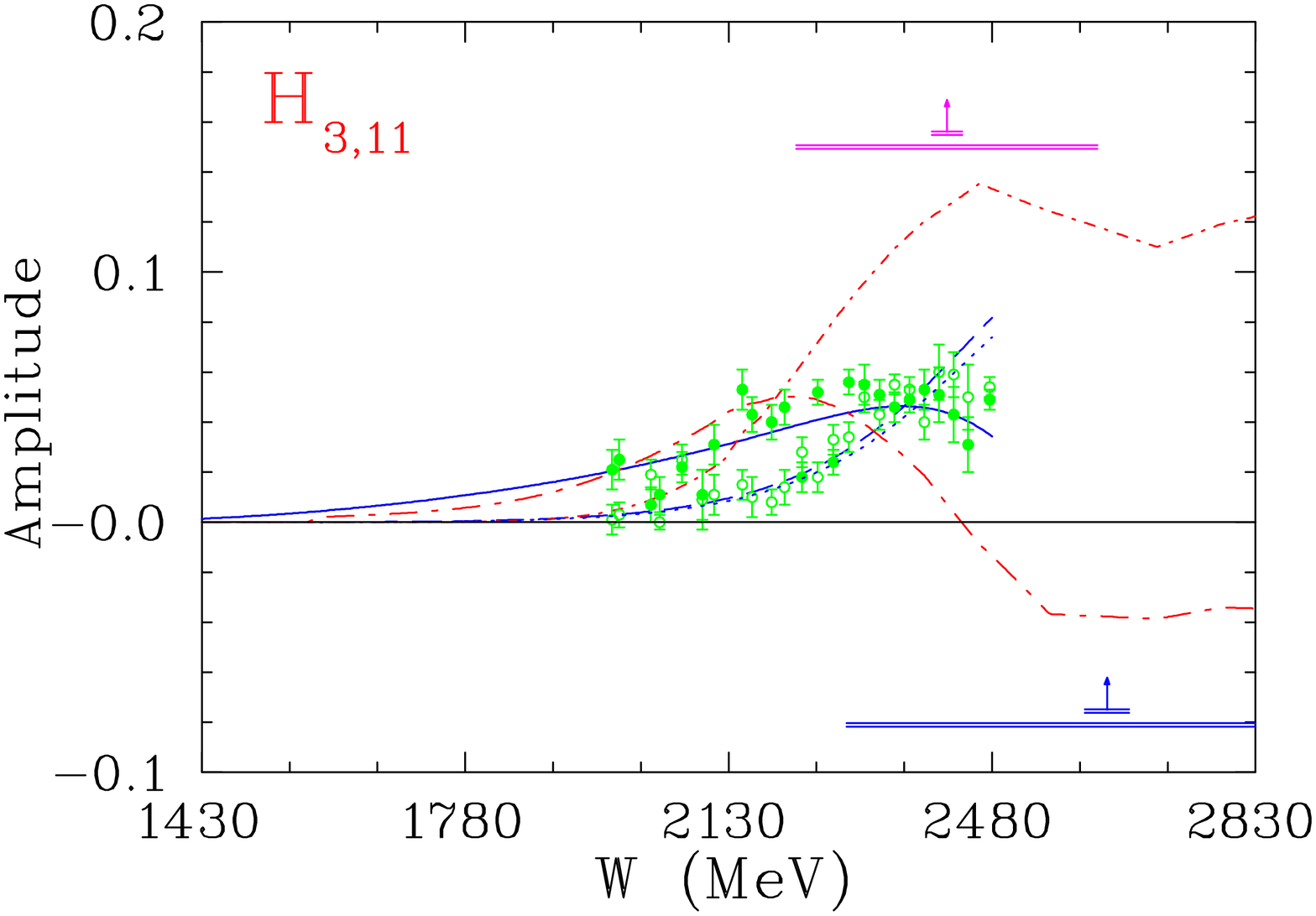}\hfill
\includegraphics[height=0.42\textwidth, angle=90]{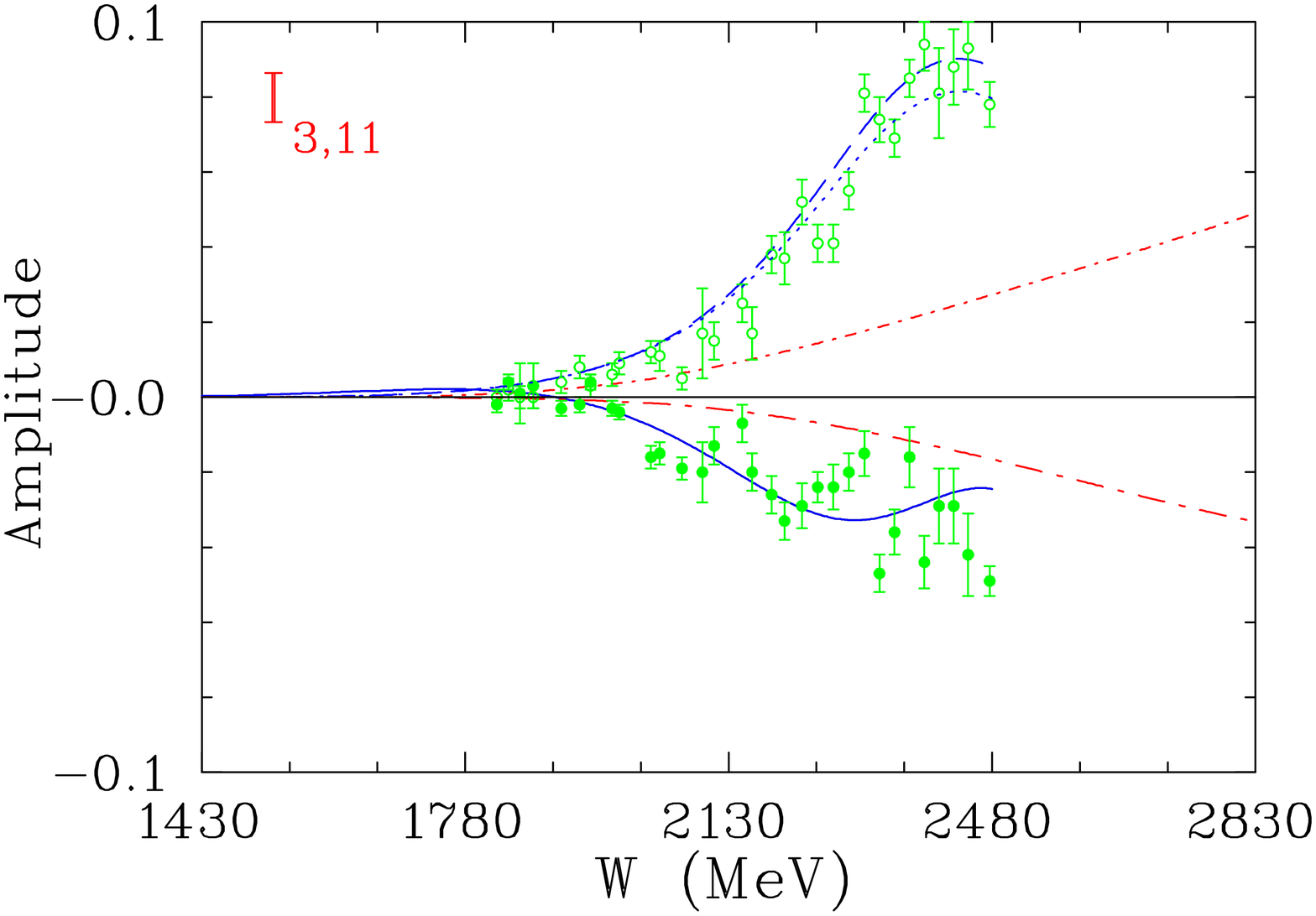}
\includegraphics[height=0.42\textwidth, angle=90]{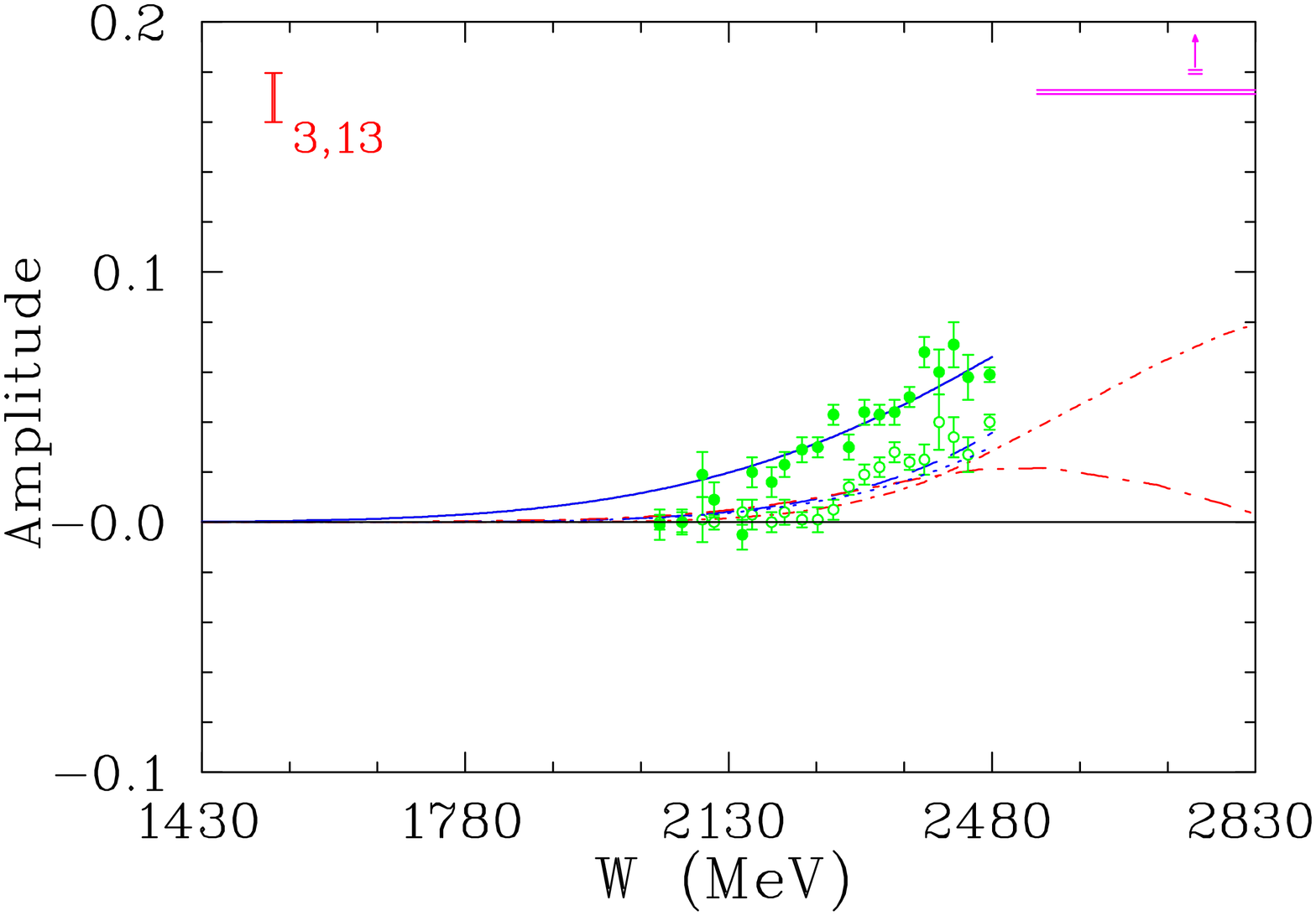}\hfill
}\caption{Isospin $3/2$ partial-wave amplitudes
          $J > 3$ ($L_{2I, 2J}$) from $T_{\pi}$ = 0 
          to 2.6~GeV.  Notation as in Fig.
          ~\protect\ref{fig:g6}. \label{fig:g7}}
\end{figure}

Some resonance structures, at the limit of our FA02 
analysis, are now better defined, while new structures 
have appeared in the G-, H- and I-waves. Both the SP06 
and KH solutions are reasonably well within the spread 
of the isospin $1/2$ SES (as shown in Figs.~\ref{fig:g4}
and~\ref{fig:g5}).  However, the KH solutions are less 
smooth suggesting the existence of additional 
resonances weakly coupled to the $\pi N$ channel. In 
our opinion, such states should be established in 
reactions where they couple more strongly; the $\pi N$ 
database can be fitted without these additional 
resonances.  A comparison of SP06 and KH partial waves 
with isospin $3/2$ is more interesting.  The $P_{31}$, 
$D_{33}$, and $D_{35}$ waves show large deviations, 
some of which have been qualitatively explained in 
Ref.~\cite{alekseev}. In other isospin $3/2$ waves 
there is better agreement. For example, the 2-star 
$\Delta(2400)G_{39}$ appears at the upper end of our 
analysis, with a mass, width and elasticity in 
reasonable agreement with the KH values. 

\begin{figure}[th]
\centering{
\includegraphics[height=0.4\textwidth, angle=90]{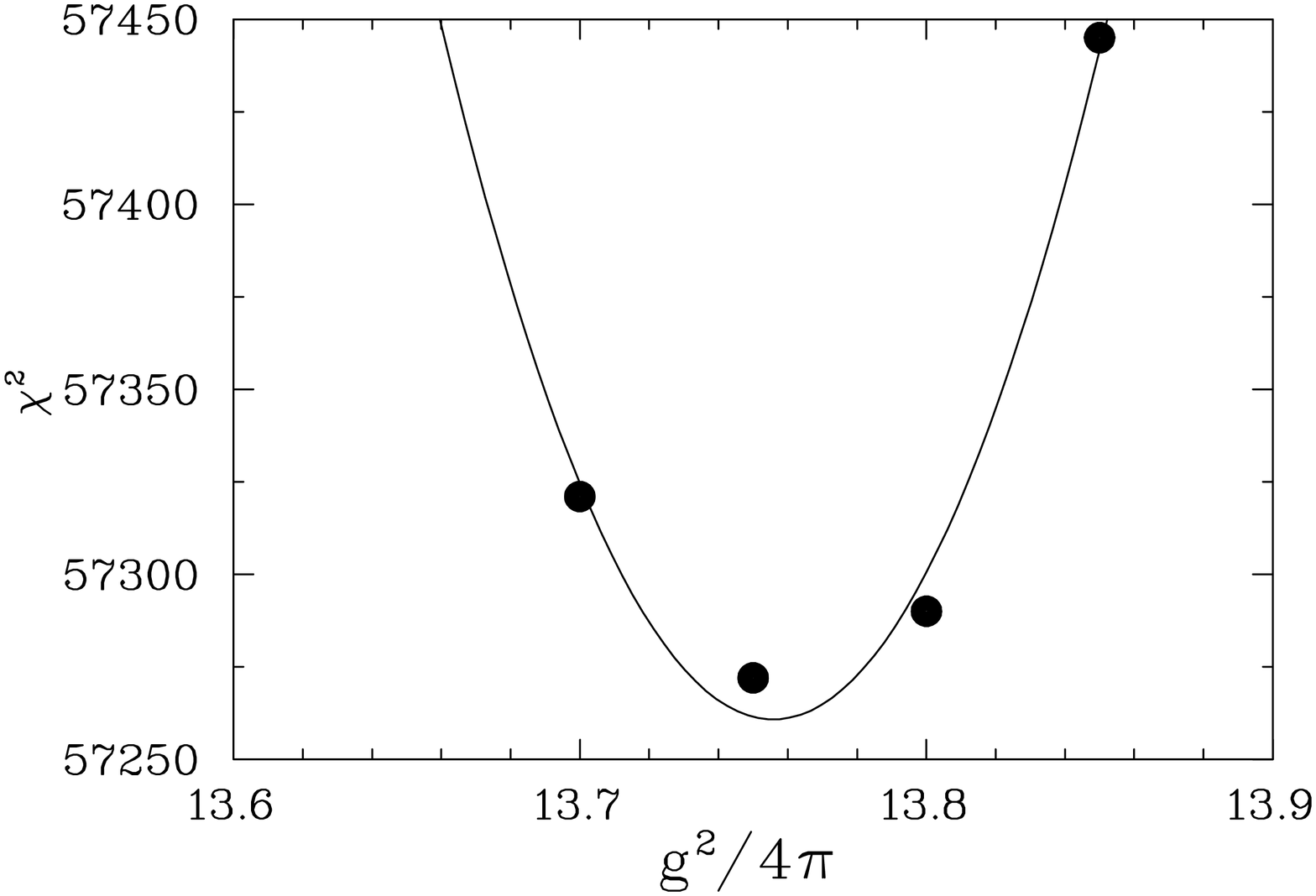}
}\caption{Best-fit $\chi^2$ as a function of the
          coupling constant $g^2/4\pi$, where
          all other parameters were fixed to
          their optimal (best-fit) values.  The
          solid curve gives the best-fit of
          $\chi^2$ vs $g^2/4\pi$ assuming $\chi^2
          = a + (\frac{g^2/4\pi - b}{c})^2$, where
          $a$, $b$, and $c$ are free parameters.
          \label{fig:g11}}
\end{figure}

Other quantities of interest, such as the scattering 
lengths, the $\pi N$ coupling constant and sigma term 
are consistent with values obtained in the FA02 fit. 
In Fig.~\ref{fig:g11} we show a quadratic fit to 
$\chi^2$ values from solutions with $g^2/4\pi$ ranging 
from 13.70 to 13.85, yeilding the value $g^2/4\pi$ = 
13.76$\pm$0.01, in agreement with the FA02 result.  
Finally, we note that the sigma 
term was extracted from our FA02 solution, using 
interior dispersion relations, in Ref.~\cite{hite}. We 
find this quantity has changed by less than 2~MeV between 
FA02 and SP06.

\acknowledgments

The authors express their gratitude to I.~Alekseev, 
B.~Johannes, R.~Meier, S.~Prakhov, G.~R.~Smith, 
A.~Starostin, D.~Svirida, and G.~Wagner for providing 
experimental data prior to publication or for 
clarification of information already published.  This 
work was supported in part by the U.~S.~Department of 
Energy under Grant DE--FG02--99ER41110.  The authors 
(R.~A., I.~S., and R.~W.) acknowledge partial support 
from Jefferson Lab and the Southeastern Universities 
Research Association under DOE contract 
DE--AC05--84ER40150.


\newpage
\begin{table}[th]
\caption{Comparison of present (SP06) and previous
         (FA02~\protect\cite{fa02}, SM95
         ~\protect\cite{sm95}, FA93~\protect\cite{fa93}, 
         SM90~\protect\cite{sm90}, and 
         FA84~\protect\cite{fa84}) energy-dependent 
         partial-wave analyses of elastic $\pi^\pm p$, 
         charge-exchange ($\pi^0n$), and $\pi^-p\to\eta 
         n$ ($\eta n$) scattering data.  For both SP06 
         and FA02 solutions, $\eta N$ data have been 
         included to 800~MeV.  The older Karlsruhe 
         KA84 and KH80 results~\protect\cite{kh80}
         are included 
         for comparison.  $N_{prm}$ is the number of 
         parameters ($I = 1/2$ and $3/2$) varied in 
         the fit. SP06$^\ast$ gives the SP06 result 
         evaluated over the energy range of our previous 
         fits. \label{tbl1}}
\begin{tabular}{ccccccc}
\colrule
Solution & Range~(MeV) &
$\chi^2$/$\pi^+p$ & $\chi^2$/$\pi^-p$ & $\chi^2$/$\pi^0n$ &
$\chi^2$/$\eta n$ & $N_{prm}$  \\
\colrule
SP06       &2600& 27155/13344 & 22702/11967& 6084/2933 & 626/257 & 93/81\\
KA84       &2600& 48394/13344 & 61845/11967& 9410/2933 &         &      \\
KH80       &2600& 32468/13344 & 40634/11967& 8005/2933 &         &      \\
SP06$^\ast$&2100& 22879/11842 & 18701/10561& 4945/2640 & 626/257 & 93/81\\
FA02       &2100& 21735/10468 & 18932/9650 & 4136/1690 & 439/173 & 86/70\\
SM95       &2100& 23593/10197 & 18855/9421 & 4442/1625 &         & 94/80\\
FA93       &2100& 23552/10106 & 20747/9304 & 4834/1668 &         & 83/77\\
SM90       &2100& 24897/10031 & 24293/9344 &10814/2132 &         & 76/68\\
FA84       &1100&  7416/ 3771 & 10658/4942 & 2062/ 717 &         & 64/57\\
\colrule
\end{tabular}
\end{table}
\begin{table}[th]
\caption{Single-energy (binned) fits of combined 
         elastic $\pi^\pm p$, charge-exchange, and 
         $\pi^-p\to\eta n$ scattering data.  
         $N_{prm}$ gives the number of parameters 
         varied in each single-energy fit and 
         $\chi^2_E$ is given by the energy-dependent 
         fit, SP06, over the same energy interval. 
         $\delta\chi^2 = 
         {[}\chi^2(SP06)-\chi^2(SES){]}$/data 
         quantifies the agreement between 
         individual SES and SP06. \label{tbl2}}
\begin{tabular}{cccccc|cccccc}
\colrule
T$_{\pi}$~(MeV)&Range~(MeV)&$N_{prm}$&$\chi^2$/data&$\chi^2_E$& $\delta\chi^2$ &
T$_{\pi}$~(MeV)&Range~(MeV)&$N_{prm}$&$\chi^2$/data&$\chi^2_E$& $\delta\chi^2$ \\
\colrule
 20&  19$-$  21&  4 & 163/ 85 & 194 & 0.36 & 930& 920$-$ 940& 28 & 338/287 & 534 & 0.68\\
 30&  26$-$  34&  4 & 291/231 & 329 & 0.16 & 960& 950$-$ 970& 32 & 350/332 & 570 & 0.66\\
 47&  45$-$  50&  4 & 181/124 & 238 & 0.46 &1000& 985$-$1015& 36 & 688/442 & 839 & 0.34\\
 66&  61$-$  70&  4 & 204/161 & 213 & 0.06 &1030&1020$-$1040& 38 & 533/400 & 661 & 0.32\\
 90&  87$-$  92&  4 & 126/121 & 149 & 0.19 &1045&1040$-$1050& 40 & 301/210 & 406 & 0.50\\
112& 107$-$ 117&  8 & 131/114 & 148 & 0.15 &1075&1070$-$1080& 40 & 220/217 & 402 & 0.84\\
124& 121$-$ 127&  8 &  82/ 63 & 101 & 0.30 &1100&1095$-$1105& 40 & 266/229 & 362 & 0.42\\
142& 139$-$ 146&  9 & 211/160 & 225 & 0.09 &1150&1140$-$1160& 42 & 665/446 & 863 & 0.44\\
170& 165$-$ 175&  9 & 174/163 & 200 & 0.16 &1180&1165$-$1185& 44 & 644/444 & 801 & 0.35\\
193& 191$-$ 195&  9 &  97/107 & 117 & 0.19 &1210&1200$-$1220& 44 & 299/274 & 378 & 0.29\\
217& 214$-$ 221&  9 & 106/109 & 145 & 0.36 &1245&1230$-$1260& 44 & 690/420 & 830 & 0.33\\
238& 235$-$ 241&  9 & 111/115 & 143 & 0.28 &1320&1300$-$1340& 46 & 824/567 &1036 & 0.37\\
266& 263$-$ 271&  9 & 152/123 & 181 & 0.24 &1370&1365$-$1375& 46 & 456/286 & 668 & 0.74\\
292& 291$-$ 294& 10 & 155/129 & 208 & 0.41 &1400&1385$-$1415& 46 & 587/423 & 871 & 0.67\\
309& 306$-$ 311& 10 & 158/140 & 180 & 0.16 &1460&1450$-$1470& 50 & 878/562 &1377 & 0.89\\
334& 332$-$ 336& 11 &  93/ 58 & 139 & 0.79 &1480&1465$-$1495& 50 & 626/409 & 861 & 0.57\\
352& 351$-$ 352& 11 &  64/109 &  84 & 0.18 &1570&1555$-$1585& 54 & 568/478 & 826 & 0.54\\
390& 370$-$ 410& 11 & 259/119 & 318 & 0.50 &1595&1580$-$1610& 55 & 507/405 & 755 & 0.61\\
425& 420$-$ 430& 12 & 170/162 & 215 & 0.28 &1660&1645$-$1675& 56 & 695/496 & 976 & 0.57\\
465& 450$-$ 480& 14 & 266/178 & 358 & 0.52 &1720&1705$-$1735& 58 & 391/286 & 511 & 0.42\\
500& 490$-$ 510& 15 & 382/245 & 444 & 0.25 &1755&1740$-$1770& 58 & 716/457 & 880 & 0.36\\
520& 511$-$ 529& 17 & 132/125 & 176 & 0.35 &1840&1825$-$1855& 58 & 423/323 & 741 & 0.98\\
535& 530$-$ 540& 19 & 270/247 & 321 & 0.21 &1870&1860$-$1880& 58 & 642/441 &1005 & 0.82\\
560& 555$-$ 565& 20 & 387/270 & 601 & 0.79 &1930&1915$-$1945& 58 & 757/549 &1021 & 0.48\\
580& 570$-$ 590& 20 & 439/401 & 542 & 0.26 &1970&1960$-$1980& 58 & 532/271 & 730 & 0.73\\
600& 595$-$ 605& 20 & 275/274 & 414 & 0.51 &2025&2010$-$2040& 58 & 397/339 & 714 & 0.94\\
625& 620$-$ 630& 21 & 182/164 & 234 & 0.32 &2075&2050$-$2100& 58 & 928/425 &1270 & 0.80\\
660& 645$-$ 675& 23 & 573/426 & 727 & 0.36 &2125&2100$-$2150& 58 & 773/492 &1366 & 1.21\\
720& 700$-$ 740& 26 & 383/307 & 597 & 0.70 &2175&2150$-$2200& 58 &1025/486 &1373 & 0.72\\
745& 735$-$ 755& 26 & 362/257 & 609 & 0.96 &2225&2200$-$2250& 58 & 915/513 &1299 & 0.75\\
765& 755$-$ 775& 26 & 375/381 & 549 & 0.46 &2275&2250$-$2300& 58 & 473/271 & 704 & 0.85\\
782& 776$-$ 788& 27 & 170/116 & 353 & 0.72 &2325&2300$-$2350& 58 & 662/419 & 870 & 0.50\\
800& 790$-$ 810& 27 & 634/441 & 747 & 0.26 &2375&2350$-$2400& 58 & 602/388 & 950 & 0.90\\
820& 813$-$ 827& 28 & 431/393 & 518 & 0.22 &2425&2400$-$2450& 58 & 205/186 & 679 & 2.55\\
875& 865$-$ 885& 28 & 661/444 & 880 & 0.49 &2475&2450$-$2500& 58 & 192/136 & 372 & 1.32\\
890& 886$-$ 894& 28 & 238/203 & 456 & 1.07 &2525&2500$-$2550& 58 & 497/171 & 889 & 2.29\\
900& 895$-$ 905& 28 & 515/409 & 776 & 0.64 &2575&2550$-$2600& 58 & 385/139 & 911 & 3.78\\
\colrule
\end{tabular}
\end{table}
\begin{table}[th]
\caption{Comparison of $\chi^2$/data for normalized
         (Norm) and unnormalized (Unnorm) data used
         in the SP06 and FA02~\protect\cite{fa02}
         solutions.  Karlsruhe KA84 and KH80 
         results~\protect\cite{kh80} are
         included for comparison.  Values for FA02
         correspond to a 2.1~GeV energy limit.  SP06,
         KH80, and KA84 are evaluated up to 2.6~GeV.
         \label{tbl3}}
\begin{tabular}{ccccccccc}
\colrule
Reaction          & SP06&      & FA02 &      & KA84 &      & KH80 &      \\
                  & Norm&Unnorm& Norm &Unnorm& Norm &Unnorm& Norm &Unnorm\\
\colrule
$\pi^+p\to\pi^+p$ & 2.0 &  6.7 & 2.1  & 9.3  & 3.6  & 10.0 & 2.4  &  8.5 \\
$\pi^-p\to\pi^-p$ & 1.9 &  6.2 & 2.0  & 7.1  & 5.2  & 13.0 & 3.4  & 10.2 \\
$\pi^-p\to\pi^0n$ & 2.1 &  4.5 & 2.4  & 9.5  & 3.2  &  7.8 & 2.7  &  5.9 \\
$\pi^-p\to\eta n$ & 2.4 & 10.1 & 2.5  & 4.6  &      &      &      &      \\
\colrule
\end{tabular}
\end{table}
\begin{table}[th]
\caption{Pole positions from the solution SP06,
         our previous solution 
         FA02~\protect\cite{fa02}, and a range 
         from the Particle Data Group 
         {[}RPP{]}~\protect\cite{rpp} (in
         square brackets).  Real ($W_R$) and
         imaginary ($-2 W_I$) parts are listed 
         for isospin $1/2$ baryon resonances.
         The second sheet pole is
         labeled by a $\dagger$.  Modulus and
         phase values are listed for the $\pi
         N$ elastic pole residue. \label{tbl4}}
\begin{tabular}{cccccc}
\colrule
Wave     & $W_R$ &$-2W_I$&Modulus& Phase  & Ref \\
         & (MeV) & (MeV) & (MeV) & (deg)  &     \\
\colrule
$S_{11}$ & 1502  &   95  &  16   & $-$16  & SP06\\
         & 1526  &  130  &  33   & $+$14  & FA02\\
         &[1490$-$1530]  &[90$-$250]&  &  & RPP\\
$S_{11}$ & 1648  &   80  &  14   & $-$69  & SP06\\
         & 1653  &  182  &  69   & $-$55  & FA02\\
         &[1640$-$1670]  &[150$-$180]&  & & RPP\\
$P_{11}$ & 1359  &  162  &  38   & $-$98  & SP06\\
         & 1357  &  160  &  36   &$-$102  & FA02\\
         &[1350$-$1380]  &[160$-$220]&  & & RPP\\
$P_{11}^\dagger$
         & 1388  &  165  &  86   & $-$46  & SP06\\
         & 1385  &  166  &  82   & $-$51  & FA02\\
         &       &       &       &        & RPP\\
$P_{13}$ & 1666  &  355  &  25   & $-$94  & SP06\\
         & 1655  &  278  &  20   & $-$88  & FA02\\
         &[1660$-$1690]  &[115$-$275]& &  & RPP\\
$D_{13}$ & 1515  &  113  &  38   & $-$5   & SP06\\
         & 1514  &  102  &  35   & $-$6   & FA02\\
         &[1505$-$1515]  &[105$-$120]& &  & RPP\\
$D_{15}$ & 1657  &  139  &  27   & $-$21  & SP06\\
         & 1659  &  146  &  29   & $-$22  & FA02\\
         &[1655$-$1665]  &[125$-$150]&  & & RPP\\
$F_{15}$ & 1674  &  115  &  42   & $-$4   & SP06\\
         & 1678  &  120  &  43   & $+$1   & FA02\\
         &[1665$-$1680]  &[110$-$135]& &  & RPP\\
$F_{15}$ & 1807  &  109  &  60   &  $-$67 & SP06\\
         &       &       &       &        & FA02\\
         &       &       &       &        & RPP\\
$G_{17}$ & 2070  &  520  &  72   & $-$32  & SP06\\
         & 2076  &  502  &  68   & $-$32  & FA02\\
         &[2050$-$2100]  &[400$-$520]&  & & RPP\\
$H_{19}$ & 2199  &  372  &  33   & $-$33  & SP06\\
         & 2209  &  564  &  96   & $-$71  & FA02\\
         &[2130$-$2200]  &[400$-$560]&  & & RPP\\
$G_{19}$ & 2217  &  431  &  21   & $-$20  & SP06\\
         & 2238  &  536  &  33   & $-$25  & FA02\\
         &[2150$-$2250]  &[350$-$550]&  & & RPP\\
$H_{1,11}$& 2203 &  133  &   1   & $-$12  & SP06\\
         &       &       &       &        & FA02\\
         &       &       &       &        & RPP\\
\colrule
\end{tabular}
\end{table}
\begin{table}[th]
\caption{Pole positions for isospin $3/2$ baryon
         resonances.  Notation as
         in Table~\protect\ref{tbl4}. $^\dagger$A 
         second $P_{33}$ state was not reported 
         in FA02.  $^{\dagger\dagger}$No RPP 
         average given. \label{tbl5}}
\begin{tabular}{cccccc}
\colrule
Wave     & $W_R$ &$-2W_I$&Modulus& Phase  & Ref \\
         & (MeV) & (MeV) & (MeV) & (deg)  &     \\
\colrule
$S_{31}$ & 1595  &  135  &  15   &$-$92   & SP06\\
         & 1594  &  118  &  17   &$-$104  & FA02\\
         &[1590$-$1610]  &[115$-$120]&  & & RPP\\
$P_{31}$ & 1771  &  479  &  45   &$+$172  & SP06\\
         & 1748  &  524  &  48   &$+$158  & FA02\\
         &[1830$-$1880]  &[200$-$500]& &  & RPP\\
$P_{33}$ & 1211  &   99  &  52   & $-$47  & SP06\\
         & 1210  &  100  &  53   & $-$47  & FA02\\
         &[1209$-$1211]  &[98$-$102]&  &  & RPP\\
$P_{33}$ & 1457  &  400  &  44   &$+$147  & SP06\\
         &       &       &       &        & FA02$^\dagger$\\
         &[1500$-$1700]  &[200$-$400]& &  & RPP\\
$D_{33}$ & 1632  &  253  &  18   & $-$40  & SP06\\
         & 1617  &  226  &  16   & $-$47  & FA02\\
         &[1620$-$1680]  &[160$-$240]& &  & RPP\\
$D_{35}$ & 2001  &  387  &   7   & $-$12  & SP06\\
         & 1966  &  364  &  16   & $-$21  & FA02\\
         &[1840$-$1960]  &[175$-$360]&  & & RPP\\
$F_{35}$ & 1819  &  247  &  15   & $-$30  & SP06\\
         & 1825  &  270  &  16   & $-$25  & FA02\\
         &[1825$-$1835]  &[265$-$300]&  & & RPP\\
$F_{37}$ & 1876  &  227  &  53   & $-$31  & SP06\\
         & 1874  &  236  &  57   & $-$34  & FA02\\
         &[1870$-$1890]  &[220$-$260]&  & & RPP\\
$G_{39}$ & 1983  &  878  &  24   & $-$139 & SP06\\
         &       &       &       &        & FA02\\
         &       &       &       &        & RPP$^{\dagger\dagger}$\\
$H_{3,11}$& 2529 &  621  &  33   & $-$45  & SP06\\
         &       &       &       &        & FA02\\
         &[2260$-$2400]  &[350$-$750]&  & & RPP\\
\colrule
\end{tabular}
\end{table}
\begin{table}[th]
\caption{Resonance couplings from a Breit-Wigner fit 
         to the SP06 solution, our previous solution 
         FA02~\protect\cite{fa02}, and a range from 
         the {[}RPP{]}~\protect\cite{rpp} (in square 
         brackets).  Masses W$_R$, widths $\Gamma$, 
         and partial width $\Gamma_{\pi 
         N}$/$\Gamma$ are listed for isospin $1/2$ 
         baryon resonances. $\Gamma_{\pi 
         N}$/$\Gamma$ for $N(1650)S_{11}$ is not 
         varied in the BW fit. 
\label{tbl6}}
\begin{tabular}{ccccc}
\colrule
Resonance      & W$_R$ & $\Gamma$    & $\Gamma_{\pi N}/ \Gamma$ & Ref\\
               & (MeV) & (MeV)       &                          &    \\
\colrule
N(1440)$P_{11}$& 1485.0$\pm$1.2&284$\pm$18    &0.787$\pm$0.016    & SP06\\
               & 1468.0$\pm$4.5&360$\pm$26    &0.750$\pm$0.024    & FA02\\
               &[1420$-$1470]  &[200$-$450]   &[0.55$-$0.75]      & RPP\\
N(1520)$D_{13}$& 1514.5$\pm$0.2&103.6$\pm$0.4 &0.632$\pm$0.001    & SP06\\
               & 1516.3$\pm$0.8&98.6$\pm$2.6  &0.640$\pm$0.005    & FA02\\
               &[1515$-$1525]  &[100$-$125]   &[0.55$-$0.65]      & RPP\\
N(1535)$S_{11}$& 1547.0$\pm$0.7&188.4$\pm$3.8 &0.355$\pm$0.002    & SP06\\
               & 1546.7$\pm$2.2&178.0$\pm$11.6&0.360$\pm$0.009    & FA02\\
               &[1525$-$1545]  &[125$-$175]   &[0.35$-$0.55]      & RPP\\
N(1650)$S_{11}$& 1634.7$\pm$1.1&115.4$\pm$2.8 &1.000              & SP06\\
               & 1651.2$\pm$4.7&130.6$\pm$7.0 &1.000              & FA02\\
               &[1645$-$1670]  &[145$-$185]   &[0.60$-$0.95]      & RPP\\
N(1675)$D_{15}$& 1674.1$\pm$0.2&146.5$\pm$1.0 &0.393$\pm$0.001    & SP06\\
               & 1676.2$\pm$0.6&151.8$\pm$3.0 &0.400$\pm$0.002    & FA02\\
               &[1670$-$1680]  &[130$-$165]   &[0.35$-$0.45]      & RPP\\
N(1680)$F_{15}$& 1680.1$\pm$0.2&128.0$\pm$1.1 &0.701$\pm$0.001    & SP06\\
               & 1683.2$\pm$0.7&134.4$\pm$3.8 &0.670$\pm$0.004    & FA02\\
               &[1680$-$1690]  &[120$-$140]   &[0.65$-$0.70]      & RPP\\
N(1720)$P_{13}$& 1763.8$\pm$4.6&210$\pm$22    &0.094$\pm$0.005    & SP06\\
               & 1749.6$\pm$4.5&256$\pm$22    &0.190$\pm$0.004    & FA02\\
               &[1700$-$1750]  &[150$-$300]   &[0.10$-$0.20]      & RPP\\
N(2000)$F_{15}$& 1817.7        &117.6         &0.127              & SP06\\
               &               &              &                   & FA02\\
               &[2000]         &              &                   & RPP\\
N(2190)$G_{17}$& 2152.4$\pm$1.4&484$\pm$13    &0.238$\pm$0.001    & SP06\\
               & 2192.1$\pm$8.7&726$\pm$62    &0.230$\pm$0.002    & FA02\\
               &[2100$-$2200]  &[300$-$700]   &[0.1$-$0.2]        & RPP\\
N(2220)$H_{19}$& 2316.3$\pm$2.9&633$\pm$17    &0.246$\pm$0.001    & SP06\\
               & 2270$\pm$11   &366$\pm$42    &0.200$\pm$0.006    & FA02\\
               &[2200$-$2300]  &[350$-$500]   &[0.1$-$0.2]        & RPP\\
N(2245)$H_{1,11}$& 2247.2$\pm$6.2&225$\pm$23  &0.014$\pm$0.001    & SP06\\
               &               &              &                   & FA02\\
               &               &              &                   & RPP\\
N(2250)$G_{19}$& 2302$\pm$6    &628$\pm$28    &0.089$\pm$0.001    & SP06\\
               & 2376$\pm$43   &924$\pm$178   &0.110$\pm$0.004    & FA02\\
               &[2200$-$2350]  &[230$-$800]   &[0.05$-$0.15]      & RPP\\
N(2600)$I_{1,11}$&2623$\pm$197 &1311$\pm$996  &0.050$\pm$0.018    & SP06\\
               &               &              &                   & FA02\\
               &[2550$-$2750]  &[500$-$800]   &[0.05$-$0.10]      & RPP\\
\colrule
\end{tabular}
\end{table}
\begin{table}[th]
\caption{Parameters for isospin $3/2$ baryon 
         resonances.  Notation 
         as in Table~\protect\ref{tbl6}.  
         $\Gamma_{\pi N}$/$\Gamma$ for 
         $P_{33}(1232)$ is not varied in the 
         BW fit.  \label{tbl7}}
\begin{tabular}{ccccc}
\colrule
Resonance      & W$_R$ & $\Gamma$    & $\Gamma_{\pi N}/ \Gamma$ & Ref\\
               & (MeV) & (MeV)       &                          &    \\
\colrule
$\Delta(1232)P_{33}$& 1233.4$\pm$0.4 &118.7$\pm$0.6 &1.000            & SP06\\ 
                    & 1232.9$\pm$1.2 &118.0$\pm$2.2 &1.000            & FA02\\
                    &[1231$-$1233]   &[116$-$120]   &[1.0]            & RPP\\
$\Delta(1620)S_{31}$& 1615.2$\pm$0.4 &146.9$\pm$1.9 &0.315$\pm$0.001  & SP06\\
                    & 1614.1$\pm$1.1 &141.0$\pm$6.0 &0.310$\pm$0.004  & FA02\\
                    &[1600$-$1660]   &[135$-$150]   &[0.2$-$0.3]      & RPP\\
$\Delta(1700)D_{33}$& 1695.0$\pm$1.3 &375.5$\pm$7.0 &0.156$\pm$0.001  & SP06\\ 
                    & 1687.9$\pm$2.5 &364.8$\pm$16.6&0.150$\pm$0.001  & FA02\\
                    &[1670$-$1750]   &[200$-$400]   &[0.10$-$0.20]    & RPP\\
$\Delta(1905)F_{35}$& 1857.8$\pm$1.6 &320.6$\pm$8.6 &0.122$\pm$0.001  & SP06\\
                    & 1855.7$\pm$4.2 &334$\pm$22    &0.120$\pm$0.002  & FA02\\
                    &[1865$-$1915]   &[270$-$400]   &[0.09$-$0.15]    & RPP\\
$\Delta(1910)P_{31}$& 2067.9$\pm$1.7 &543.0$\pm$10.1&0.239$\pm$0.001  & SP06\\
                    & 2333$\pm$36    &1128$\pm$238  &0.390$\pm$0.019  & FA02\\
                    &[1870$-$1920]   &[190$-$270]   &[0.15$-$0.30]    & RPP\\
$\Delta(1930)D_{35}$& 2233$\pm$53    &773$\pm$187   &0.081$\pm$0.012  & SP06\\
                    & 2046$\pm$45    &402$\pm$198   &0.040$\pm$0.014  & FA02\\
                    &[1900$-$2020]   &[220$-$500]   &[0.05$-$0.15]    & RPP\\
$\Delta(1950)F_{37}$& 1921.3$\pm$0.2 &271.0$\pm$1.1 &0.471$\pm$0.001  & SP06\\
                    & 1923.3$\pm$0.5 &278.2$\pm$3.0 &0.480$\pm$0.002  & FA02\\
                    &[1915$-$1950]   &[235$-$335]   &[0.35$-$0.45]    & RPP\\
$\Delta(2400)G_{39}$& 2643$\pm$141   &895$\pm$432   &0.064$\pm$0.022  & SP06\\
                    &                &              &                 & FA02\\
                    &[2400]          &              &                 & RPP\\
$\Delta(2420)H_{3,11}$& 2633$\pm$29  &692$\pm$47    &0.085$\pm$0.008  & SP06\\
                    &                &              &                 & FA02\\
                    &[2300$-$2500]   &[300$-$500]   &[0.05$-$0.15]    & RPP\\
\colrule
\end{tabular}
\end{table}
\begin{table}[th]
\caption{Comparison of SP06 and BW plus background 
         representations for isospin $1/2$ baryon 
         resonance fits (see text and associated 
         Table~\protect\ref{tbl6}).  ``Data" refers 
         to the number of scattering data between
         Wmin and Wmax.  \label{tbl8}}
\begin{tabular}{cccccc}
\colrule
Resonance      & Wmin & Wmax & BW fit   & SP06     & Data\\
               & (MeV)& (MeV)& $\chi^2$ & $\chi^2$ &     \\
\colrule
$N(1440)P_{11}$& 1350 & 1550 & 5437     & 5377     & 3104\\
$N(1520)D_{13}$& 1480 & 1560 & 3350     & 3399     & 2068\\
$N(1535)S_{11}$& 1490 & 1590 & 3451     & 3481     & 2195\\
$N(1650)S_{11}$& 1620 & 1770 & 8658     & 8558     & 4678\\
$N(1675)D_{15}$& 1610 & 1730 & 7072     & 7093     & 3932\\
$N(1680)F_{15}$& 1620 & 1730 & 6326     & 6317     & 3443\\
$N(1720)P_{13}$& 1620 & 1820 &10701     &10743     & 5837\\
$N(2190)G_{17}$& 2050 & 2250 &10414     &10549     & 4908\\
$N(2220)H_{19}$& 2150 & 2350 &11649     &11690     & 4660\\
$N(2245)H_{1,11}$& 2050 & 2380 &17451     &17508     & 7573\\
$N(2250)G_{19}$& 2050 & 2350 &16073     &16095     & 6895\\
$N(2600)I_{1,11}$& 2070 & 2460 &19554     &19414     & 7590\\
\colrule
\end{tabular}
\end{table}
\begin{table}[th]
\caption{Comparison of SP06 and BW plus background 
         representations for isospin $3/2$ baryon 
         resonance fits (see text and associated 
         Table~\protect\ref{tbl7}).  \label{tbl9}}
\begin{tabular}{cccccc}
\colrule
Resonance      & Wmin & Wmax & BW fit   & SP06     & Data\\
               & (MeV)& (MeV)& $\chi^2$ & $\chi^2$ &     \\
\colrule
$\Delta(1232)P_{33}$& 1180 & 1270 & 1283     & 1278     & 1016\\
$\Delta(1620)S_{31}$& 1570 & 1680 & 4696     & 4715     & 2705\\
$\Delta(1700)D_{33}$& 1550 & 1750 & 9959     & 9992     & 5490\\
$\Delta(1905)F_{35}$& 1770 & 1920 & 7545     & 7567     & 4039\\
$\Delta(1910)P_{31}$& 1650 & 2150 &26540     &25363     &13258\\
$\Delta(1930)D_{35}$& 1770 & 2100 &16241     &16176     & 8442\\
$\Delta(1950)F_{37}$& 1800 & 2000 &10842     &10890     & 5437\\
$\Delta(2400)G_{39}$& 2140 & 2460 &16855     &16626     & 6134\\
$\Delta(2420)H_{3,11}$& 2150 & 2460 &16149   &16138     & 5970\\
\colrule
\end{tabular}
\end{table}

\begin{thebibliography}{99}
\bibitem{rpp} S.~Eidelman \textit{et al.} \textit{Review 
              of Particle Physics}, Phys.\ Lett.\ B\ 
              \textbf{592}, 1 (2004) and 2005 partial 
              update for edition 2006; 
	      \hbox{http://pdg.lbl.gov}. 
\bibitem{kh80}G.~H\"ohler, \textit{Pion--Nucleon
              Scattering}, Landoldt--B\"ornstein Vol.
              \textbf{I/9b2}, edited by H.~Schopper
              (Springer Verlag, 1983);
              R.~Koch, Z.\ Phys.\ C\ \textbf{29}, 597
              (1985); R.~Koch, private communication.  We
              have use Karlsruhe solution KA84, generated
              through a FORTRAN subroutine suplied by
              R.~Koch.
\bibitem{cmb} R.~E.~Cutkosky \textit{et al.}, Phys.\
              Rev.\ D\ \textbf{20}, 2839 (1979);
              R.~E.~Cutkosky in \textit{Proceedings of
              the 4th Conference ob Baryon Resonances,
              Toronto, 1980}, edited by N.~Isgur (World
              Scientific, Singapore, 1981), p.~19.
\bibitem{fa02}R.~A.~Arndt, W.~J.~Briscoe, I.~I.~Strakovsky, 
              R.~L.~Workman, and M.~M.~Pavan, Phys.\ Rev.\ 
              C\ \textbf{69}, 035213 (2004) 
              [nucl--th/0311089].
\bibitem{KHnew}P.~Piirola, E.~Pietarinen, and M.~E.~Sainio,
              $\pi N$ Newsletter \textbf{16}, 121 (2002)
              [hep--ph/0110044].
\bibitem{km_paper} R.~A.~Arndt, W.~J.~Briscoe, T.~W.~
              Morrison, I.~I.~Strakovsky, R.~L.~Workman,
              A.~B.~Gridnev, Phys.\ Rev.\ C\ \textbf{72},
              045202 (2005) [nucl--th/0507024].
\bibitem{sm95}R.~A.~Arndt, I.~I.~Strakovsky, R.~L.~
              Workman, and M.~M.~Pavan, Phys.\ Rev.\ C\
              \textbf{52}, 2120 (1995)
              [nucl--th/9505040].
\bibitem{fa93}R.~A.~Arndt, R.~L.~Workman, and M.~M.~
              Pavan, Phys.\ Rev.\ C\ \textbf{49}, 2729
              (1994).
\bibitem{sm90}R.~A.~Arndt, L.~Zhujun, L.~D.~Roper, R.~L.~
              Workman, and J.~M.~Ford, Phys.\ Rev.\ D\
              \textbf{43}, 2131 (1991).
\bibitem{fa84}R.~A.~Arndt, J.~M.~Ford, and L.~D.~Roper,
              Phys.\ Rev.\ D\ \textbf{32}, 1085 (1985).
\bibitem{hep} HEPDATA, The Durham RAL Database compiled
              by the Durham Database Group (UK) with help
              from the COMPAS group (Russia);
              \hbox{http://www.slac.stanford.edu/spires/hepdata/reac.html}.
\bibitem{flag}The 1-, 2-, and 3-star rating of data is
              described in Ref.~\protect\cite{sm90}.
              Only 2- and 3-star data are included in
              analyses.  Unrated data are included unless
              they have been {}``flagged\char`\"{} for
              deletion from analyses.  It should be noted
              that these flagged data are still retained
              in our database.  We have followed this
              recipe since 1990~\cite{flag1}.
\bibitem{flag1}Historically, B.~M.~K.~Nefkens,  
              W.~J.~Briscoe, M.~M.~Sadler, R.~A.~Arndt,
              and G.~H\"ohler analysed and classified  
              all $\pi N$ measurements completed before 
              1983.  0-, 1-, 2-, and 3-star ratings
              were assigned to the $\pi N$ database 
              entries.  This star rating system was 
              described at Few Body'83 and the 1st 
              $\pi N$ International Workshop by   
              B.~M.~K.~Nefkens, in {[}\textit{Few Body
              Problems in Physics (Contr.\ Papers) 
              (Proceedings of 10th International 
              Conference on Few Body Problems in 
              Physics, Karlsruhe, Germany, 1983)} edited 
              by B.~Zeitnitz, 1984, p.~137; Nucl.\ Phys.\ 
              \textbf{A416}, 193 (1984); Proceedings of 
              the First Workshop on $\pi N$ scattering,
              Karlsruhe, Germany, Aug.~1984, edited by 
              G.~H\"ohler{]}.
\bibitem{said}The full database and numerous PWAs can be
              accessed via an ssh call to the SAID
              facility \hbox{gwdac.phys.gwu.edu}, with
              userid: said (no password), or a link to
              the website \hbox{http://gwdac.phys.gwu.edu}.
\bibitem{de06}H.~Denz \textit{et al.} (CHAOS Collaboration),
              Phys.\ Lett.\ B\ \textbf{633}, 209 (2006)
              [nucl--ex/0512006];
\bibitem{me04}R.~Meier \textit{et al.}, Phys.\ Lett.\ B\
              \textbf{588}, 155 (2004) [nucl--ex/0402018].
\bibitem{psi} J.~Breitschopf \textit{et al.},
              nucl-ex/0605017.
\bibitem{st05}A.~Starostin \textit{et al.} (Crystal Ball 
              Collaboration), Phys.\ Rev.\ C\ \textbf{72},
              015205 (2005).
\bibitem{pr05}S.~Prakhov \textit{et al.} (Crystal Ball 
              Collaboration), Phys.\ Rev.\ C\ \textbf{72},
              015203 (2005).
\bibitem{al05}I.~G.~Alekseev \textit{et al.}, Eur.\ Phys.\
              J\ C\ \textbf{45}, 383 (2005)
              [hep--ex/0510010].
\bibitem{al95}I.~G.~Alekseev \textit{et al.}, Phys.\ 
              Lett.\ B\ \textbf{351}, 585 (1995). 
\bibitem{alekseev}I.~G.~Alekseev, V.~P.~Kanavets, B.~V.~
              Morozov, D.~N.~Svirida, S.~P.~Kruglov, A.~A.
              ~Kulbardis, V.~V.~Sumachev, R.~A.~Arndt,
              I.~I.~Strakovsky, and R.~L.~Workman, Phys.\
              Rev.\ C\ \textbf{55}, 2049 (1997)
              [nucl--th/9608043].
\bibitem{pdgXX}G.~H\"ohler and R.~Workman in 
              Ref.~\protect\cite{rpp}.
\bibitem{hite}G.~E.~Hite, W.~B.~Kaufmann, and R.~J.~Jacob,
              Phys.\ Rev.\ C\ \textbf{71} 065201 (2005).
\end{thebibliography}
\end{document}